\documentclass[manuscript,screen]{acmart}

\usepackage{verbatim}
\usepackage{moresize}%Not supported for JACM

\usepackage{multicol}%Not supported for JACM
\usepackage{pifont}%Not supported for JACM
\newcommand{\cmark}{\text{\ding{51}}}
\newcommand{\xmark}{\text{\ding{55}}}

\usepackage{multirow}
\usepackage[inline]{enumitem}
\usepackage{bm}
\usepackage{graphicx}
\usepackage{array}
\usepackage{nicefrac}%Not supported for JACM
\usepackage{titlecaps}%Not supported for JACM
\usepackage{aliascnt}%Not supported for JACM
\usepackage{xpatch}%Not supported for JACM
\usepackage{amsthm}
\usepackage{hyperref}

\usepackage[ruled, linesnumbered, longend, lined, nofillcomment,noresetcount]{algorithm2e} % For algorithms
\SetAlFnt{\fontsize{8pt}{8pt}\selectfont}

\SetCommentSty{mycommentfont}

\newcommand{\addprefix}[1][]{%%
   \setcounter{AlgoLine}{0}
   \renewcommand{\theAlgoLine}{{#1}\arabic{AlgoLine}} %%
}

\newcommand{\tableFontSize}{\scriptsize}

\usepackage{tikz,circuitikz}%Not supported for JACM
\usetikzlibrary{shapes.geometric, shapes.multipart, arrows, chains, positioning, fit, backgrounds, decorations.pathreplacing, decorations.markings, calc}

\tikzstyle{qnode} = [rectangle split, rectangle split parts=2, rectangle split horizontal, draw=black]
\tikzstyle{node} = [draw=black]

\setcopyright{none}
\acmBooktitle{Adaptive and Fair Transformation for Recoverable Mutual Exclusion}

\usepackage{cleveref}%Not supported for JACM

\newtheorem{claim}{Claim}[section]

\newcommand{\bigO}[1]{\mathcal{O}(#1)}
\newcommand{\bigOmega}[1]{{\Omega}(#1)}
\newcommand{\smallOmega}[1]{\omega(#1)}

\newcommand{\smallO}[1]{o(#1)}
\newcommand{\card}[1]{\bigm| #1 \bigm|\xspace}

\newcommand{\CAS}{\textsf{CAS}}
\newcommand{\FAS}{\textsf{FAS}}

\newcommand{\F}{F}
\newcommand{\n}{n}
\newcommand{\pc}{\ddot{c}}

\newcommand{\NCS}{\texttt{NCS}}
\newcommand{\CS}{\texttt{CS}}
\newcommand{\Recover}{\texttt{Recover}}
\newcommand{\Enter}{\texttt{Enter}}
\newcommand{\Exit}{\texttt{Exit}}
\newcommand{\segment}{segment}

\newcommand{\consequence}{consequence}
\newcommand{\hazard}{failure-density}

\newcommand{\inprogress}{in-progress}

\newcommand{\true}{\textsf{true}}
\newcommand{\false}{\textsf{false}}

\newcommand{\PC}{PM}
\newcommand{\nonadaptive}{non-adaptive}
\newcommand{\cadaptive}{semi-adaptive}
\newcommand{\adaptive}{adaptive}
\newcommand{\sadaptive}{super-adaptive}
\newcommand{\unbounded}{unbounded}
\newcommand{\bounded}{bounded}
\newcommand{\wbounded}{well-bounded}
\newcommand{\fadaptive}{unbounded adaptive}
\newcommand{\badaptive}{bounded-adaptive}

\newcommand{\sefficient}{\wbounded{} \sadaptive}

\newcommand{\SEfficient}{\sefficient}

\newcommand{\sen}{sensitive}
\newcommand{\ins}{instruction}
\newcommand{\senins}{\sen{} \ins{}}

\newcommand{\responsive}{responsive}

\newcommand{\safe}{safe}
\newcommand{\unsafe}{unsafe}

\newcommand{\locality}{locality}

\newcommand{\mynull}{\textbf{null}}

\newcommand{\mytail}{tail}

\newcommand{\weakMCS}{\textsc{WR-Lock}}
\newcommand{\weakMCSFair}{\textsc{WR-Lock-MR}}

\newcommand{\FnEnter}{\textsc{Enter}}
\newcommand{\FnExit}{\textsc{Exit}}
\newcommand{\FnCleanup}{\textsc{Cleanup}}
\newcommand{\FnRecover}{\textsc{Recover}}

\newcommand{\InNCS}{\textsc{Free}}
\newcommand{\InRecover}{\textsc{Initializing}}
\newcommand{\InEnter}{\textsc{Trying}}
\newcommand{\InCS}{\textsc{InCS}}
\newcommand{\InExit}{\textsc{Leaving}}

\newcommand{\varstate}{\mathit{state}}
\newcommand{\varnode}{\mathit{mine}}
\newcommand{\varpred}{\mathit{pred}}
\newcommand{\varpath}{\mathit{type}}
\newcommand{\varside}{\mathit{side}}
\newcommand{\varnext}{\mathit{next}}
\newcommand{\varlocked}{\mathit{locked}}
\newcommand{\vartail}{\mathit{tail}}
\newcommand{\vartemp}{\mathit{result}}

\newcommand{\admissible}{admissible}

\newcommand{\inadmissible}{inadmissible}
\newcommand{\vital}{active} % Can be reconsidered
\newcommand{\relieve}{relieve} % Can be reconsidered
\newcommand{\successor}[1]{succ(#1)}
\newcommand{\owner}[1]{owner(#1)}
\newcommand{\link}[2]{edge(#1,#2)}
\newcommand{\front}{front\xspace}
\newcommand{\back}{rear\xspace}

\newcommand{\core}{core}
\newcommand{\splitter}{splitter}
\newcommand{\arbitrator}{arbitrator}
\newcommand{\filter}{filter}
\newcommand{\target}{target}

\newcommand{\FAST}{\textsc{FAST}}
\newcommand{\SLOW}{\textsc{SLOW}}

\newcommand{\LEFT}{\textsc{Left}}
\newcommand{\RIGHT}{\textsc{Right}}

\newcommand{\theleftside}{the {\LEFT} side}
\newcommand{\therightside}{the {\RIGHT} side}

\newcommand{\base}{base}

\newcommand{\fast}{fast}
\newcommand{\slow}{slow}
\newcommand{\medium}{medium-slow}
\newcommand{\normal}{normal}

\newcommand{\nalock}{\textsc{NA-Lock}}  %% Non-adaptive
\newcommand{\salock}{\textsc{SA-Lock}}  %% Semi-adaptive
\newcommand{\balock}{\textsc{BA-Lock}}  %% Bounded adaptive

\newcommand{\symfilter}{\mathcal{F}} %% symbol for filter lock
\newcommand{\symcore}{\mathcal{C}} %% symbol for core lock
\newcommand{\symarbitrator}{\mathcal{A}} %% symbol for arbitrator lock

\newcommand{\YA}{\arbitrator} %% Variable name for instance of 2-process lock
\newcommand{\sr}{\core}  %% Variable name for instance of strongly recoverable lock
\newcommand{\x}{\mathit{owner}}  %% Variable name for instance of selector lock

\newcommand{\fails}{k} %Variable for number of failures
\newcommand{\levels}{m} %Total number of levels of recursion

\newcommand{\procinlock}{\ensuremath{\mathbb{P}}}
\newcommand{\failinlock}{\ensuremath{\mathbb{UF}}}
\newcommand{\Super}{\ensuremath{\Pi}}

\newcommand{\setP}[2]{\procinlock(#1, #2)}
\newcommand{\setF}[2]{\failinlock(#1, #2)}
\newcommand{\setSuper}[2]{\Super(#1, #2)}

\newcommand{\allF}{\Phi} %set of all failures

\newcommand{\CIFCFS}{CI-FCFS}
\newcommand{\CIFC}{CI-concurrent}
\newcommand{\singular}{exclusive}

\newcommand{\broadcast}{\textsc{Broadcast}}
\newcommand{\bSet}{\textsc{Set}}
\newcommand{\bWait}{\textsc{Wait}}
\newcommand{\bRead}{\textsc{Read}}

\newcommand{\wellformed}{well-formed}
\newcommand{\legal}{legal}

\newcommand{\complete}[1][]{complete{#1}-successfully}
\newcommand{\run}{run}
\newcommand{\onlyWait}[2]{\mathcal{W}(#1,#2)}

\newcommand{\broadcastcc}{\broadcast-\textsc{CC}}
\newcommand{\broadcastdsm}{\broadcast-\textsc{DSM}}

\newcommand{\varcount}{count}
\newcommand{\varbefore}{\mathit{A}}
\newcommand{\varafter}{\mathit{B}}
\newcommand{\varannounce}{\mathit{announce}}
\newcommand{\vartarget}{\mathit{target}}
\newcommand{\varwakeup}{\mathit{wakeup}}
\newcommand{\varlast}{\mathit{last}}

\newcommand{\attempt}{attempt}
\newcommand{\useful}{useful}
\newcommand{\useless}{useless}

\newcommand{\snapshot}{snapshot}
\newcommand{\checkpoint}{checkpoint}
\newcommand{\rendezvous}{catch-up}
\newcommand{\barrier}{yield}
\newcommand{\switch}{switch}  %% Track switching of pool buffers

\newcommand{\allowance}{allowance period}
\newcommand{\GPTD}{APTD}

\newcommand{\FnNewNode}{\textsc{GetNewNode}}
\newcommand{\FnAdvanceOne}{\textsc{ExecuteOneStride}}
\newcommand{\FnRetireNode}{\textsc{RetireLastNode}}

\newcommand{\varpool}{\mathit{pool}}
\newcommand{\varcurrentpool}{\mathit{currentPool}}
\newcommand{\varbackuppool}{\mathit{backupPool}}
\newcommand{\varstart}{\mathit{start}}
\newcommand{\varcheckpoint}{\mathit{checkpoint}}
\newcommand{\varfinish}{\mathit{finish}}
\newcommand{\varstep}{\mathit{stride}}
\newcommand{\varlaststep}{\mathit{latest}}
\newcommand{\varpenalty}{\mathit{penalty}}
\newcommand{\varsnapshot}{\mathit{snapshot}}
\newcommand{\varindex}{\mathit{j}}

\newcommand{\inevitable}{inevitable}

\newcommand{\atTime}[2]{t(#1,#2)}
\newcommand{\atTimeFor}[3]{t(#1,#2,#3)}

\newcommand{\JJJ}{Jayanti, Jayanti and Joshi}
\newcommand{\KM}{Katzan and Morrison}

\newcommand{\minitab}[2][l]{\begin{tabular}{@{}#1@{}}#2\end{tabular}}

\title{Adaptive and Fair Transformation for Recoverable Mutual Exclusion}

\titlenote{Parts of this work have appeared earlier in the \emph{Proceedings of the 39th ACM Symposium on Principles of Distributed Computing (PODC)}, 2020.}

\author{Sahil Dhoked}
\affiliation{
  \institution{The University of Texas at Dallas}
  \state{TX}
  \postcode{75080}
  \country{USA}}
\email{sahil.dhoked@utdallas.edu}
\author{Neeraj Mittal}
\affiliation{
  \institution{The University of Texas at Dallas}
  \state{TX}
  \postcode{75080}
  \country{USA}}
\email{neerajm@utdallas.edu}

\usepackage{subfiles}%Not supported for JACM

    \begin{abstract}
        Mutual exclusion is one of the most commonly used techniques to handle contention in concurrent systems. Traditionally, mutual exclusion algorithms have been designed under the assumption that a process does not fail while acquiring/releasing a lock or while executing its critical section. However, failures do occur in real life, potentially leaving the lock in an inconsistent state. This gives rise to the problem of \emph{recoverable mutual exclusion (RME)} that involves designing a mutual exclusion (ME) 
        algorithm that can tolerate failures, while maintaining safety and liveness properties.
        
        In this work, we present a framework that transforms any algorithm that solves the RME problem into an algorithm that can also simultaneously \emph{adapt} to 
        \begin{enumerate*}[label=(\alph*)]
        \item the number of processes competing for the lock, \emph{as well as}
        \item the number of failures that have occurred in the recent past,
        \end{enumerate*}
        while
        maintaining the correctness and performance properties
        of the underlying RME algorithm. Additionally, the algorithm constructed as a result of this transformation adds certain desirable properties like fairness (a variation of FCFS) and bounded recovery.
        We further extend our framework by presenting a novel memory reclamation algorithm to bound the worst-case space complexity of the RME algorithm. The memory reclamation techniques maintain the fairness, performance and correctness properties of our transformation. The technique is general enough that is may also be employed to bound the space of other RME algorithms.
        
        One of the important measures of performance of any ME algorithm, including an RME algorithm, is the number of \emph{remote memory references (RMRs)} made by a process---for acquiring and releasing a lock as well as recovering the lock structure after a failure.
        
        Assume that the worst-case RMR complexity of a critical section request in the underlying RME algorithm is $R(n)$, where $n$ denotes the number of processes in the system. Then, our framework yields an RME algorithm for which the worst-case RMR complexity of a critical section request is given by $\bigO{\min\{\pc,\sqrt{\F+1},\ R(\n) \}}$, where $\pc$ denotes the \emph{point contention} of the request and  $\F$ denotes the number of failures in the \emph{recent past} of the request. 
         
        In addition to read and write instructions, our algorithm uses compare-and-swap (\CAS{}) and fetch-and-store (\FAS{}) hardware instructions, both of which are commonly available in most modern processors. 
        
    \end{abstract}



\begin{document}

\maketitle

    \section{Introduction}
    
    %% MUTUAL EXCLUSION

    One of the most commonly used techniques to handle contention in a concurrent system is to use \emph{mutual exclusion (ME)}. The mutual exclusion problem was first defined by Dijkstra more than half a century ago in~\cite{Dij:1965:CACM}. Using locks that provide mutual exclusion enables a process to execute its \emph{critical section} (part of the program that involves accessing shared resources) in isolation without worrying about interference from other processes. This avoids race conditions, thereby ensuring that the system always stays in a consistent state and produces correct outcome under all scenarios.

    %% FAILURES

    Generally, algorithms for mutual exclusion are designed with the assumption that failures do not occur, especially while a process is accessing a lock or a shared resource. However, such failures can occur in the real world.
    A power outage or network failure might create an unrecoverable situation causing processes to be unable to continue. If such failures occur, traditional mutual exclusion algorithms, which are not designed to operate properly in the presence of failures, may fail to guarantee important safety and/or liveness properties (\emph{e.g.}, system may deadlock).  
    In many cases, such failures may have disastrous consequences.
    This gives rise to the \emph{recoverable mutual exclusion (RME) problem}. The RME problem involves designing an 
    algorithm that ensures mutual exclusion under the assumption that process failures may occur at \emph{any} point during their execution, but the system is able to recover from such failures and proceed without any adverse consequences.

    %% NVRAM
    
    Traditionally, concurrent algorithms use checkpointing and logging to tolerate failures by regularly saving relevant portion of application state to a persistent storage such as hard disk drive (HDD). Accessing a disk is orders of magnitude slower than accessing main memory. As a result, checkpointing and logging algorithms are often designed to minimize disk accesses.
    \emph{Non-volatile random-access memory (NVRAM)} is a new class of memory technologies that combines the low latency and high bandwidth of traditional random access memory with the density, non-volatility, and economic characteristic of traditional storage media (\emph{e.g.}, hard disk drive). Existing checkpointing and logging algorithms can be modified to use NVRAMs instead of disks to yield better performance, but, in doing so,we would not be leveraging the true power of NVRAMs \cite{NarHod:2010:ASPLOS, GolRam:2019:DC}. NVRAMs can be used to directly store implementation specific variables and, as such, have the potential for providing near-instantaneous recovery from failures. 
    
    Most of the application data can be easily recovered after failures by directly storing implementation variables on NVRAMs. However, recovery of implementation variables alone is not enough. Processor state information such as contents of general and special purpose CPU registers (\emph{e.g.}, program counter, condition code register, stack pointer, \emph{etc.}) as well as contents of cache cannot always be recovered fully. In other words, 
    recovery may be lossy, and, if not handled properly, a failure may cause the system to behave erroneously upon recovery. Due to this reason, there is a renewed interest in developing fast and dependable algorithms for solving many important computing problems in software systems vulnerable to process failures using NVRAMs. Using innovative methods, with NVRAMs in mind, we aim to design efficient and robust fault-tolerant algorithms for solving mutual exclusion and other important concurrent problems.
   
    The RME problem in the current form was formally defined a few years ago by Golab and Ramaraju in~\cite{GolRam:2016:PODC}. Several algorithms have been proposed to solve this problem~\cite{GolRam:2019:DC, GolHen:2017:PODC, JayJos:2017:DISC,JayJay+:2019:PODC, ChaWoe:2020:PODC, JayJos:2019:NETSYS, KatMor:2020:OPODIS}. One of the most important measures of performance of an RME algorithm is the maximum number of \emph{remote memory references (RMRs)} made by a process per critical section request in order to acquire and release the lock as well as recover the lock after a failure. 
    Intuitively, RMR complexity captures the number of ``expensive'' steps performed by a process.
    Whether or not a memory reference is considered an RMR depends on the underlying memory model. The two most common memory models used to analyze the performance of an RME algorithm are \emph{cache-coherent (CC)} and \emph{distributed shared memory (DSM)} models.  
    Roughly speaking, a step is considered to incur an RMR in the CC model if it causes a memory location to be cached or a cached copy to be invalidated, and in the DSM model if it accesses data stored on a remote memory module. The CC model captures the working of the caching system used by hardware manufacturers to mask the high latency of memory, whereas the DSM model captures the NUMA (non-uniform memory access) effect observed in machines with large number of cores when memory is partitioned into multiple modules. 
    Hereafter, any RMR complexity bounds mentioned in the text is assumed to hold for both CC and DSM models \emph{unless otherwise stated}.
    
    Different RME algorithms provide different trade-offs in performance guarantees under different failure scenarios. 
    For example, one of the RME algorithms proposed in~\cite{GolRam:2019:DC} has an RMR complexity that grows linearly with the number of failures. Specifically, it has \emph{optimal} RMR complexity of $\bigO{1}$ in the absence of failures, but its RMR complexity may grow \emph{unboundedly} if failures occur repeatedly.
    On the other hand, the RME algorithm in~\cite{JayJay+:2019:PODC} has an RMR complexity of $\bigO{\nicefrac{\log \n}{\log \log \n}}$, where $\n$ is the number of processes in the system, irrespective of how many failures have occurred in the system (including the case when the system has not experienced any failures).
	Recently, Chan and Woelfel have proved a lower bound of $\bigOmega{\nicefrac{\log \n}{\log \log \n}}$ on the worst-case RMR complexity of any RME algorithm using currently available hardware instructions and practical word size of $\Theta(\log \n)$~\cite{ChaWoe:2021:PODC}.
	A more detailed description of the related work is given later in~\cref{sec:related}.

    \begin{table}
        \renewcommand{\arraystretch}{1.25}
        \caption{Comparison of known solutions to recoverable mutual exclusion problem with respect to RMR complexity under three different scenarios.}
        \tableFontSize
        \begin{tabular}{||>{\raggedright}m{16em}|c|c|c|c||} \hline
        \multicolumn{1}{||c|}{\multirow{2}{*}{\textbf{Algorithm}}} & \multicolumn{3}{|c|}{\textbf{RMR Complexity}} & \multirow{2}{*}{\minitab[c]{\textbf{Adaptive to} \\ \textbf{Contention}}} \\ \cline{2-4}
        & {\textbf{No failures}} & {\textbf{Limited failures}} & 
        \multicolumn{1}{m{12em}|}{\centering \textbf{Arbitrarily number of failures}}  & \\
        \hline \hline
        Golab and Ramaraju's transformation for recoverability \mbox{\cite[Section~4.1]{GolRam:2019:DC}} using MCS lock & $\bigO{1}$ & $\bigO{\n \F+1}$ & unbounded & No \\
        \hline
        Golab and Ramaraju's transformation for bounding RMR complexity \mbox{\cite[Section~4.2]{GolRam:2019:DC}} using MCS lock & $\bigO{1}$ & $\bigO{\n}$ & $\bigO{\n}$ & No \\
        \hline
        Golab and Hendler's arbitration tree using $k$-port MCS lock$^\ast\dagger$ \cite{GolHen:2017:PODC} & $\bigO{\nicefrac{\log \n}{\log \log \n}}$ & $\bigO{\nicefrac{\log \n}{\log \log \n}}$ & $\bigO{\nicefrac{\log \n}{\log \log \n}}$ & No \\
        \hline
        Jayanti and Joshi's wait-free recoverable lock \cite{JayJos:2017:DISC} & $\bigO{\log \n}$ & $\bigO{\log \n}$ & $\bigO{\log \n}$ & No \\
        \hline
        \JJJ{'s} arbitration tree using $k$-port MCS lock \cite{JayJay+:2019:PODC} & $\bigO{\nicefrac{\log n}{\log \log \n}}$ & $\bigO{\nicefrac{\log \n}{\log \log \n}}$ & 
        $\bigO{\nicefrac{\log \n}{\log \log \n}}$ & No \\
        \hline
        Chan and Woelfel's array based recoverable lock$^\ddagger$ \cite{ChaWoe:2020:PODC} & $\bigO{1}$ & $\bigO{F+1}$ & unbounded & No \\ \hline
        Katzan and Morrison's abortable and recoverable lock \cite{KatMor:2020:OPODIS} & $\bigO{\nicefrac{\log n}{\log \log \n}}$ & $\bigO{\nicefrac{\log \n}{\log \log \n}}$ & 
        $\bigO{\nicefrac{\log \n}{\log \log \n}}$ & Yes \\ 
        \hline
        \hline
        Our recoverable lock [this work] & $\bigO{1}$ & $\bigO{\sqrt{\F+1}}$ & $\bigO{\nicefrac{\log \n}{\log \log \n}}$ & Yes \\
        \hline
        \multicolumn{5}{l}{$\n$: the number of processes in the system} \\
        \multicolumn{5}{l}{$\F$: the number of failures in the system} \\
        \multicolumn{5}{l}{$\ast$: It has been recently shown in \cite{JayJay+:2019:PODC} that the algorithm is prone to deadlocks} \\
        \multicolumn{5}{l}{$\dagger$: RMR complexity measures only hold for the CC model} \\
        \multicolumn{5}{l}{$\ddagger$: RMR complexity is constant in the amortized case}
        \end{tabular}
        \label{table:RMRComparisons}
    \end{table}

    \subsection{Our Contributions}
    
     %% POLISH: see the abstract comments
     %%
     Our main contribution in this work is a \emph{novel} framework that transforms any algorithm that solves the RME problem into an algorithm that is simultaneously \emph{adaptive} to 
     \begin{enumerate*}[label=(\alph*)]
     \item the number of processes competing for the lock, \emph{as well as}
     \item the number of failures that have occurred in the recent past,
     \end{enumerate*}
     while having the same
     %(asymptotic) 
     worst-case RMR complexity as that of the base RME algorithm.
     In particular, assume that the worst-case RMR complexity of a critical section request in the underlying RME algorithm is $R(\n)$, where $\n$ denotes the number of processes in the system. Then, our framework yields an RME algorithm for which the worst-case RMR complexity of a critical section request is given by $\bigO{\min\{\pc,\sqrt{\F+1},\ R(\n) \}}$, where $\pc$ denotes the maximum number of requests that are simultaneously active while the given request is outstanding (referred to as \emph{point contention}) and $\F$ denotes the number of failures that have occurred in the \emph{recent past} of the given request (referred to as \emph{\hazard}). Note that the RMR complexity of a request in our algorithm is high only when both point contention and \hazard{} of the request are high.
   
     In addition to preserving the safety and liveness properties of the underlying RME algorithm (mutual exclusion, starvation freedom and critical section re-entry), our transformation also maintains its other desirable properties
     such as \emph{bounded exit} and \emph{bounded recovery} as applicable. 
     Roughly speaking, an RME algorithm satisfies the bounded exit property if a process is able to leave its critical section within a bounded number of its own steps unless it fails. It satisfies the bounded recovery property if a process is able to recover from a failure within a bounded number of its own steps unless it fails again. 
     %%
     \begin{comment}
     
     It satisfies the the critical section reentry property if, when a process $p$ fails inside its critical section, then no other process enters its critical section until $p$ has (re)entered its critical section. 
     
     \end{comment}

     The key idea behind our approach is to use a solution to a \emph{weaker} variant of the RME problem, which we refer to as \emph{weakly RME problem}, in which a failure may cause the mutual exclusion property to be violated temporarily albeit in a controlled manner, repeatedly as a ``filter'' to limit contention and achieve adaptability. To that end, we present an efficient algorithm for a weakly RME algorithm that has \emph{optimal} RMR complexity of only $\bigO{1}$.

     We also show that our RME algorithm is fair; in particular, it satisfies a variant of the first-come-first-served (FCFS) property, which we refer to as \emph{CI-FCFS}. Intuitively, CI-FCFS guarantees first-come-first-served (FCFS) order among requests provided their recent past is failure-free.

     Finally, we design a \emph{novel} memory reclamation algorithm that enables us to bound the  space complexity of our weakly RME algorithm by $\bigO{\n^2}$. This, in turn, enables us to bound the space complexity of the algorithm generated by our framework by $\bigO{\n^2 \cdot R(\n) + S(\n)}$, where $S(\n)$ denote the space complexity of the underlying RME algorithm.

    \subsection{Organization of the Text}
    The rest of the text is organized as follows. 
    
    We describe our system model and formally define the RME problem in \cref{sec:model|problem}. 
    We define the weaker variant of the RME problem and its properties in \cref{sec:weak_recoverability}.
    
    We present a highly efficient solution to the weaker variant of the RME problem with constant RMR complexity in \cref{sec:weak|MCS}. 
    In \cref{sec:framework|basic}, we present a framework to transform any given RME algorithm into a new RME algorithm that preserves the worst-case RMR complexity of the original RME algorithm but has lower RMR complexity in the absence of failures. This transformation uses a solution to the weaker variant of the RME problem as a building block. Applying this transformation recursively, we create a new transformation in \cref{sec:framework|recursive} that preserves the worst-case RMR complexity of the original RME algorithm and whose performance degrades sub-linearly  with the number of ``recent'' failures (specifically, in proportion to $\sqrt{F}$). This transformation achieves the desired RMR complexity for each of the three scenarios mentioned earlier (as shown in \cref{table:RMRComparisons}).
    In \cref{sec:fairness}, we discuss the fairness guarantee provided by our RME algorithms (weakly as well as strongly recoverable).

    In \cref{sec:broadcast|object}, we present a recoverable broadcast object with constant RMR complexity that allows a process to notify other processes that are waiting on it to reach a certain point in its execution. 
    We use the broadcast object to design a recoverable memory reclamation algorithm that bounds the space-complexity of our (recursive) framework while maintaining its RMR complexity in \cref{sec:memory}.
    
    A detailed description of the related work is given in \cref{sec:related}.
    Finally, in \cref{sec:conclusion}, we present our conclusions and outline directions for future research.

    \section{System Model and Problem Formulation}
    \label{sec:model|problem}
    
        We follow the same model as used by Golab and Ramaraju in their work on recoverable mutual exclusion (RME) \cite{GolRam:2019:DC}. 
        
    \subsection{System model}
    
        We consider an asynchronous shared-memory system consisting of $\n$ unreliable processes labeled $p_1, p_2, \ldots, p_\n$. Shared memory is used to store variables that can be accessed by any process. Besides shared memory, each process also has its own private memory that is used to store variables that can only be accessed by that process (\emph{e.g.}, program counter, CPU registers, execution stack, \emph{etc.}). Processes can only communicate by performing read, write and read-modify-write (RMW) instructions on shared variables. Processes are not assumed to be reliable and may fail.
        
        A system execution is modeled as a sequence of process steps. In each step, some process either performs some local computation affecting only its private variables or executes one of the available instructions (read, write or RMW) on a shared variable or fails. Processes may run at arbitrary speeds and their steps may interleave arbitrarily. In any execution, between two successive steps of a process, other processes can perform an unbounded but finite number of steps.
        %% Processes are not assumed to be reliable and may fail.
    
        %%
        %% SHOULD REMOVE THE NEXT SENTENCE
        %%
        %% Using the variables in private and shared memory, processes compete with each other to acquire a lock in order to execute their respective critical sections.
        %%
        
        \begin{comment}
        To avoid race conditions resulting from multiple processes trying to access the same shared resource simultaneously, processes synchronize their accesses to shared resources using a \emph{lock} that provides mutual exclusion (ME); at most one process can hold the lock at any time.
        \end{comment}
        
   \subsection{Failure model}
        We assume the \emph{crash-recover} failure model. A process may fail at any time during its execution by crashing. A crashed process recovers eventually and restarts its execution. A crashed process does not perform any steps until it has restarted. A process may fail multiple times, and multiple processes may fail concurrently.
        
        Note that, upon restarting after a failure, the state of the lock as well as the underlying application utilizing the lock needs to be restored to a proper condition. In this work, we focus only on the recovery of the internal structure of the lock. Restoring the application state to its proper condition (using logs and/or persistent memory) is assumed to be the responsibility of the programmer and is beyond the scope of this work~\cite{GolRam:2019:DC,GolHen:2017:PODC,JayJay+:2019:PODC}.
        
        We assume that, upon crashing, a process loses the contents of its private variables, including but not limited to the contents of its program counter, CPU registers (general and special purpose) and execution stack. However, the contents of the shared variables remain unaffected and are assumed to persist despite any number of failures. When a crashed process restarts, all its private variables are reset to their initial values.
        
        Processes that have crashed are difficult to distinguish from processes that are running arbitrarily slow. However, we assume that every process is \emph{live} in the sense that a process that has not crashed eventually executes its next step and a process that has crashed eventually recovers. In this work, we consider a failure to be associated with a single process. If a failure causes multiple processes to crash, unless otherwise stated, we treat each process crash as a separate failure. 
        
        %% In this work, we only consider benign crash failures. Processes may either crash, or run as 
        %% expected. A non-crashed process cannot perform any malicious activities.

        %Mention how the following example violates fairness
        %
        %   lock1.lock()
        %   lock2.lock()
        %   lock1.unlock()
        %   lock2.unlock()
        %

    \subsection{Process execution model}
        A process execution is modeled using two types of computations, namely \emph{non-critical section} and \emph{critical section}. A critical section refers to the part of the application program in which a process needs to access shared resources in isolation. A non-critical section refers to the remainder of the application program. 
       
        If multiple processes access and modify shared resource(s) concurrently, it may lead to race conditions which may prevent the application from working properly and may possibly have disastrous consequences. To avoid such race conditions, a lock (or a mutual exclusion (ME) algorithm) is used to enable each process to execute its critical section in isolation. At most one process can hold the lock at any time, and a process can execute its critical section only if it is holding the lock. The lock can be granted to another request only after the process (more specifically, request) holding the lock releases it after completing its critical section.
        Hereafter, we use the terms ``mutual exclusion algorithm'', ``ME algorithm'' and ``lock'' interchangeably.

    \addprefix[E.]
    \begin{algorithm}[t]
            \DontPrintSemicolon
            \While {true} 
            {
                Non-Critical Section (NCS)\;
                Recover\;
                Enter\;
                Critical Section (CS)\;
                Exit\;
            }
            \caption{Process execution model.}
            \label{algo:PEM}
    \end{algorithm}

        The execution model of a process with respect to a lock is depicted in \cref{algo:PEM}. As shown, a process repeatedly executes the following five  \segment{s} in order: \NCS{}, \Recover{}, \Enter{}, \CS{} and \Exit{}.
        The first \segment{}, referred to as \NCS, models the steps executed by a process in which it only accesses variables outside the lock.
        The second \segment{}, referred to as \Recover, models the steps executed by a process to perform any cleanup required due to past failures and restore the internal structure of the lock to a consistent state. 
        The third \segment{}, referred to as \Enter, models the steps executed by a process to acquire the lock so that it can execute its critical section in isolation.
        The fourth \segment{}, referred to as \CS, models the steps executed by a process in the critical section where it accesses shared resources in isolation.
        Finally, the fifth \segment{}, referred to as \Exit, models the steps executed by a process to release the lock it acquired earlier in \Enter{} \segment. 
       
        We assume that, in the \NCS{} \segment{}, a process does not access any part of the lock or execute any computation that could potentially cause a race condition. Moreover, in the \Recover, \Enter{} and \Exit{} \segment{s}, a process accesses shared variables pertaining to the lock (and the lock only).
        \emph{Our execution model only considers steps taken by a process during its \Recover{}, \Enter{} or \Exit{} \segment{s}.}
        
        A process may crash at any point during its execution, including while executing the \NCS, \Recover, \Enter, \CS{} or \Exit{} \segment{}.  
        We assume that a crashed process upon restarting starts its execution from the beginning of the loop shown in \cref{algo:PEM}, specifically from the beginning of \NCS{} \segment{}. 
        Note that any steps executed by a process to recover the application state are not explicitly modeled here. Specifically, both \NCS{} and \CS{} \segment{s} may consist of code in the beginning to recover relevant portions of the application state. 
        
		In the rest of the text, by the phrase ``acquiring a recoverable lock,'' we mean ``executing \Recover{} and \Enter{} \segment{s} (in order) of the associated RME algorithm.'' Likewise, by the phrase ``releasing a recoverable lock,'' we mean ``executing \Exit{} \segment{} of the associated RME algorithm.''

        \begin{definition}[passage]
            A \emph{passage} of a process is defined as the sequence of steps executed by the process from when it begins executing \Recover{} \segment{} to either when it finishes executing the corresponding \Exit{} \segment{} or experiences a failure, whichever occurs first.
        \end{definition}
        
        %A process is said to be in cleanup if it is executing the \texttt{Recover} section of its first passage after it experienced a failure. 
        
        \begin{definition}[failure-free passage]
            A passage of a process is said to be \emph{failure-free} if the process has successfully executed \Recover, \Enter{} and \Exit{} \segment{s} of that passage without experiencing any failures. 
        \end{definition}

        \begin{definition}[super-passage]
            A \emph{super-passage} of a process is a maximal non-empty sequence of consecutive passages 
            executed by the process, where only the last passage of the process in the sequence can be failure-free.
        \end{definition}
        
        \begin{definition}[failure-free super-passage]
            A super-passage of a process is said to be \emph{failure-free} if it consists of exactly one passage.
        \end{definition}

        For ease of exposition, if a process has a super-passage \inprogress{} (\emph{i.e.}, not completed), then we say that the process has a pending or outstanding \emph{request} for a critical section or super-passage. Note that a process may execute multiple failure-free \CS{} \segment{s} during its . This is because a super-passage is considered to be complete only after the process has completed a failure-free passage, which includes \Exit{} \segment. We consider all these \CS{} \segment{s} to be associated with the \emph{same} request.

    \subsection{Problem definition}
    \label{sec:problem}
    
        A \emph{history} is a collection of steps taken by processes. 
        A process $p$ is said to be \emph{live} in a history $H$ if $H$ contains at least one step by $p$.
        We assume that every critical section is finite.

        \begin{definition}[fair history]
        A history $H$ is said to be \emph{fair} if 
        \begin{enumerate*}[label=(\alph*)]
            \item it is finite, or 
            \item if it is infinite and every live process in $H$ either executes infinitely many steps or stops taking steps after a failure-free passage.
        \end{enumerate*}
        \end{definition}
        
        Designing a recoverable mutual exclusion (RME) algorithm involves designing \Recover, \Enter{} and \Exit{} \segment{s} such that the following correctness properties are satisfied.  
        
        \begin{description}
        
            \item[Mutual Exclusion (ME)] For any  history $H$, at most one process is in its \CS{} at any point in $H$. 
        
            \item[Starvation Freedom (SF)] Let $H$ be an infinite fair history in which every every process fails only a finite number of times during each of its super-passage. Then, if a process $p$ leaves the \NCS{} \segment{} in some step of $H$, then $p$ eventually enters its \CS{} \segment{}.
            
            \item[Bounded Critical Section Reentry (BCSR)] For any history $H$, if a process $p$ crashes inside its \CS{} \segment{}, then, until $p$ has reentered its \CS{} \segment{} at least once, any subsequent execution of \Enter{} \segment{} by $p$ either completes within a bounded number of $p$'s own steps or ends with $p$ crashing.
            
        \end{description}
        
        Note that mutual exclusion is a safety property, and starvation freedom is a liveness property. The bounded critical section reentry is a safety as well as a liveness property. If a process fails inside its \CS{}, then a shared object or resource (\emph{e.g.}, a shared data structure) may be left in an inconsistent state. The bounded critical section reentry property allows such a process to ``fix'' the shared resource if needed before any other process can enter its \CS{} (\emph{e.g.}, \cite{GolRam:2019:DC,GolHen:2017:PODC,JayJay+:2019:PODC}). 
        This property assumes that a \CS{} is idempotent in the sense that, within the same super-passage, executing a \CS{} multiple times, possibly partially in some cases, is equivalent to executing it once.
        %%
        \begin{comment}
        This property assumes that the \CS{} is idempotent in the sense that, until a process , \emph{i.e.}, a \CS{} is designed such that, in a super passage, multiple executions of the \CS{} by a process is equivalent to a same execution of the \CS{}.
        \end{comment} 
        %%
        Our correctness properties are the same as those used in \cite{GolRam:2019:DC,GolHen:2017:PODC,JayJay+:2019:PODC}. We have stated them here for the sake of completeness.
        In addition to the correctness properties, it is also desirable for an RME algorithm to satisfy the following additional properties.
        
        \begin{description}
        
            \item[Bounded Exit (BE)] For any infinite history $H$, any execution of the \Exit{} \segment{} by any process $p$ either completes in a bounded number of $p$'s own steps or ends with $p$ crashing.
         
            \item[Bounded Recovery (BR)] For any infinite history $H$, any execution of \Recover{} \segment{} by process $p$ either completes in a bounded number of $p$'s own steps or ends with $p$ crashing.
        
        \end{description}
        
    \subsection{Performance measures}
    
        We measure the performance of RME algorithms in terms of the number of \emph{remote memory references (RMRs)} incurred by the algorithm during a \emph{single} passage (\emph{i.e.}, \Recover, \Enter{} and \Exit{} \segment{s}). The definition of a remote memory reference depends on the memory model implemented by the underlying hardware architecture. In particular, we consider the two most popular shared memory models:
        
       \begin{description}
            \item[Cache Coherent (CC)]
                The CC model assumes a \emph{centralized} main memory that acts as a global, shared store of variables. In addition, each process has a \emph{local} cache memory.  Whenever a process accesses (reads or writes) a variable, a copy of the variable is stored in the cache of that process. Any subsequent access to that variable is serviced using that cached copy as long as the copy is still valid. A write access causes other cached copies of the variable to be either updated or invalidated. In addition, it may also cause the copy stored in the main memory to be updated.
                In this model, a step incurs an RMR if it causes the contents of any of the caches to be modified (including invalidation).  
                %%
                \begin{comment}
                The CC model assumes a centralized main memory. Each process has access to the central shared memory in addition to its local cache memory. The shared variables, when needed, are cached in the local memory. These variables may be invalidated if updated by another process. Reading from an invalidated variable causes a cache miss and requires the variable value to be fetched from the main memory. Similarly, write on shared variables is performed on the main memory. Under this model, a remote memory reference occurs each time there is a fetch operation from or a write operation to the main memory.
                \end{comment}
                %%
            \item[Distributed Shared Memory (DSM)]
                The DSM model assumes that the main memory is \emph{partitioned} into multiple memory modules with one module attached to every process. A process can access a variable stored on any memory module, be it local or remote. However, accessing a variable stored on its local module is much faster than accessing the one stored on a remote module. In this model, a step incurs an RMR if it involves accessing a variable stored on a remote memory module.
                %%
                \begin{comment}
                the has no centralized memory. Shared variables reside on individual process nodes. These variables may be accessed by processes either via the interconnect or a local memory read, depending on where the variable resides. Under this model, a remote memory reference occurs when a process needs to perform \textit{any} operation on a variable that does not reside in its own node's memory.
                \end{comment}
                %%
        \end{description}
        
        In the rest of the text, if not explicitly specified,  
        the RMR complexity measure of an algorithm applies to \emph{both CC and DSM models}.
        
        We analyze the RMR complexity of an RME algorithm under three scenarios:
        \begin{enumerate*}[label=(\alph*)]
        \item in the absence of failures (failure free RMR complexity),
        \item in the presence of $\F$ failures (limited failures RMR complexity), and
        \item in the presence of an unbounded number of failures (arbitrary failures RMR complexity).
        \end{enumerate*}

        Let $g(\n, \F)$ be a function of $\F$ and $\n$, where $\n \geq 1$ is the number of processes in the system and $\F \geq 0$ is the number of failures that have occurred so far. We assume that $g(\n, \F)$ is a monotonically non-decreasing function of $\n$ and $\F$ since the function is used to represent the worst-case RMR complexity of an RME algorithm. 
        %%
        \begin{comment}
        A number $k$ is said to be \emph{covered} by a (possibly infinite) set of numbers $S$, denoted by $k \vDash S$, if it lies between two numbers belonging to $S$. Formally,
        %%
        \[
        \exists u, v : \{ u, v \} \subseteq S : u \leq k \leq v
        \]
        \end{comment}
        %%
        We identify several desirable performance measures applicable to an RME algorithm. To that end, we first define the following concepts for the function $g(\n, \F)$.

        \begin{description}
            \item[\PC~1. (Constantness)] In the absence of failures, the function has a constant value independent of $\n$. Formally, $g(\n, 0) = \bigO{1}$.
            \item[\PC~2. (Adaptiveness)] In order to capture adaptiveness, we define a function $\Delta(\n)$ as: 
            \[ \Delta(\n) = \card{ \{\F \mid  g(\n, \F) < g(\n, \F+1)\} } \]
            With limited number of failures, we identify three different cases. 
            \begin{enumerate}[label=(\alph*)]
                \item The function has a non-trivial dependence on $\F$. Formally, $\Delta(\n) = \bigOmega{1}$.
                \item The function has a strong dependence on $\F$. Formally, $\Delta(\n) = \smallOmega{1}$.
                \item The function has strong and sub-linear dependence on $\F$.
                Formally, $\Delta(\n) = \smallOmega{1}$ and $g(\n, \F) = \smallO{\F}$.
            \end{enumerate}
            \item[\PC~3. (Boundedness)] With arbitrary number of failures, we identify two different cases:
            \begin{enumerate}[label=(\alph*)]
                \item The function is finite-valued even as $\F$ tends to infinity. 
                Formally, $\lim\limits_{\F \rightarrow \infty}  g(\n, \F)$ is finite-valued. 
                \item The function is bounded by a sub-logarithmic function of $\n$. 
                Formally, $\forall \F : g(\n, \F) = \bigO{\nicefrac{\log \n}{\log \log \n}}$.
            \end{enumerate}
        \end{description}

    \begin{table}
    \renewcommand{\arraystretch}{1.25}
    \caption{Comparison of known solutions to recoverable mutual exclusion problem with respect to the five performance measures.}
    \tableFontSize
    \begin{tabular}{||>{\raggedright}m{18em}|c|c|c|c|c|c|m{6.5em}||} \hline
        \multicolumn{1}{||c|}{\multirow{2}{*}{\textbf{Algorithm}}} & \multicolumn{6}{|c|}{\textbf{Performance Measure}} &
        \multicolumn{1}{|c||}{\multirow{2}{*}{\textbf{Classification}}}
        \\ \cline{2-7}
        & \textbf{\PC~1} & \textbf{\PC~2(a)} & \textbf{\PC~2(b)} & \textbf{\PC~2(c)} & \textbf{\PC~3(a)} & \textbf{\PC~3(b)} & \\
        \hline \hline
        Golab and Ramaraju's transformation for recoverability \mbox{\cite[Section~4.1]{GolRam:2016:PODC}} using MCS lock & \cmark & \cmark & \cmark & \xmark & \xmark & \xmark & \unbounded{} \linebreak \adaptive{} \\
        \hline
        Golab and Ramaraju's transformation for bounding RMR complexity \mbox{\cite[Section~4.2]{GolRam:2016:PODC}} using MCS lock & \cmark &  \cmark  &  \xmark  & \xmark & \cmark & \xmark & \bounded{}  \linebreak \cadaptive{} \\
        \hline
        Golab and Hendler's arbitration tree using $k$-port MCS lock$^\ast$ \cite{GolHen:2017:PODC} &   \xmark  &  \xmark  & \xmark & \xmark &  \cmark  & \cmark & \wbounded{} \linebreak \nonadaptive{} \\
        \hline
        Jayanti and Joshi's wait-free recoverable lock \cite{JayJos:2017:DISC} &   \xmark &  \xmark  &  \xmark & \xmark &  \cmark & \xmark & \bounded{} \linebreak \nonadaptive{} \\
        \hline
        Jayanti and Joshi's arbitration tree using $k$-port MCS lock \cite{JayJay+:2019:PODC} &  \xmark  &  \xmark  & \xmark &  \xmark & \cmark & \cmark & \wbounded{} \linebreak \nonadaptive{} \\ \hline
        Chan and Woelfel's array based recoverable lock$^\dagger$ \cite{ChaWoe:2020:PODC} &  \xmark & \cmark & \cmark & \xmark & \xmark & \xmark & \unbounded{} \linebreak \adaptive{}  \\ \hline
        Katzan and Morrison's abortable and recoverable lock \cite{KatMor:2020:OPODIS} & \xmark  &  \xmark  & \xmark &  \xmark & \cmark & \cmark & \wbounded{} \linebreak \nonadaptive{} \\ 
        \hline
        \hline
        Our recoverable lock [this work] & \cmark &  \cmark &  \cmark & \cmark & \cmark & \cmark &
        \wbounded{} \linebreak \sadaptive{} \\
        \hline
        \multicolumn{7}{l}{$\ast$: it has been recently shown in \cite{JayJay+:2019:PODC} that the algorithm is prone to deadlocks} \\
        \multicolumn{7}{l}{$\dagger$: RMR complexity is constant in the amortized case} 
        \end{tabular}
    \label{table:PerformanceCharacteristics}
    \end{table}

        Note that \PC~2(b) implies \PC~2(a), \PC~2(c) implies \PC~2(b) and \PC~3(b) implies \PC~3(a).
        
        Consider an RME algorithm $\mathcal{A}$ and let $g(\n, \F)$ denote the \emph{best known} bound on the worst-case RMR complexity of $\mathcal{A}$. We call $\mathcal{A}$ based on the performance measures satisfied by $g(\n, \F)$ as follows:
        
        \begin{enumerate}
            \item Based on adaptiveness
                \begin{itemize}
                
                \item \emph{\nonadaptive} if it does not satisfy \PC~2(a).
                
                \item \emph{\cadaptive} if it satisfies \PC~2(a).
                
                \item \emph{\adaptive} if it satisfies \PC~2(b) (hence also \PC~2(a).
                
                \item \emph{\sadaptive} if it satisfies \PC~2(c) (hence also \PC~2(b) and \PC~2(a).
                
                \end{itemize}
                
            \item Based on boundedness 
                \begin{itemize}
                
                \item \emph{\unbounded} if it does not satisfy \PC~3(a).
                               
                \item \emph{\bounded} if it satisfies \PC~3(a).
                
                \item \emph{\wbounded} if it satisfies \PC~3(b).
                
                \end{itemize}
        
        \end{enumerate}
        
        A comparison of the known RME algorithms with respect to the above performance measures \PC~1 to \PC~3  
        is shown in \cref{table:PerformanceCharacteristics}. As shown in \cref{table:PerformanceCharacteristics}, all existing RME algorithm are either \nonadaptive, \cadaptive{} or \fadaptive. To our knowledge, there is no \badaptive, let alone \wbounded{} \sadaptive{} RME algorithm currently for either memory model. 
        Note that our taxonomy may not be able to classify all possible RME algorithms (or recoverable algorithms in general), but it is sufficient for classifying and comparing all existing RME algorithms.

       \begin{comment}
       POLISH: revise next paragraph
       
       It should follow from adaptiveness and boundedness
       
       \end{comment}    
       
       To quantify the impact of failures on the performance of an RME algorithm, 
       we use the notion of \emph{\consequence{} interval} of a failure. Roughly speaking, we use it to capture the maximum duration for which the impact of the failure may be felt in the system. It is related to, but different from, the notion of $k$-failure-concurrent passage defined by Golab and Ramaraju in~\cite{GolRam:2019:DC}. A more detailed comparison of the two notions is deferred to \cref{subsec:fairness|comparison}.
       
           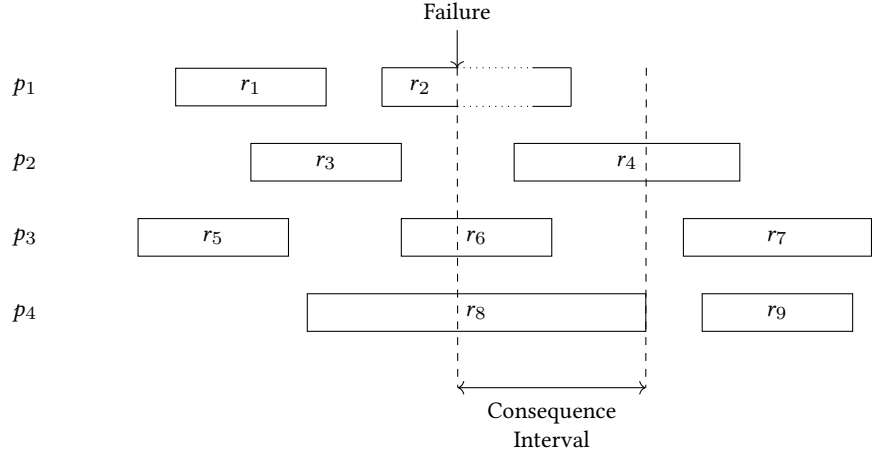
\begin{figure}[t]
        \centering
    \begin{tikzpicture}[draw, minimum height=0.5cm]
        \node (p1) at (2,10) {$p_1$};
        \node (p2) at (2,9) {$p_2$};
        \node (p3) at (2,8) {$p_3$};
        \node (p4) at (2,7) {$p_4$};
    
        \node (r1) [draw = black, rectangle, minimum width=2cm] at (5,10) {$r_1$};
        \node (r2) [rectangle, minimum width=1cm] at (7.25,10) {$r_2$};
        \node (r2p) [rectangle, minimum width=1cm] at (8.25,10) {};
        \node (r2r) [rectangle, minimum width=0.5cm] at (9,10) {};
        
        \draw[-] (r2.north west) -- (r2.north east);
        \draw[-] (r2.north west) -- (r2.south west);
        \draw[-] (r2.south west) -- (r2.south east);
        \draw[-, dotted] (r2p.north west) -- (r2p.north east);
        \draw[-, dotted] (r2p.south west) -- (r2p.south east);
        \draw[-] (r2r.north west) -- (r2r.north east);
        \draw[-] (r2r.north east) -- (r2r.south east);
        \draw[-] (r2r.south west) -- (r2r.south east);
        
        \node (r3) [draw = black, rectangle, minimum width=2cm] at (6,9) {$r_3$};
        \node (r4) [draw = black, rectangle, minimum width=3cm] at (10,9) {$r_4$};
        
        \node (r5) [draw = black, rectangle, minimum width=2cm] at (4.5,8) {$r_5$};
        \node (r6) [draw = black, rectangle, minimum width=2cm] at (8,8) {$r_6$};
        \node (r7) [draw = black, rectangle, minimum width=2.5cm] at (12,8) {$r_7$};
        
        \node (r8) [draw = black, rectangle, minimum width=4.5cm] at (8,7) {$r_8$};
        \node (r9) [draw = black, rectangle, minimum width=2cm] at (12,7) {$r_9$};
        
        \node (failure) [above =0.5cm of r2p.north west] {Failure};
        \draw[->] (failure.south -| r2p.west) -- (r2p.north west);
        
        \draw[-, dashed] (r2p.north west) -- (7.75,6);
        {\draw[-, dashed] (r2p.north -| r8.east) -- (10.25,6);
        \draw[<->] (7.75,6) -- (10.25,6);
        \node (CI)[text width=2cm,align=center] at (9, 5.5) {Consequence\\Interval};}
    \end{tikzpicture}
    \captionof{figure}{Illustration of the consequence interval of a failure}
    \label{fig:CIexample}
    \end{figure}
    
        \begin{definition}[\consequence{} interval]
            The \emph{\consequence{} interval} of a failure
            %% with respect to a lock
            in a history $H$ is defined as the interval in time that starts from the onset of the failure and extends to the point in time when either every super-passage that started
            %% in the lock
            before this failure occurred in $H$ has completed or the last step in $H$ is performed, whichever happens earlier.
        \end{definition}
        
        An illustration of the consequence interval of a failure is provided in \cref{fig:CIexample}.
        Intuitively, we consider a failure to be \emph{recent} with respect to a time $t$ if $t$ is contained in the \consequence{} interval of the failure. 
        %%
        %The \consequence{} interval of  failure is said to be \emph{\inprogress} if it overlaps with a super-passage that started before the failure occurred.
        %%
        The following notion captures the ``concentration'' of failures in the recent past of a request.
        
        \begin{definition}[\hazard]
            The \emph{\hazard{}} of a request is defined as the number of failures whose \consequence{} interval overlaps with the super-passage of the request.
        \end{definition}

        We show that our framework yields an RME algorithm that is adaptive not only to \hazard{} (our main focus) but also to contention. To show the latter, we use the well-known notion of point contention defined as follows:

        \begin{definition}[point contention]
        The \emph{point contention} of a request is defined as the maximum number of super-passages  that are simultaneously \inprogress{} at any point during the super-passage associated with the request.
        \end{definition}
        
        In the remainder of the text, unless stated otherwise, the term ``adaptive'' is used in the context of \hazard{}.

 \subsection{Synchronization primitives}
       
        We assume that, in addition to read and write instructions, the system also supports atomic \emph{fetch-and-store (\FAS)} and \emph{compare-and-swap (\CAS{})} read-modify-write (RMW) instructions.

        A fetch-and-store instruction takes two arguments: $address$ and $new$; it replaces the contents of a memory location ($address$) with a given value ($new$) and returns the old contents of that location.

        A compare-and-swap instruction takes three arguments: $address$, $old$ and $new$; it compares the contents of a memory location ($address$) to a given value ($old$) and, only if they are the same, modifies the contents of that location to a given new value ($new$). It returns \true{} if the contents of the location were modified and \false{} otherwise.

        Both instructions are commonly available in many modern processors such as Intel~64~\cite{Intel64Manual} and AMD64~\cite{AMD64Manual}.

        \begin{comment}
        
        We assume a \emph{strong}  CC model in which \CAS{} instruction updates or invalidates the cached location even if fails. Thus, any process spinning on a memory location incurs an RMR if another process performs a failed \CAS{} on that location. 
        
       \end{comment}

%\subfile{Sections/BuildingBlocks}

%\subfile{Sections/GenericAlgo}

    \section{Weak Recoverability}
        \label{sec:weak_recoverability}
        To design a \wbounded{} \sadaptive{} RME algorithm, we use a solution to the \emph{weaker} variant of the RME problem as a \emph{building block} in which a failure may cause the ME property to be violated albeit only temporarily and in a controlled manner. 
        We refer to this variant as the \emph{weakly recoverable mutual exclusion problem}.
        
        %%
 	    %%
 	    %% Should the properties be defined in terms of a history
 	    %%
 	    %%    
           
        To formally define how long a violation of the ME property may last, we use the notion of \consequence{} interval of a failure defined earlier.

       \begin{definition}[weakly recoverable mutual exclusion]
            An algorithm is a \emph{weakly recoverable mutual exclusion algorithm} if, in addition to starvation freedom, it satisfies the following property: for any history $H$, if two or more processes are in their critical sections simultaneously at some point in $H$, then that point overlaps with the \consequence{} interval of some failure.
            \label{def:wRME}
        \end{definition}
       
       Roughly speaking, a weakly RME algorithm satisfies the ME property as long as no failure has occurred in the ``recent'' past.
       Hereafter, to avoid confusion, we sometimes refer to the traditional recoverable mutual exclusion problem (respectively, algorithm) as defined in \autoref{sec:problem} as \emph{strongly recoverable mutual exclusion} problem (respectively, algorithm).
       
       %% POLISH - what about fairness
       %% Will become moot once discussion about fairness is moved to a separate section
       
       The bounded exit, bounded recovery and bounded critical section reentry properties defined earlier in \autoref{sec:problem} are applicable to weakly RME problem as well.
              
       We demonstrate that it is possible to design an optimal weakly recoverable mutual exclusion algorithm using existing hardware instructions whose worst-case RMR complexity is only $\bigO{1}$ under both CC and DSM models. 
       In contrast, as proven in~\cite{ChaWoe:2021:PODC}, 
       the worst-case RMR complexity of any strongly RME algorithm is $\bigOmega{\nicefrac{\log \n}{\log \log \n}}$ under both CC and DSM models. 
       We exploit this \emph{gap} to design an RME algorithm that is adaptive and bounded. 
       To prove that our algorithm is also  \sefficient, we utilize some additional properties of our weakly RME algorithm.
       
       Note that not all failures may cause the ME property to be violated when using a weakly RME algorithm. To that end, we define the notion of \senins{} of an algorithm. 
       
       \begin{definition}[\senins{}]
           An \ins{} $\sigma$ of a weakly RME algorithm is said to be \emph{\sen} if there exists a finite history $H$ that satisfies the following conditions:
           \begin{enumerate*}[label=(\alph*)]
               \item it contains exactly one failure in which a process crashes immediately after executing said \ins{} $\sigma$ and 
               \item two or more processes are in \CS{} at the end of $H$;
           \end{enumerate*}
           it is said to be \emph{non-\sen{}} otherwise.
       \end{definition}

       \begin{definition}[\unsafe{} failure]
            A failure is said to be \emph{\unsafe} with respect to a weakly RME algorithm if it involves a process crashing while (immediately before or after) performing a \senins{} with respect to the algorithm; it is said to be \emph{\safe} otherwise.
       \end{definition}
       
       Note that, by definition, every \ins{} of a strongly RME algorithm is a non-\senins{}. As a result, every failure is \safe{} with respect to a strongly RME algorithm. 
       
       The next notion limits the ``degree'' of violation (of the ME property) by a weakly RME algorithm if and when it occurs.

        \begin{definition}[\responsive{} weakly recoverable mutual exclusion]
            We say that a weakly recoverable mutual exclusion algorithm is \emph{\responsive{}} if, for all $k \geq 1$, it satisfies the following property: for any history $H$, if at least $k+1$ processes are in their critical sections simultaneously at some point in $H$, then that point overlaps with the \consequence{} intervals of at least $\bigOmega{k}$ (\unsafe) failures.
            \label{def:wRME|responsive}
        \end{definition}

        \subsection{Composite recoverable locks}
        
        The properties defined above are with respect to a \emph{single} weakly recoverable lock. In order to construct a \wbounded{} \sadaptive{} (strongly) recoverable lock with desired performance characteristics, we use multiple weakly recoverable locks. 
        We call a lock as \emph{composite} if it is employs one or more (weakly or strongly recoverable) locks. Composite locks might have several possible structures. For instance, the \Enter{} \segment{} of one lock could be contained in the \Enter{} or \CS{} \segment{} of another lock or the \CS{} \segment{} of one lock may be contained in the \NCS{} \segment{} of another lock. An example of a composite lock is a lock based on the tournament algorithm \cite{GolRam:2019:DC}.
        
        Note that, when we have multiple locks, the notions defined  earlier, namely \consequence{} interval, \senins{} and \unsafe{} failure, become \emph{relative} to the specific lock.
        For example, a failure will have a different \consequence{} interval with respect to each lock. An \ins{} may be \sen{} with respect to one lock but non-\sen{} with respect to another.
        Thus, in a composite lock,  a failure may be \unsafe{} with respect to one or more weakly recoverable locks.

        \begin{definition}[\locality{} property]
        A composite (weakly or strongly) recoverable lock  is said to satisfy the \emph{\locality{} property} if, for any \ins{} $\sigma$, $\sigma$ is \sen{} with respect to \emph{at most one} of its component weakly recoverable locks.
        \end{definition}
        
        A composite lock whose component locks are all strongly recoverable trivially satisfies satisfies the \locality{} property.

    \section{An Optimal Weakly Recoverable Lock}
		\label{sec:weak|MCS}
    
        In this section, we present a weakly recoverable lock whose RMR complexity is $\bigO{1}$ per passage for all three failure scenarios under both CC and DSM models.
        Our lock is based on the well-known MCS queue-based (non-recoverable) lock~\cite{MelSco:1991:TrCS}. 
        The original lock did not satisfy the bounded exit property. Dvir and Taubenfeld proposed an extension to the original algorithm in~\cite{DviTau:2018:DISC} to make the \Exit{} \segment{} wait-free. 
        We extend the augmented MCS lock, which satisfies bounded-exit property, to make it weakly recoverable.
        
        \subsection{Original MCS queue based lock}
        
            Processes in the MCS mutual exclusion algorithm use queue nodes to synchronize their executions of \CS{} \segment{s}.  The algorithm maintains a first-come-first-served (FCFS) queue of outstanding requests using a linked-list of their associated nodes. A node contains two fields:
            \begin{enumerate*}[label=(\alph*)]
               \item $\varnext$, a reference to its successor node in the queue (if any), and
               \item $\varlocked$, a boolean variable used by a process to spin while waiting for its turn to enter its critical section. 
            \end{enumerate*}
            The queue itself is represented using a shared variable $\vartail$ that contains a reference to the last node in the queue if non-empty and \mynull{} otherwise.

        	To acquire the lock, a process first initializes its queue node by setting its $\varnext$ and $\varlocked$ fields to \mynull{} and \true{}, respectively. It then appends the node to the queue by performing an \FAS{} instruction on $\vartail$ using the reference to its own node as an argument (to the instruction). Note that the instruction returns the contents of $\vartail$ just before it is modified. 
        	If the return value is \mynull{}, then it indicates that the lock is free and the process has successfully acquired the lock. 
        	If not, then it indicates that the lock is not free and the return value is the reference to the predecessor of the process' own node in the queue. In that case, it notifies the owner of the predecessor node of its presence. To that end, it stores the reference to its own node in the $\varnext$ field of the predecessor node, thereby creating a forward link between the two nodes. It then starts spinning on the $\varlocked$ field of its own node waiting for it to be reset to \false{} by the owner of the predecessor node as part of 
        	releasing the lock.
            
            To release the lock, a process first tries to reset the $\vartail$ variable to \mynull{} (if $\vartail$ still contains the reference to this process' node) using a \CAS{} instruction. If the instruction returns true, then it implies that the queue does not contain any more outstanding requests and the lock is now free. On the other hand, if the instruction returns false, then it implies that the queue contains at least one outstanding request and its own node is guaranteed to have a successor. It then waits until the $\varnext$ field of its own node contains a valid reference (a non-null value) indicating that a link has been created between its own node and its successor. Finally, it follows this link and resets the $\varlocked$ field in its successor node to \false{}.
        
        \subsection{Adding bounded exit property}
        
            The original algorithm as described above does not satisfy the bounded-exit property since a process leaving its critical section may have to wait until a link between its own node and its successor has been created. 
            
            To achieve the bounded-exit property, the original algorithm is augmented with a mechanism that allows a leaving process to notify the process next in line to acquire the lock that the lock is now free. To that end, a process, on leaving its critical section, attempts to store a special value (\emph{e.g.}, reference to its own node) in the $\varnext$ field of its own node using a \CAS{} instruction. Likewise, a link is also created using a \CAS{} instruction instead of a simple write instruction as in the original algorithm.  Both \CAS{} instructions are designed to succeed only if the $\varnext$ field contains \mynull{} value, thereby ensuring that the $\varnext$ field can only be modified once.  
            
            Thus, if the \CAS{} instruction performed  by a process leaving its critical section returns false, then that process can conclude that the forward link has already been created and it then follows this link and resets the $\varlocked$ field of its successor node. On the other hand, if the \CAS{} instruction performed by a process trying to create the link returns false, then that process can infer that the lock is free and that it now holds the lock. 
            
            With this modification, unlike in the original algorithm,  
            a process cannot always reuse its own node for the next request after releasing the lock.  
            
        \subsection{Adding weak recoverability}

    \addprefix[W.]
        
    \begin{algorithm}[t]
        \SetKw{LNot}{not}
           \begin{multicols}{2}
            \SetKw{Shared}{shared variables}
            \SetKw{Local}{local variables}
            \SetKw{Struct}{struct}
            \SetKw{Integer}{int}
            \SetKw{Boolean}{bool}
            \SetKw{Array}{array}
            \SetKw{Await}{await}
            \SetKwProg{initialization}{initialization}{\\ begin}{end}
            \SetKwProg{function}{Function}{\\ begin}{end}

            \tcc{Data structure used by the algorithm}
            \Struct QNode 
            \{ \\
            \Indp
            \tcc{location used for spinning while waiting to enter \CS{}} 
            $\varlocked$: boolean variable\;
            \tcc{reference to the successor node}
            $\varnext$: reference to QNode\;
            \Indm
            \}\;
            
            \BlankLine
            
            \Shared \\
            \Indp
            \tcc{reference to the last node in the queue} 
            $\vartail$: reference to QNode\;
            \tcc{state of the process with respect to the lock; in the DSM model, the $i$-{th} entry is local to process $p_i$}
            $\varstate$: \Array $[1{\dots}\n]$ of integer variables\;
            \tcc{reference to my own node; in the DSM model, the $i$-{th} entry is local to process $p_i$} 
            $\varnode$: \Array $[1{\dots}\n]$ of references to QNode\;
            \tcc{reference to the predecessor node; in the DSM model, the $i$-{th} entry is local to process $p_i$} 
            $\varpred$: \Array $[1{\dots}\n]$ of references to QNode\;
           
            \Indm
            
            \BlankLine
  
            \initialization{}{
            
                $\vartail$ $\leftarrow$ \mynull\tcp*[r]{queue is initially empty}
                \ForEach{$j \in \{ 1, 2, \dots, \n\}$}
                {
                   \tcc{reset node references}
                   $\varpred[j]$ $\leftarrow$ \mynull\;
                   $\varnode[j]$ $\leftarrow$ \mynull\;
                   \tcc{reset state}
                   $\varstate[j]$ $\leftarrow$ \InNCS\;
                }
           	}

           \columnbreak
            \function{\FnEnter(~)}
            {
                \If{($\varstate[i]$ = \InRecover)}
                {
                    \label{line:enter:doorway|begin}
                    \label{line:enter:recover|if}
                    \If{($\varnode[i]$ = \mynull)}
                    {
                        $\varnode[i]$ $\leftarrow$ \FnNewNode{(~)}\;
                        \label{line:MCS:enter:newnode}
                    }
                    \tcc{initialize fields of my own node}
                    $\varnode[i].\varnext$ $\leftarrow$ \mynull\; %% \tcp*[r]{initialize $\mynext$ field}
                    \label{line:MCS:enter:setnext}
                    $\varnode[i].\varlocked$ $\leftarrow$ \true\; %% \tcp*[r]{initialize $\mylocked$ field}
                    \label{line:MCS:enter:setlocked}
                    \tcc{the next initialization step helps to determine if \FAS{} has been performed}
                    $\varpred[i]$ $\leftarrow$ $\varnode[i]$\;
                    \label{line:MCS:enter:setpred}
                    $\varstate[i]$ $\leftarrow$ \InEnter\tcp*[r]{advance the state}  
                  
                }
                \label{line:enter:recover|endif}

                \If{($\varstate[i]$ = \InEnter)}
                {
                    \label{line:enter:enter|if}
                    \If{($\varpred[i]$ = $\varnode[i]$)}
                    {
                        \tcc{append my own node to the queue}
                        QNode $\vartemp$ $\leftarrow$ \FAS( $\vartail$, $\varnode[i]$ )\;
                        \label{line:enter:FAS}
                        \tcc{persist the result of \FAS{}}
                        $\varpred[i]$ $\leftarrow$ $\vartemp$\;
                        \label{line:enter:persist}
                    }
                }
                \label{line:enter:doorway|end}
                
                \If{($\varstate[i]$ = \InEnter)}
                {
                    \label{line:enter:waitingroom|begin}
                    \If{($\varpred[i]$ $\neq$ \mynull)}
                    {
                        \label{line:enter:pred|if}
                        \tcc{have a predecessor; create the link}
                        \CAS( $\varpred[i].\varnext$, \mynull, $\varnode[i]$ )\;
                        \label{line:enter:CAS}
                        \If{($\varpred[i].\varnext$ = $\varnode[i]$)}
                        {
                            \tcc{wait for the predecessor to complete}
                     	    \Await \LNot{}($\varnode[i].\varlocked$)\tcp*[r]{spin}
                     	    \label{line:enter:await}
                        }
                    }
                    \label{line:enter:pred|endif}
                    $\varstate[i]$ $\leftarrow$ \InCS\tcp*[r]{advance the state}
                }
                \label{line:enter:waitingroom|end}
                \label{line:enter:enter|endif}
            }

           \end{multicols}
           \caption{Pseudocode of process $p_i$ for a weakly recoverable MCS-based lock with wait-free exit.}
           \label{algo:weak|MCS}
        \end{algorithm}

        \begin{algorithm}[t]
        \SetKw{LNot}{not}
           \begin{multicols}{2}
            \SetKw{Shared}{shared variables}
            \SetKw{Local}{local variables}
            \SetKw{Struct}{struct}
            \SetKw{Integer}{int}
            \SetKw{Boolean}{bool}
            \SetKw{Array}{array}
            \SetKw{Await}{await}
            \SetKwProg{initialization}{initialization}{\\ begin}{end}
            \SetKwProg{function}{Function}{\\ begin}{end}

            \function{\FnCleanup(~)}
            {
                \label{line:exit:first}
                $\varstate[i]$ $\leftarrow$ \InExit\tcp*[r]{advance the state}
                \If{($\varnode[i] \neq$ \mynull)}
                {
                
                    \tcc{remove my node from the queue if it has no successor}          
                    \CAS( $\vartail$, $\varnode[i]$, \mynull{} )\;
                    \label{line:exit:CAS|tail}
                    \tcc{may have a successor; signal it to enter \CS{}}
                    \CAS( $\varnode[i].\varnext$, \mynull, $\varnode[i]$ )\;   
                    \label{line:exit:CAS|next}
                  
                    \If{($\varnode[i].\varnext$ $\neq$ $\varnode[i]$)}
                    {
                  	    \tcc{link already created; tell the successor to stop spinning}
                        $\varnode[i].\varnext.\varlocked$ $\leftarrow$ \false\;
                        \label{line:exit:CAS|next|locked}
                    }
                    \label{line:exit:successor|stop}
                    $\varpred[i]$ $\leftarrow$ \mynull\;
                    $\varnode[i]$ $\leftarrow$ \mynull\;
                }
				\FnRetireNode{(~)}\;
				\label{line:exit:retire}
                $\varstate[i]$ $\leftarrow$ \InNCS\tcp*[r]{advance the state}   
                \label{line:exit:last}        
             
            }
            
            \columnbreak
            \function{\FnRecover(~)}
            {
             
               \uIf{($\varstate[i]$ = \InEnter)}
               {
                  \label{line:recover:enter|if}
                  \If{($\varpred[i]$ = $\varnode[i]$)}
                  {
                     \tcc{may have failed earlier while performing \FAS{} instruction; abort the attempt}
                     \tcc{once \FAS{} step has been performed without any interruption, the two references are guaranteed to be different}
                     \FnCleanup(~)\tcp*[r]{execute cleanup method}
                  }
                  \label{line:recover:enter|endif}
               } \ElseIf{($\varstate[i]$ = \InExit)}
               {
                  \label{line:recover:exit|if}
                  \FnCleanup(~)\tcp*[r]{finish executing cleanup method}
                 
               }
                \label{line:recover:exit|endif}
                
               \If(\tcp*[f]{initialize lock}){($\varstate[i]$ = \InNCS)}
               {
                  \label{line:recover:ncs|if}
                  %% $\varnode[i]$ $\leftarrow$ \mynull\tcp*[r]{reset reference to own node}
                  %% \label{line:recover:minenull}
                  %% $\varpred[i]$ $\leftarrow$ \mynull\;
                  $\varstate[i]$ $\leftarrow$ \InRecover\tcp*[r]{advance the state}
                  \label{line:recover:state:advance}
                 
               }
               \label{line:recover:ncs|endif}
            }
            
           \BlankLine
            
            \function{\FnExit(~)}
            { 
               \FnCleanup(~)\;
            }

            \end{multicols}
            \caption{Pseudocode of process $p_i$ for a weakly recoverable MCS-based lock with wait-free exit (continued).}
            \label{algo:weak|MCS|2}
     \end{algorithm}

            A pseudocode of the weakly recoverable lock is given in \cref{algo:weak|MCS,algo:weak|MCS|2}. Our pseudocode uses the following shared variables. 
            The first variable, $\vartail$, contains the address of the last node in the queue if the queue is non-empty and \mynull{} otherwise. The next three variables, $\varstate$, $\varnode$ and $\varpred$, are arrays with one entry for each process.   
            The $i$-{th} entry of $\varstate$, denoted by $\varstate[i]$, 
            contains process $p_i$'s current state with respect to the lock (explained later). 
            The $i$-{th} entry of $\varnode$, denoted by $\varnode[i]$, contains the address of the queue node associated with process $p_i$'s most recent request.
            The $i$-{th} entry of $\varpred$, denoted by $\varpred[i]$, contains the address of the predecessor node, if any, of process $p_i$ after its node has been appended to the queue.
     
            The state of a process with respect to a lock has five possible values, namely \InNCS{}, \InRecover{}, \InEnter{}, \InCS{} and \InExit{}. 
            At the beginning, the state of a process, say $p_i$, is set to \InNCS{}.
            It is changed to \InRecover{} at the end of the \Recover{} \segment{} (\cref{line:recover:state:advance}).
            It is changed to \InEnter{} after 
            \begin{enumerate*}
                \item $p_i$ has initialized $\varnode[i]$ with the address of a new node (\cref{line:MCS:enter:newnode}),
                \item initialized the two fields of $\varnode[i]$ (\cref{line:MCS:enter:setnext,line:MCS:enter:setlocked}) and finally 
                \item initialized $\varpred[i]$ by setting it equal to $\varnode[i]$ (\cref{line:MCS:enter:setpred}).
            \end{enumerate*}
            It is changed to \InCS{} after $p_i$ has acquired the lock. 
            It is changed to \InExit{} when $p_i$ starts releasing the lock, and is changed to 
            \InNCS{} again after $p_i$ finishes releasing the lock.

            Our algorithm has only one \senins{}, namely the one involving the \FAS{} instruction (\cref{line:enter:FAS}). 
            Recall that a process uses this \ins{} to append its own node to the queue and also obtain the address of its predecessor node. 
            If a process fails immediately after executing this \ins{} (\cref{line:enter:FAS}) but before it is able to store its return value to the shared memory (\cref{line:enter:persist}), 
            there is no easy way to recover this address (of the predecessor) based on the current knowledge of the failed process.
            The queue continues to grow beyond this node, but it would be disconnected from the previous part of the queue, thereby creating one more sub-queue. For an example, please refer to \cref{fig:subqueue}.
            Thus, an \unsafe{} failure occurs if a process fails immediately after executing the \ins{} at \cref{line:enter:FAS}.
            
            If a process detects that it may have failed while executing the (\FAS) \ins{}, it ``relinquishes'' its current node, informs its successor (if any) that the lock is now ``free'' using the wait-free signalling mechanism described earlier, retires the current node and retries acquiring the lock using a new node. This potentially creates multiple queues (or sub-queues) which may allow multiple processes to execute their critical sections concurrently, thereby violating the ME property.
            A node is \relieve{d} (either after failure or completion of critical section) by executing the cleanup function (\crefrange{line:exit:first}{line:exit:last}). The function can be invoked from either \Recover{} or \Exit{} \segment. 
            All other \ins{s} of our algorithm, except for \FAS{}, are non-\sen{}. We achieve that by using the following ideas.

                \begin{footnotesize}
        \begin{figure}
            \centering
            \begin{tikzpicture}[draw, minimum width=1cm, minimum height=0.5cm, start chain=A going right, start chain=B going right, start chain=C]
    
                    \node (P1) [qnode, on chain=A] {$p_1$};
                    \node (P2) [qnode, on chain=A] {$p_2$};
                    \node (P3) [qnode, on chain=A] {$p_3$};
                    \node (gndA) [ground, right of=P3] {};
                    \node (P4) [qnode, below of=P1, on chain=B] {$p_4$};
                    \node (P5) [qnode, on chain=B] {$p_5$};
                    \node (P6) [qnode, on chain=B] {$p_6$};
                    \node (gndB) [ground, right of=P6] {}; 
                    \node (P7) [qnode, below of=P4, on chain=C] {$p_7$};
                    \node (P8) [qnode, on chain=C] {$p_8$};
                     \node (gndC) [ground, right of=P8] {};

                    \draw[->] (P1) -- (P2);
                    \draw[dashed, ->] (P2) -- (P3);
                    %\draw[->] (P3) -- (P4);
                    \draw[->] (P4) -- (P5);
                    \draw[->] (P5) -- (P6);
                    %\draw[->] (P6) -- (P7);
                    \draw[->] (P7) -- (P8);
                    
                    \draw[-] (P3) -- (gndA);
                    \draw[-] (P6) -- (gndB);
                    \draw[-] (P8) -- (gndC);
                    
                    \node (T) [below of=P8] {$\vartail$};
                    \draw[->] (T) -- (P8);
                    
            \end{tikzpicture}
            \captionof{figure}{Processes $p_1$ \dots $p_8$ successfully append their nodes to the tail of the queue using an \FAS{} instruction. Processes $p_4$ and $p_7$ failed to store the outcome of the \FAS{} instruction to persistent memory, and are thus unable to set the next field of the nodes of $p_3$ and $p_6$. Process $p_3$ has captured the address of the node of $p_2$ and is about to set the corresponding next field on the node of $p_2$. Effectively, three sub-queues are created due to failures of $p_4$ and $p_7$.} 
            \label{fig:subqueue}
        \end{figure}
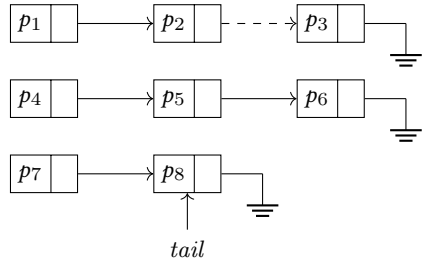
    \end{footnotesize}
            
            First, a process does not use the outcome of the \CAS{} instruction used to modify the $\varnext$ field of a node (\cref{line:enter:CAS} and \cref{line:exit:CAS|next}). 
            After performing the \CAS{} instruction on the $\varnext$ field, it reads the contents of the field again and determines its next step based on what it read. Note that, once initialized, the $\varnext$ field can only be modified once. This makes the two steps involving the \CAS{} instruction on the $\varnext$ field 
            as \emph{idempotent}; the effect of performing the \CAS{} instruction multiple times if interrupted due to failures is the same as performing it once. 
            
            Second, portions of \Recover{} and \Enter{} \segment{s} are enclosed in if-blocks to be executed conditionally.
            Intuitively, the guard of an if-block represents the pre-condition that needs to hold before its body can be executed. The outermost if-blocks use guards based on the current state of the process, which is advanced only at the end of the block. The inner if-blocks use guards based on other variables. Except for the if-block containing the \FAS{} instruction (which constitutes a \senins{}), all other if-blocks are idempotent and can be executed repeatedly if interrupted due to failures without any adverse impact starting from the evaluation of the guard
            (\crefrange{line:recover:enter|if}{line:recover:enter|endif},
            \crefrange{line:recover:exit|if}{line:recover:exit|endif},
            \crefrange{line:recover:ncs|if}{line:recover:ncs|endif},
            \crefrange{line:enter:recover|if}{line:enter:recover|endif} and
            \crefrange{line:enter:pred|if}{line:enter:pred|endif}).
            Note that if the guard of an if-block does not hold, its body is not executed.
           
            Third, similar to the case of the $\varnext$ field, a process does not use the outcome of the \CAS{} instruction used to modify the $\vartail$ pointer of the queue in the cleanup function (\cref{line:exit:CAS|tail}). After performing the \CAS{} instruction on the $\vartail$ pointer, irrespective of the outcome of the instruction, it blindly executes the remainder of the steps pertaining to signalling the successor node (\crefrange{line:exit:CAS|next}{line:exit:successor|stop}). If the node has no successor, then the steps are redundant, but have no adverse impact even if the node has already been removed from the queue by an earlier \CAS{} instruction.
            
            The functions \FnNewNode{} and \FnRetireNode{} are ``hooks'' for the memory reclamation algorithm described later.
            
            %% POLISH - RetireNode is run to completion at least once before 
            %% NewNode is invoked
            %% Should be mentioned when discussing memory reclamation
            
            We refer to the algorithm for a weakly recoverable lock described in this section as \weakMCS{}.
            Note that \crefrange{line:enter:doorway|begin}{line:enter:doorway|end}, which do not contain any loop, constitute the \emph{doorway} of \weakMCS{} whereas \crefrange{line:enter:waitingroom|begin}{line:enter:waitingroom|end} constitute its \emph{waiting} room.

    \subsection{Correctness proof}
    \label{sec:MCS:proof}

          We prove that \weakMCS{} is a \responsive{} weakly recoverable mutual exclusion algorithm. 
          We first define some concepts and notations used in our proofs. All definitions use the notion of \emph{current} time.
          
          The \weakMCS{} uses a queue node to synchronize among processes and serialize their critical section executions. 
          In the absence of failures, a process \emph{allocates} a new node at the beginning of a passage and \emph{\relieve{s}}  it at the end of that passage. 
          However, in the presence of failures, a node allocated during one passage may be \relieve{d} during a different passage.
          Every node is \emph{owned} by a (unique) process. Let $\owner{u}$ denote the owner process of the node $u$.
          A node becomes \emph{\vital} once it has been \emph{appended} to the queue
          by storing its address in the \mytail{} pointer using an \FAS{} instruction (\cref{line:enter:FAS}). It follows from code inspection that:
          
          \begin{proposition}
          A node is appended to the queue at most once. 
          \end{proposition}

         Note that \vital{} nodes can be \emph{ordered} based on the sequence in which they were appended to the queue.
         We refer to the last \vital{} node to be appended to the queue as the \emph{tail} node. Note that

         If a node $v$ is appended to the queue immediately after a node $u$, then then we say that $v$ is the \emph{successor} of $u$ and there is an \emph{edge} from $v$ to $u$.
         We use $\successor{u}$ to denote the successor of node $u$ and $\link{v}{u}$ to the denote the edge from node $v$ to node $u$ provided $v = \successor{u}$.  
         Intuitively, the edge $\link{v}{u}$ captures the \emph{wait-for} dependency that $v$ has on $u$. In particular, $\owner{v}$  cannot enter its  \CS{} \segment{} until $\owner{u}$ has \relieve{d} $u$.
         The set of \vital{} nodes along with the set of edges, as defined above, form a chain referred to as \emph{global chain}.

         %% POLISH - remove active node - use only admissible and release
         
         A node is said to have been \emph{\relieve{d}} if its owner has executed either the \CAS{} instruction at \cref{line:exit:CAS|next} successfully (to signal its successor process) or the write instruction at  \cref{line:exit:CAS|next|locked} (to unlock the successor node).
 
         A node is said to be \emph{\admissible} if it is \vital{} but not \relieve{d}.
         
         The following proposition follows from code inspection:
                   
         \begin{proposition}
         \label{proposition:one:progress}
         A process can own at most one \admissible{} node in the global chain.  Further, if a process owns an \admissible{} node in the global chain, then it has a super-passage in progress.
         \end{proposition}

         Consider two nodes $u$ and $v$ such that $v = \successor{u}$ and let $p = \owner{v}$. The $\link{v}{u}$ is said to be \emph{\admissible} if the following conditions hold.
         First, $u$ is \admissible{}. 
         Second, when $p$ appended  $v$ to the queue, the \FAS{} instruction returned the address of $u$. 
         Third, since $p$ appended $u$ to the queue,  either $p$ has not failed  or has stored the address of $u$ in the persistent memory by executing the write instruction at \cref{line:enter:persist}.
         Intuitively, if the edge $\link{v}{u}$ is not \admissible{}, then either $\owner{u}$ has already \relieve{d} $u$ or $\owner{v}$ has lost the address of $u$.
    
         We call any \emph{non-empty} connected set of nodes in the global chain as a \emph{fragment}. 
         A fragment is said to be \emph{\admissible} if all its nodes as well as all its edges are \admissible{}. An \admissible{} fragment is said to be \emph{maximal} if it is not strictly contained in another \admissible{} fragment.
          
         We define \emph{sub-queue} as a maximal \admissible{} fragment. Note that each sub-queue has a \emph{\front} node (no outgoing \admissible{} edge) and a \emph{\back} node (no incoming \admissible{} edge).

    \begin{footnotesize}
        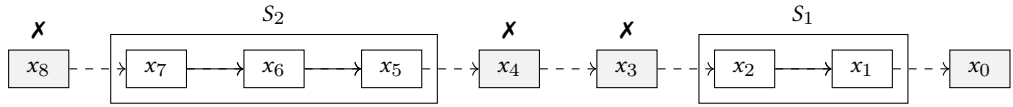
\begin{figure}
            \centering
            \begin{tikzpicture}[draw, minimum width=8mm, minimum height=5mm, start chain=A going left, node distance=7.5mm]
    
            \foreach \x/\c in {0/5,1/0,2/0,3/5,4/5,5/0,6/0,7/0,8/5} {
                \node [node, on chain=A, fill=black!\c] (N\x) {$x_\x$};
            }
            
            \foreach \x in {3,4,8} {
                \node [label={above:\xmark}] at (N\x) {};
            }
            
            \foreach \x in {0,1,2,3,4,5,6,7} {
                \pgfmathsetmacro\y{\x+1}
                \draw[->, dashed] (N\y) -- (N\x);
            }
            
            \foreach \x in {1,5,6} {
                \pgfmathsetmacro\y{\x+1}
                \draw[->] (N\y) -- (N\x);
            }
            
            \node [draw, fit=(N1)(N2), inner sep=2mm, label={above:$S_1$}] {};
            \node [draw, fit=(N5)(N7), inner sep=2mm, label={above:$S_2$}] {};
            
            \end{tikzpicture}
            \captionof{figure}{An illustration of the notions used in the correctness proof of \weakMCS. The global chain contains nine nodes, labeled $x_0$ to $x_8$. In the figure,
            (a)~unshaded nodes represent \admissible{} nodes while shaded nodes represent non-\admissible{} nodes, and 
            (b)~solid lines represent \admissible{} edges while dashed lines represent non-\admissible{} edges.
            A node that was abandoned by its owner due to an
            \unsafe{} failure has been labeled with `\xmark'. 
            Nodes that belong to the same maximal \admissible{} fragment or sub-queue have been enclosed in a rectangle. As shown, the global chain contains two sub-queues, given by $S_1 = \langle x_1, x_2 \rangle$ and $S_2 = \langle x_5, x_6, x_7 \rangle$. Nodes $x_5$  and $x_6$ are the \front{} and \back{} nodes, respectively, of the sub-queue $S_2$. Finally, node $x_8$ is the tail node of the global chain.} 
            \label{fig:globalchain}
        \end{figure}
    \end{footnotesize}

         \Cref{fig:globalchain} gives an illustration of the concepts defined so far.
         
         The following propositions follow from the above concepts.
          
         \begin{proposition}
         \label{proposition:distinct:disjoint}
         Any two distinct sub-queues are node disjoint. 
         \end{proposition}
         
         \begin{proposition}
         \label{proposition:CS:front}
         If a process owns a node in some sub-queue, then it has a super-passage in progress. Moreover, if a process is in its \CS{}, then it owns the \front node of some sub-queue.
         \end{proposition}

         \begin{proposition}
         \label{proposition:front:release}
          Assume that every process fails only a finite number of times during each of its super-passage.  If a process owns the \front node of some sub-queue, then it eventually \relieve{s} the node.
         \end{proposition}

         \begin{proposition}
         \label{proposition:front:complete}
          If a process owns the \front node of some sub-queue and never fails thereafter in its super-passage currently in progress, then the super-passage eventually completes.
         \end{proposition}
          
         Note that every \vital{} non-tail node has an \emph{incoming} edge. Consider an \admissible{} non-tail node $u$ and let $v = \successor{u}$. Note that, just after $v$ is appended to the queue, the edge from $v$ to $u$ is \admissible{}. Thus, if the edge from $v$ to $u$ is not \admissible{} now, then it implies that $\owner{v}$ failed before it was able to store the address of $u$ in the persistent memory.  If this happens, then we say that the edge from $v$ to $u$ (whose 
         status changed from \admissible{} to non-\admissible) has been \emph{disrupted} by the \unsafe{} failure. 
         
         The following proposition follows from above concepts.
          
         \begin{proposition}
         \label{proposition:disrupt:unsafe}
         The incoming edge to an \admissible{} non-tail node can only be disrupted by an \unsafe{} failure. Moreover, an \unsafe{} failure can disrupt at most one edge in the global chain.  
         \end{proposition}

         Based on the above propositions, we prove the following results about \weakMCS.
          
         \begin{lemma}
         \label{lemma:subqueues}
         Given a history $H$ and a non-negative integer $\fails$, if at least $\fails$ processes are  in their critical sections simultaneously, then the global chain contains at least $\fails$  distinct sub-queues.
         \end{lemma}
         \begin{proof}
         The lemma follows from \cref{proposition:distinct:disjoint,proposition:CS:front}.  
         \end{proof}

          %% We use the above proposition to argue that \weakMCS{} is \responsive. 
        
        %% DONE

        \begin{theorem}
        \label{theorem:MCS_safety}
        Given a history $H$, and a non-negative integer $\fails$, if $\fails + 1$ processes are in their critical sections simultaneously, then there exists at least $\fails$ \unsafe{} failures  whose \consequence{} interval is still \inprogress. 
        \end{theorem}
        \begin{proof}
        %%
        %% In this proof, unless otherwise stated, the ``current time'' refers to time $t$.
        %%
        It follows from \cref{lemma:subqueues} that the global chain contains at least $\fails + 1$ distinct sub-queues. Clearly, the \back node of all sub-queues, except possibly for one, is a \emph{non-tail} node. 
        It follows from \cref{proposition:disrupt:unsafe} that the incoming edge of every non-tail \back node was disrupted by a \emph{unique} \unsafe{} failure. 
        
        Now, consider an arbitrary sub-queue whose \back node, say $u$, is a non-tail node. Note that the global chain contains at least $\fails$ such nodes.  Let $v = \successor{u}$ and let $f$ be the unsafe failure that disrupted $\link{v}{u}$.  
        Clearly, $u$ is appended to the queue \emph{before} $v$, and $v$ is appended to the queue \emph{before} $f$ occurred. Using \cref{proposition:one:progress}, it implies that $\owner{u}$ has a super-passage currently in progress that started before $f$ occurred. In other words, the \consequence{} interval of $f$ is still active.  
        \end{proof}

          %% MODIFY
          
          \begin{theorem}
          \label{theorem:MCS_SF}
          \weakMCS{} satisfies the SF property.
          \end{theorem}
          \begin{proof}
          Assume that every process fails only a finite number of times during each of its super-passage. Assume, on the contrary, that some super-passage, say $\Super$, belonging to a process, say $p$, never completes. 
          Let $\sigma$ denote the first passage that $p$ starts after recovering from its last failure in $\Super$. By assumption, $p$ never fails during $\sigma$. Clearly, $p$ eventually advances to \cref{line:enter:doorway|end} during $\sigma$ at which point it is guaranteed to own an \admissible{} node, say $u$, in the global chain.
          By applying \cref{proposition:front:release} repeatedly, we can infer that $u$ eventually becomes the \front node of its sub-queue. By applying \cref{proposition:front:complete}, we can infer that $\Super$ eventually completes---a contradiction.
          \end{proof}

          %% DONE
          
          \begin{theorem}
            \label{theorem:MCS_BCSR}
            \weakMCS{} satisfies the BCSR property.
          \end{theorem}
          \begin{proof}
            Assume that some process, say $p_i$, fails while executing its \CS{} \segment{} implying that 
            $\varstate[i] = \InCS$ at the time of failure. Upon restarting, $p_i$ executes \NCS{}, \Recover{} and \Enter{} \segment{s} in that order. As the code inspection shows, since $state[i] = \InCS{}$, $p_i$ only evaluates a constant number of if-conditions, all of which evaluate to \false, and is therefore able to proceed to the \CS{} \segment{} in a bounded number of its own steps.  
          \end{proof}
          
          It follows from \cref{theorem:MCS_safety,theorem:MCS_BCSR,theorem:MCS_SF} that

          \begin{theorem}
            \label{theorem:MCS_responsive}
            \weakMCS{} is a \responsive{} weakly recoverable mutual exclusion algorithm.
          \end{theorem}

          \begin{theorem}
          \weakMCS{} satisfies the BR and BE properties.
          \end{theorem}
          \begin{proof}
            As the code inspection shows, \Recover{} and \Exit{} \segment{s} do not involve any loops. Thus, a process can execute these \segment{s} within a bounded number of its own steps. Hence, \weakMCS{} satisfies the BR and BE properties.
          \end{proof}

    \subsection{Complexity analysis}   
          
          %% CHECKED
          
          \begin{theorem}
          The RMR complexity of each of \Recover{}, \Enter{} and \Exit{} \segment{s} of \weakMCS{} is $\bigO{1}$.
          \end{theorem}
          \begin{proof}
            As the code inspection shows, \Recover{} and \Exit{} \segment{s} do not contain any loop and only contain a constant number of steps. The \Enter{} \segment{}, however has one loop at \cref{line:enter:await} of \cref{algo:weak|MCS}, but otherwise contains a constant number of steps. This loop involves waiting on a boolean variable until it becomes \true{} and the variable can be written to only once. This incurs only $\bigO{1}$ RMRs in the CC model. In the DSM model, this variable is mapped to a location in local memory module. Hence, the RMR complexity of the \Enter{} \segment{} is also $\bigO{1}$ in both CC and DSM models.
          \end{proof}
   %%%%%%%%%%%%%%%%%%%%%%%%%%%%%%%%%%%%%%%%%%%%%%%%%%%     

\section{A Strongly Recoverable \SEfficient{} Lock}
\label{sec:framework}
    In this section, we describe a framework that uses other types of recoverable locks with certain properties as building blocks to construct a lock that is not only strongly recoverable but also \sefficient{} under both CC and DSM models. We describe our (\sefficient{}) lock in two steps. We first describe a basic framework to transform a \bounded{} \nonadaptive{} strongly recoverable lock to a \bounded{} \cadaptive{} strongly recoverable lock. We then extend this framework to make the lock \sadaptive{} while ensuring that it stays strongly recoverable and \bounded{}. Finally, instantiating the framework with an appropriate \wbounded{} \nonadaptive{} lock yields the desirable \sefficient{} lock.
    
    The basic framework is based on the one used by Golab and Ramaraju in~\cite[Section~4.2]{GolRam:2019:DC} to construct a strongly recoverable lock that is \cadaptive. Specifically, in their framework, Golab and Ramaraju use two different types of strongly recoverable locks, referred to as base lock and auxiliary lock, along with two other components to build another strongly recoverable lock, referred to as target lock. The target lock constructed is \bounded{} \cadaptive{} based on the base lock that is \fadaptive{} and the auxiliary lock that is \nonadaptive. They achieve this by customizing the base lock so that, upon detecting a failure, processes can abort their attempts and reset the (base) lock. 
    In the presence of failures (even a single failure), the RMR complexity of the target lock is dominated by the overhead of aborting the attempt to acquire the base lock and then resetting the base lock, thereby making the lock \cadaptive. In the rest of the text, we use the term ``target lock'' to refer to the (strongly recoverable) lock we want to build.
   
    \subsection{A \wbounded{} \cadaptive{} RME algorithm}
    \label{sec:framework|basic}
        
        \subsubsection{Building blocks}
        We use four different components as building blocks. 
        
        \begin{itemize}
            \item \emph{\titlecap{\filter{}} lock:} A \responsive{} weakly recoverable lock that provides mutual exclusion in the absence of failures. We use an instance of the lock proposed in \autoref{sec:weak|MCS}, which has $\bigO{1}$ RMR complexity for all three failure scenarios.
        
            \item \emph{\titlecap{\splitter}:} Used to split processes into \emph{fast} or \emph{slow} paths. If multiple processes navigate the \splitter{} concurrently (which would happen only if an \unsafe{} failure has occurred with respect to the \filter{} lock), only one of them is allowed to take the fast path and the rest are diverted to the slow path. In other words, the \splitter{} is \emph{biased}. Intuitively, it can be viewed as a strongly recoverable \emph{try} lock. It is implemented using an atomic integer and a \CAS{} instruction, which has $\bigO{1}$ RMR complexity for all three failure scenarios.
        
            \item \emph{\titlecap{\arbitrator{}} lock:} A \emph{dual-port} strongly recoverable lock. Each port corresponds to a side. We refer to the two sides as \emph{\LEFT} and \emph{\RIGHT}. At any time, at most one process should be allowed to attempt to acquire the lock from any side. However, \emph{any two} of the $\n$ processes can compete to acquire the lock. We use the implementation of the dual-port RME algorithm proposed by Golab and Ramaraju in~\cite[Section~3.1]{GolRam:2019:DC} (a transformation of Yang and Anderson's mutual exclusion algorithm to add recoverability), which has $\bigO{1}$ RMR complexity for all three failure scenarios.
       
            \item \emph{\titlecap{\core{}} lock:} a (presumably \nonadaptive{}) strongly recoverable lock that assures mutual exclusion among processes taking the slow path. We may use an instance of any of the existing RME algorithms.

        \end{itemize}

        Note that our target lock satisfying BCSR, BR and BE properties is contingent upon \filter{}, \arbitrator{} and \core{} locks satisfying BCSR, BR and BE properties. 
        
        \subsubsection{The execution flow}
        
        In order to acquire the \target{} lock, a process proceeds as follows. It first waits to acquire the \filter{} lock. Once granted, it navigates through the \splitter{} trying to enter the fast path. If successful, it then attempts to acquire the \arbitrator{} lock from \theleftside{}. If one or more failures occur that are \unsafe{} with respect to the \filter{} lock, then multiple processes may acquire the \filter{} lock simultaneously. If this results in contention at the \splitter{}, then all but one processes are diverted to the slow path. If a process is forced to take the slow path, it attempts to acquire the \core{} lock. Once granted, it then waits to acquire the \arbitrator{} lock from \therightside{}. Finally, once the process has successfully acquired the \arbitrator{} lock, it is deemed to have acquired the \target{} lock as well, and is now in the \CS{} of the \target{} lock.
        
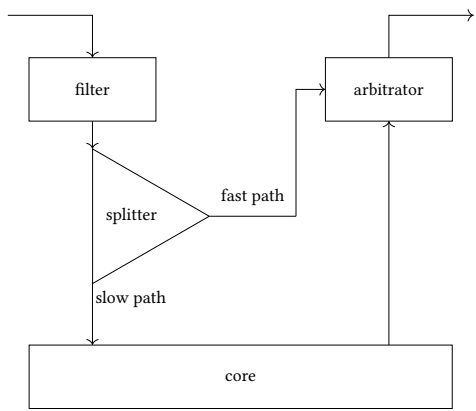
\begin{figure}[t]
	\scalebox{0.225}{
		\centering
		\begin{tikzpicture}[draw, 
							minimum width=7.5cm, 
							minimum height=3.75cm, 
							every node/.style={transform shape, font=\HUGE},
							triangle/.style = {regular polygon, regular polygon sides=3},
							border rotated/.style = {shape border rotate=270},
							decoration={markings, mark=at position 1 with {\arrow[scale=4,black]{>}}}]

		   \node (W) [rectangle, draw=black] {\filter};
		   \node (B) [rectangle, draw=black, right=10cm of W] {\arbitrator};
		   \node (X) [triangle,  draw=black, border rotated, minimum size=1cm, below=7.5cm of W.center, anchor=west] {\splitter};
		   \node (IA) [minimum size=0cm, below=2.5cm of X.corner 3] {};
		   \node (IB) [minimum size=0cm, right=5cm of X.corner 1] {};

		   \node (Box)  [transform shape=false, fit=(W)(B)(IA.center), inner sep=0mm] {};
			 
		   \draw[postaction={decorate}] (W) -- (X.corner 2);
		   \draw[-] (X.corner 1) -- (IB.center) node[pos=0.5, above, yshift=-0.75cm]{fast path};
		   \draw[postaction={decorate}] (IB.center) |- (B.west);
		   
		   \node (S) [rectangle, draw=black, below=1.0cm of Box, minimum width=25cm] {\core};

		   \draw[postaction={decorate}] (X.corner 3) -- ([shift={(3.75cm,0)}]S.north west) node[pos=0.25, right, xshift=-1.5cm]{slow path};
		   \draw[postaction={decorate}] ([shift={(-3.75cm,0)}]S.north east) -- (B);   
			 
		   \draw[postaction={decorate}] ([shift={(-5cm,2.5cm)}] W.north) -| (W.north);     
		   \draw[postaction={decorate}] (B.north) |- ([shift={(5cm,2.5cm)}] B.north);
		\end{tikzpicture}
		}
		\captionof{figure}{A pictorial representation of the framework.} 
		\label{fig:framework2}
	\end{figure}

        A pictorial representation of the execution flow is depicted in \autoref{fig:framework2}. Note that the pictorial representation depicts the two sides of the \arbitrator{} lock as left and bottom, which actually correspond to \theleftside{} and \therightside{} of the \arbitrator{} lock respectively.
        
        In the absence of failures, every process takes the fast path, albeit one at a time. However, some processes do take the fast path even if their super-passage overlaps with the \consequence{} interval of an \unsafe{} failure with respect to the \filter{} lock.
        Note that at most one process can take the fast path at a time and at most one process can hold the \core{} lock at a time. Any process that takes the fast path will always attempt to acquire the \arbitrator{} lock from \theleftside{}. Any process that takes the slow path and acquires the \core{} lock will always attempt to acquire the \arbitrator{} lock from \therightside{}. Since the \core{} lock is strongly recoverable, at most one process will try to acquire the \arbitrator{} lock from each side at a time.
        
        In order to release the \target{} lock, a process simply releases its component locks in the reverse order in which it acquired them: the \arbitrator{} lock, followed by the \core{} lock (in case the process took the slow path), followed by the \splitter{} and finally the \filter{} lock.
         
        The RMR complexity of the fast path is given by the sum of the RMR complexities of the \filter{} lock, the \splitter{} and the \arbitrator{} lock. On the other hand, the RMR complexity of the slow path is given by the sum of the RMR complexities of the \filter{} lock, the \splitter{}, the \core{} lock and the \arbitrator{} lock. 
            
        For ease of exposition, we use the following terminology. Before a process is assigned a particular path, we refer to it as a \emph{\normal} process. It is classified as a \emph{\fast} process if it takes the fast path and a \emph{\slow} process otherwise. A \slow{} process becomes a \emph{\medium} process once it acquires the \core{} lock.
        
     \addprefix[F.]
    
     \begin{algorithm}[t]
            \begin{multicols}{2}
            \SetKw{Shared}{shared variables}
            \SetKw{Local}{local variables}
            \SetKw{Struct}{struct}
            \SetKw{Integer}{int}
            \SetKw{Boolean}{bool}
            \SetKw{Array}{array}
            
            \SetKw{Definitions}{definitions}
            \SetKwProg{initialization}{initialization}{\\ begin}{end}
            \SetKwProg{function}{Function}{\\ begin}{end}
            	
            \Shared \\
            \Indp
            
            $\symfilter$: $\n$-process weakly recoverable lock\;
            \tcc{to implement \splitter{} - used to store the identifier of the process currently occupying the fast path}
            $\x$: integer variable\;
            \tcc{\core{} lock}
            $\symcore$: $\n$-process strongly recoverable lock\;
            \tcc{\arbitrator{} lock }
            $\symarbitrator$: $\n$-process dual-port strongly recoverable lock\;
            \tcc{path of the process; in the DSM model, the $i$-{th} entry is local to process $p_i$}
            $\varpath$: \parbox[t]{2in}{\Array $[1{\dots}\n]$ of boolean variables}\;
            \tcc{$\varpath[i] \in$ \{\FAST, \SLOW\}} 
            
            \Indm
            
			\BlankLine
			
			\initialization{}{
			   $\x$ $\leftarrow$ 0\tcp*[r]{fast path is empty}
			
			   \ForEach{$j \in \{ 1, 2, \dots, \n\}$}
			   {
			       $\varpath[j]$ $\leftarrow$ \FAST\tcp*[r]{default path type}
			   }
		    }
			
			\BlankLine

			\Definitions \\

			   \quad $\varside(\varpath) = 
                \begin{cases}
                    \LEFT{} & \text{if } \varpath = \FAST{} \\
                    \RIGHT{} & \text{otherwise}
                \end{cases}
               $

			\BlankLine

			\function{Recover(~)}
			{
			    
			   \tcc{In order to follow the execution model of a lock described in \autoref{sec:model|problem} 
			   (\NCS, \Recover, \Enter, \CS, \Exit{} in that order), we execute the \Recover{} \segment{}
			   of each of the recoverable locks ($\symfilter$, $\symcore$ and $\symarbitrator$) 
			   just prior to executing their respective \Enter{} \segment{s}}
			    %% \tcp{Thus, no special actions are required for recovery}
		 	}

			 \columnbreak

			 \function{Enter(~)}
			 {
			 	$\symfilter$.Recover(~)\tcp*[r]{recover the \filter{} lock}
			    $\symfilter$.Enter(~)\tcp*[r]{acquire the \filter{} lock}

			    \If(\tcp*[f]{not yet on the slow path}){($\varpath[i]$ $\neq$ \SLOW)}
			    {
			       \CAS( $\x$, 0, $i$ )\tcp*[r]{attempt to take the fast path}
			       \label{line:path|fast|CAS}
			    }

			    \If(\tcp*[f]{unable to take the fast path}){($\x$ $\neq$ $i$)}
			    {
			       $\varpath[i]$ $\leftarrow$ \SLOW\tcp*[r]{committed to take the slow path}
			       \label{line:path|slow}
			       $\symcore$.Recover(~)\tcp*[r]{recover the \core{} lock}
			       $\symcore$.Enter(~)\tcp*[r]{acquire the \core{} lock}

			    }
			    
			    $\symarbitrator$.Recover($\varside(\varpath[i]$))\tcp*[r]{recover \arbitrator{} lock}
			    $\symarbitrator$.Enter($\varside(\varpath[i])$)\tcp*[r]{acquire the \arbitrator{} lock} 
			    
			 }           
			 
			 \BlankLine

			 \function{Exit(~)}
			 { 
			    $\symarbitrator$.Exit($\varside(\varpath[i])$)\tcp*[r]{release the \arbitrator{} lock}
			    
			    %% \uIf(\tcp*[f]{took the slow path}){($\x$ $\neq$ $i$)}
			    \uIf(\tcp*[f]{took the slow path}){($\varpath[i]$ = \SLOW)}
			    {
			     	$\symcore$.Exit(~)\tcp*[r]{release the \core{} lock} 
			    	
			    } \Else(\tcp*[f]{took the fast path})
			    {
			      $\x$ $\leftarrow$ 0\tcp*[r]{the fast path is now empty}
			    }
			    $\varpath[i]$ $\leftarrow$ \FAST\tcp*[r]{reset the path type to default}
			    \label{line:path|reset}
			    $\symfilter$.Exit(~)\tcp*[r]{release the \filter{} lock}           
			 }

            \end{multicols}
            \caption{Pseudocode of process $p_i$ for the framework to design a \cadaptive{} lock.}
            \label{algo:framework|semi}
    \end{algorithm}

        The pseudocode is given in \autoref{algo:framework|semi}. The pseudocode closely follows the above description in text. A \splitter{} is implemented using an integer (shared) variable, namely $\x$. The fast path is occupied if and only if $\x$ has a non-zero value, in which case the value refers to the identifier of the process currently occupying the fast path. To take the fast path, a process attempts to store its own identifier in $\x$ using a \CAS{} instruction provided its current value is zero (\autoref{line:path|fast|CAS}). If the attempt fails, the process changes its path type to \SLOW{} (\autoref{line:path|slow}). Note that a process resets its path type from \SLOW{} to its default value of \FAST{} only after it has executed the \Exit{} \segment{} of the \core{} lock at least once without encountering any failure (\autoref{line:path|reset}). 
        
        In Golab and Ramaraju's framework, even if a process takes the fast path, it may still incur $\bigOmega{\n}$ RMR complexity in the presence of even a single failure because of the overhead of aborting an attempt and then resetting the base lock, which is an expensive operation. 
        In our framework, on the other hand, a process taking the fast path incurs only $\bigO{1}$ RMR complexity even with arbitrary number of failures because the RMR complexity of acquiring the \filter{} lock, followed by navigating the \splitter{} to take the fast path and finally acquiring the \arbitrator{} lock is only $\bigO{1}$ irrespective of the number of failures.
        %%        
        %% Also, note that a slow process is created only if an \unsafe{} failure occurs with respect to the \filter{} lock.

\subsubsection{Correctness proof and complexity analysis}
    
    We refer to the algorithm described in the previous section as \emph{\salock}.
    The doorway of \salock{} is given by the doorway of its \filter{} lock.
    When convenient, we use $\symfilter$ and $\symcore$ to refer to the \filter{} and \core{} locks, respectively, of \salock.

    %% DONE
    
    \begin{theorem}
            \salock{} satisfies the ME property.
            \label{theorem:algo_ME}
        \end{theorem}
        \begin{proof}
            A process enters the \CS{} \segment of  \salock{} after acquiring the \arbitrator{} lock from one of the sides.
            The \arbitrator{} lock satisfies the ME property as long as no more than one process attempts to acquire it from either side \LEFT{} or \RIGHT{} at any time.
            The \splitter{} ensures that, at any time, at most one process attempts to acquire the \arbitrator{} lock from \theleftside{}. 
            The \core{} lock ensures that, at any time, at most 
            one process attempts to acquire the \arbitrator{} lock from \therightside{}.
            Therefore, \salock{} satisfies the ME property.
        \end{proof}

        \begin{theorem}
             \salock{} satisfies the SF property.
            \label{theorem:algo_SF}
        \end{theorem}
        \begin{proof}
        Assume that every process fails only a finite number of times during each of its super-passage.
        
        All three locks used in the framework, namely \filter, \core{} and \arbitrator, individually satisfy SF and BCSR properties. Also, navigating a \splitter{} involves executing a constant number of instructions.
        
        Consider an infinite fair history $H$, a process $p$ that is live in $H$ and a super-passage $\Super$ of $p$ in $H$.  
        Clearly, there exists a time after which $p$ does not fail any more in $\Super$. Let $t$ denote the \emph{earliest} such time. Consider the first passage that $p$ starts after time $t$. 
        
        Note that either the SF or BCSR property of the \filter{} lock guarantees that $p$ eventually leaves the \Enter{} \segment{} of the \filter{} lock and enters its \CS{} \segment{}. 
        Process $p$ then completes navigating the \splitter{} within a bounded number of its own steps. Note that, if it was able to enter the fast (respectively, slow) path during an earlier passage of $\Super$, then it is guaranteed to take the fast (respectively, slow) path again in this passage. 
        As in the case of \filter{} lock, we can argue that $p$ is guaranteed to enter the \CS{} \segment{} of the \core{} lock if it takes the slow path during this passage.
        Likewise, we can argue that $p$ is guaranteed to enter the \CS{} \segment{} of the \arbitrator{} lock, which in turn implies that $p$ is guaranteed to enter the \CS{} segment of the target lock.
        \end{proof}

        \begin{theorem}
            \salock{} satisfies the BCSR property.
            \label{theorem:algo_BCSR}
        \end{theorem}
        \begin{proof}
            If some process $p_i$ is in the \CS{} \segment{} of \salock{}, then it currently holds 
            the \filter{} lock and it either
            \begin{enumerate*}[label={(\alph*)}] 
                \item acquired the \arbitrator{} lock from \theleftside{} by taking the fast path or
                \item acquired the \core{} lock first and then acquired the \arbitrator{} lock from \therightside{} by taking the slow path. 
            \end{enumerate*}
            
            If $p_i$ fails in the \CS{} \segment{} of \salock, it determines the path it took
            by checking the $\varpath[i]$ variable and then retraces the same steps it had executed earlier. Since the \filter{} lock, the \core{} lock as well as the \arbitrator{} lock satisfy the BCSR property, and the fact that the \splitter{} is wait-free, $p_i$ is guaranteed to be able to acquire the requisite locks and  reenter the \CS{} \segment{} of \salock{} within a bounded number of its own steps. Hence, \salock{} satisfies the BCSR property.
        \end{proof}

        %% DONE
        
        \begin{theorem}
           \salock{} satisfies the BE and BR properties.
           %BE is dependent on BE of core?
            \label{theorem:algo_BE_BR}
        \end{theorem}
        \begin{proof}
            The \Recover{} \segment{} of \salock{} is empty and hence it trivially satisfies the BR property.
            
            As part of the \Exit{} \segment{} of \salock{}, a
            process executes the \Exit{} \segment{} of
            the \arbitrator{} lock, optionally followed by the \Exit{} \segment{} of the \core{} lock, followed by the \Exit{} \segment{} of the \filter{} lock.
            Since each of three locks individually satisfy the BE property, and the \splitter{} is wait-free, it follows that \salock{} also satisfies the BE property.
        \end{proof}
        
        It follows from \cref{theorem:algo_ME,theorem:algo_SF,theorem:algo_BCSR,theorem:algo_BE_BR} that
        
        \begin{theorem}
            \label{theorem:framework|RME}
            \salock{} is a strongly recoverable lock.
        \end{theorem}
        
        %% DONE
        
        \begin{theorem}[\salock{} is \bounded{} \cadaptive{}]
            The RMR complexity of any passage in a super-passage of \salock{} is $\bigO{1}$ if the \hazard{} of the super-passage is zero and $\bigO{R(\n)}$ otherwise, where $R(\n)$ denotes the worst-case RMR complexity of the \core{} lock for $\n$ processes.
        \end{theorem}
        \begin{proof}
            In the absence of failures, \emph{only one} process can successfully acquire the \filter{} lock (\cref{def:wRME}). This process navigates the \splitter{} in $\bigO{1}$ steps, 
            takes the fast path and acquires the \arbitrator{} lock from \theleftside{} (skipping the \core{} lock along the way). The RMR complexity of the \arbitrator{} lock is $\bigO{1}$. Thus, in this case, the RMR complexity of the \target{} lock is given by $\bigO{1}$.
            
            In the presence of failures, all $\n$ processes may be able to successfully acquire the \filter{} lock and proceed to the \splitter{}. Only one of these processes is allowed to take the fast path, which then
            attempts to acquire the \arbitrator{} lock from \theleftside{}. The remaining $(\n-1)$ processes are diverted to the slow path and have to acquire the \core{} lock and then acquire the \arbitrator{} lock from \therightside{}. Thus, in this case, the RMR complexity of the \target{} lock is given by $\bigO{R(\n)}$. 
        \end{proof}
        
        %% DONE
        
        \begin{theorem}[A \wbounded{} \cadaptive{} lock]
            Assume that we use \JJJ{'s} or \KM{'s} RME algorithm~\cite{JayJay+:2019:PODC,KatMor:2020:OPODIS} to implement the \core{} lock.
            Then, the RMR complexity of any passage in a super-passage of \salock{} is $\bigO{1}$ if the \hazard{} of the super-passage is zero and $\bigO{\nicefrac{\log n}{\log\log \n}}$ otherwise. 
        \end{theorem}

        \begin{comment}
        
        Note that at any given time, each process that attempts to acquire the \core{} lock in a particular super-passage has attempted to acquire the \filter{} lock in the same super-passage (with respect to the \target{} lock). In fact, the processes that attempt to acquire the \core{} lock at any time is a proper subset of the number of processes that acquire the \filter{} lock. Moreover, the number of processes that attempt to acquire the \core{} lock corresponds to the number of failures in the \filter{} lock since the \filter{} lock is \responsive{} (\cref{theorem:MCS_responsive}). To capture this notion, we define the following terms($\procinlock, \failinlock, \Super$ and formally present this relation in \cref{theorem:failure_count|base}.
        
        \end{comment}

        In the rest of this subsection, we prove an important lemma that is crucial to establish that the lock described in the next subsection is \sadaptive. Recall that the notions of super-passage and \unsafe{} failures are relative to a specific lock. 
           
        Intuitively, the set of processes that attempt to acquire the \core{} lock is strictly smaller than the set of processes that attempt to acquire the \filter{} lock. Further, the size of the former set depends on the number of \unsafe{} failures that have occurred with respect to the \filter{} lock.
        To capture this formally, we first define some notations.
        Let $\setP{\ell}{t}$ denote the set of processes that have a super-passage in progress with respect to lock $\ell$ at time $t$.
        %%
        \begin{comment}
        
        begun executing the \Enter{} \segment{} of the lock $\ell$ before or at time $t$, but have not begun executing the corresponding \Exit{} \segment.
        
        \end{comment}
        %%
        Also, let $\setF{\ell}{t}$ denote the set of all failures that are \unsafe{} with respect to the lock $\ell$ and whose \consequence{} interval extends at least until time $t$. 
        Further, if a process $p$ has a super-passage in progress with respect to the target lock at time $t$, then we use $\setSuper{p}{t}$ to denote the super-passage of $p$ with respect to the target lock at time $t$.

        Recall that $\symfilter$ and $\symcore$ refer to the \filter{} and \core{} locks, respectively, of \salock.
        
        \begin{lemma}
        \label{lemma:failure_count|base}
            Consider a time $t_{\symcore}$ such that $\card{\setP{\symcore}{t_{\symcore}}} > 0$. Then there exists time $t_{\symfilter}$ with $t_{\symfilter} \leq t_{\symcore}$ such that the following properties hold.
            
            \begin{enumerate}[label=(\alph*)]
                \item
                $\setP{\symcore}{t_{\symcore}} \subsetneq \setP{\symfilter}{t_{\symfilter}}$,
                \label{enum:core:filter}
                \item
                $\forall p \in \setP{\symcore}{t_{\symcore}}$, $\setSuper{p}{t_{\symfilter}} = \setSuper{p}{t_{\symcore}}$,
                \label{enum:same:super}
                and
                \item $\card{\setF{\symfilter}{t_{\symfilter}}} \geq \card{\setP{\symcore}{t_{\symcore}}}$.
                \label{enum:unsafe:core}
            \end{enumerate}
        \end{lemma}
        \begin{proof}
            None of the processes in the set $\setP{\symcore}{t_{\symcore}}$ was able to take the fast path while navigating the \splitter{}. Let $q$ be the \emph{last} process in $\setP{\symcore}{t_{\symcore}}$ to read  the contents of the variable $\x$ and $t$ denote the time when it performed the read step.  Clearly, $t \leq t_{\symcore}$. Furthermore, let $r$ denote the process whose identifier was stored in $\x$ when $q$ read its contents. 
            We set $t_{\symfilter}$ to $t$. We now prove each property one-by-one.
        
            \begin{enumerate}[label=(\roman*)]
            
                \item Due to the arrangement of the locks, each process in the set $\setP{\symcore}{t_{\symcore}}$ holds the lock $\symfilter$ at time $t_{\symfilter}$, Moreover, process $r$ also holds the lock $\symfilter$ at time $t_{\symfilter}$. In other words, $\setP{\symcore}{t_{\symcore}} \subseteq \setP{\symfilter}{t_{\symfilter}}$, $r \in \setP{\symfilter}{t_{\symfilter}}$ and $r \not\in \setP{\symcore}{t_{\symcore}}$.  Thus the property \ref{enum:core:filter} holds.  
                
                \item Consider an arbitrary process $s \in \setP{\symcore, t_{\symcore}}$. Assume, by the way of contradiction, that $\setSuper{s}{t_{\symfilter}} \neq \setSuper{s}{t_{\symcore}}$. This means that process $s$ started a \emph{new} super-passage after time $t_{\symfilter}$. Since $s \in \setP{\symcore}{t_{\symcore}}$, process $s$ read the contents of the variable $\x$ some time after $t_{\symfilter}$ but before $t_{\symcore}$. This contradicts our choice of $t_{\symfilter}$.  In other words, $\setSuper{s}{t_{\symfilter}} = \setSuper{s}{t_{\symcore}}$. Since $s$ was chosen arbitrarily, it follows that for each $p \in \setP{\symcore}{t_{\symcore}}$, $\setSuper{r}{t_{\symfilter}} = \setSuper{p}{t_{\symcore}}$. Thus the property \ref{enum:same:super} holds.
        
                \item Let $\card{\setP{\symcore}{t_{\symcore}}} = \fails$. Thus, using property \ref{enum:core:filter}, we can conclude that $\card{\setP{\symfilter}{t_{\symfilter}}} \geq \fails + 1$. It follows from \cref{theorem:MCS_safety} that there exist at least $\fails$ failures that are \unsafe{} relative to the lock $\symfilter$ and whose \consequence{} interval overlaps with time $t_{\symfilter}$. We have $\card{\setF{\symfilter}{t_{\symfilter}}} \geq k$ $=\  \card{\setP{\symcore}{t_{\symcore}}}$. Thus the property \ref{enum:unsafe:core} holds.
                
            \end{enumerate}
            
            This establishes the result.
        \end{proof}
    
               %%%%%%%%%%%%%%%%%%%%%%%%%%%%%%%%%%%%%%%%%%

        \subsection{A \sefficient{} RME algorithm}
        \label{sec:framework|recursive}
        \subsubsection{The main idea}
        
        We use the \emph{gap} between the worst-case RMR complexity of implementing a weakly recoverable lock and  
        that of implementing a strongly recoverable lock to achieve our goal.
    
        The main idea is to \emph{recursively} transform the \core{} lock using instances of our \cadaptive{} lock. We transform the \core{} lock repeatedly upto a height $\levels{}$ that is equal to the worst-case RMR complexity of another strongly recoverable lock under arbitrary number of failures. The strongly recoverable lock now becomes the base case of the recursion. For ease of exposition, we refer to the \core{} lock in the base case as the \emph{\base{} lock}.
        
        Let \nalock{} be a \bounded{} (presumably \nonadaptive{} but does not have to be) strongly recoverable lock.
        whose worst-case RMR complexity is $\bigO{R(\n)}$ for $\n$ processes. 
        Let \salock{} denote an instance of the \cadaptive{} lock described in \autoref{sec:framework|basic}. And, finally, let \balock{} denote the \bounded{} \sadaptive{} lock that we wish to construct (\nalock{} is the base lock and \balock{} is the target lock). 
        The idea is to create $\levels = R(\n)$ levels of \salock{} such that the \core{} lock component of the \salock{}
        at each level is built using another instance of \salock{} for up to $\levels-1$ levels and using an instance of \nalock{} at the base level (level $\levels$). Let \salock$[i]$ denote the instance of \salock{} at level $i$.  Formally,

        \begin{align*}
           \balock & ~=~ \salock[1] \\
           \salock[i].\text{\core} & ~=~ \salock[i+1] \qquad \forall i \in  \{1, 2, \dots, \levels-1\} \\
           \salock[\levels].\text{\core} & ~=~ \nalock
        \end{align*}
    
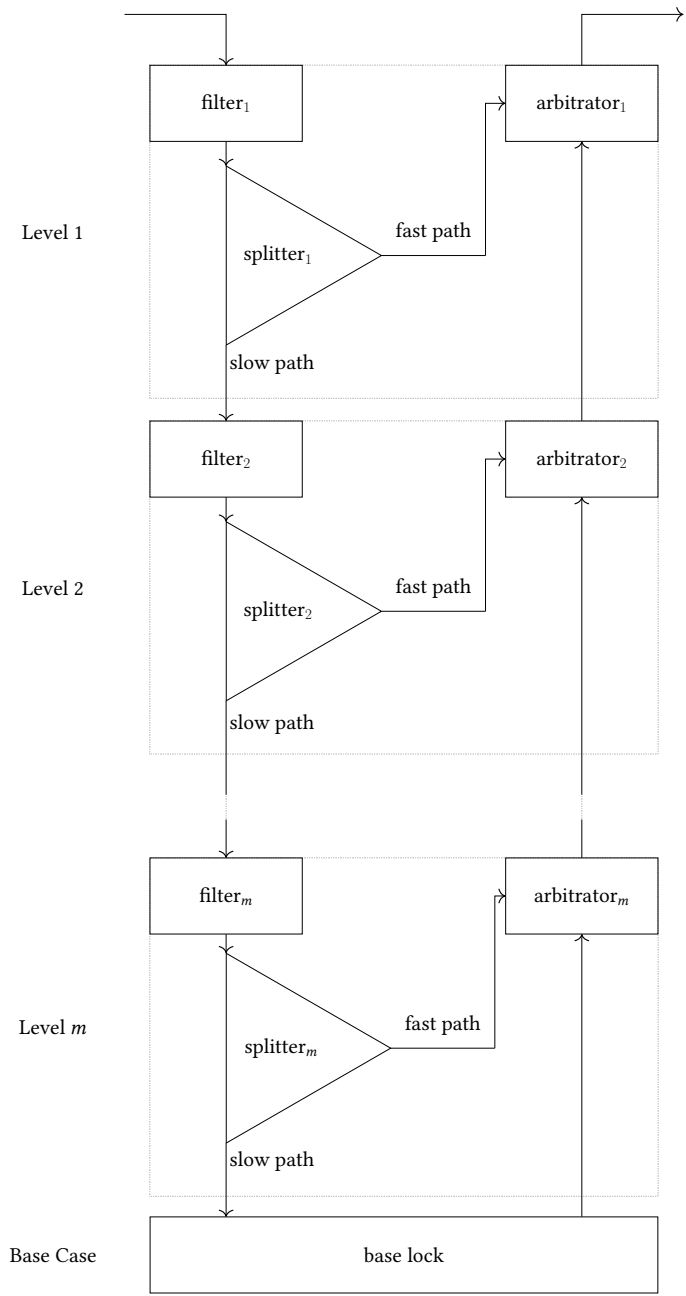
\begin{figure}[t]
        \scalebox{0.27}{
            \centering
            \begin{tikzpicture}[draw, 
            					minimum width=7.5cm, 
            					minimum height=3.75cm, 
            					every node/.style={transform shape, font=\HUGE},
            					triangle/.style = {regular polygon, regular polygon sides=3},
            					border rotated/.style = {shape border rotate=270},
            					decoration={markings, mark=at position 1 with {\arrow[scale=4,black]{>}}}]

               %% Level 1
               
               \node (W1) [rectangle, draw=black] {\filter$_1$};
               \node (B1) [rectangle, draw=black, right=10cm of W1] {\arbitrator$_1$};
               \node (X1) [triangle,  draw=black, border rotated, minimum size=1cm, below=7.5cm of W1.center, anchor=west] {\splitter$_1$};
               \node (IA1) [minimum size=0cm, below=2.5cm of X1.corner 3] {};
               \node (IB1) [minimum size=0cm, right=5cm of X1.corner 1] {};
               
               %% Box all shapes at level 1
               \node (Box1) [transform shape=false, dotted, draw,fit=(W1)(B1)(IA1.center), inner sep=0mm] {};
               
               %% Draw arrows within level 1
               \draw[postaction={decorate}] (W1) -- (X1.corner 2);
               \draw[-] (X1.corner 1) -- (IB1.center) node[pos=0.5, above, yshift=-0.75cm]{fast path};
               \draw[postaction={decorate}] (IB1.center) |- (B1.west);
               		       
               %% Level 2
               \node (W2) [rectangle, draw=black, below=1cm of IA1] {\filter$_2$};
               \node (B2) [rectangle, draw=black, right=10cm of W2] {\arbitrator$_2$};
               \node (X2) [triangle,  draw=black, border rotated, minimum size=1cm, below=7.5cm of W2.center, anchor=west] {\splitter$_2$};
               \node (IA2) [minimum size=0cm, below=2.5cm of X2.corner 3] {};
               \node (IB2) [minimum size=0cm, right=5cm of X2.corner 1] {};
              
               %% Box all shapes at level 2
               \node (Box2) [transform shape=false, dotted, draw,fit=(W2)(B2)(IA2.center), inner sep=0mm] {};
              
               %% Draw arrows within level 2
               \draw[postaction={decorate}] (W2) -- (X2.corner 2);
               \draw[-] (X2.corner 1) -- (IB2.center) node[pos=0.5, above, yshift=-0.75cm]{fast path};
               \draw[postaction={decorate}] (IB2.center) |- (B2.west);

               %% Level k
               \node (Wk) [rectangle, draw=black, below=5.0cm of IA2] {\filter$_\levels$};
               \node (Bk) [rectangle, draw=black, right=10cm of Wk] {\arbitrator$_\levels$};
               \node (Xk) [triangle,  draw=black, border rotated, minimum size=1cm, below=7.5cm of Wk.center, anchor=west] {\splitter$_\levels$};
               \node (IAk) [minimum size=0cm, below=2.5cm of Xk.corner 3] {};
               \node (IBk) [minimum size=0cm, right=5cm of Xk.corner 1] {};
                       
               %% Box all shapes at level k
               \node (Boxk)  [transform shape=false, dotted, draw, fit=(Wk)(Bk)(IAk.center), inner sep=0mm] {};
   
   			   %% Draw arrows at level k
   			  
               \draw[postaction={decorate}] (Wk) -- (Xk.corner 2);
               \draw[-] (Xk.corner 1) -- (IBk.center) node[pos=0.5, above, yshift=-0.75cm]{fast path};
               \draw[postaction={decorate}] (IBk.center) |- (Bk.west);
          
          	   %% Draw base case
		       \node (WkBk) [transform shape=false, fit=(Wk)(Bk), inner sep=0mm] {};
        	   \node (Base) [rectangle, draw=black, below=1.0cm of Boxk, minimum width=25cm] {\base{} lock};

          	   %% Create invisible nodes between level 2 and level k
      
           	   \node (EA1) [minimum size=0cm, above=3.0cm of Wk] {};
          	   \node (EA2) [minimum size=0cm, above=3.0cm of Bk] {};
          	   \node (EB1) [minimum size=0cm, below=1.0cm of EA1] {};
          	   \node (EB2) [minimum size=0cm, below=1.0cm of EA2] {};

        	   %% Draw slow paths  	   
          	   \draw[postaction={decorate}] (X1.corner 3) -- (W2) node[pos=0.25, right, xshift=-1.5cm]{slow path};
          	   \draw[-] (X2.corner 3) -- (EA1.center) node[pos=0.25, right, xshift=-1.5cm]{slow path};
          	   \draw[postaction={decorate}] (Xk.corner 3) -- ([shift={(3.75cm,0)}]Base.north west) node[pos=0.25, right, xshift=-1.5cm]{slow path};
          	   
          	   %% Draw ellipses between level 2 and level k
          	   
          	   \draw[-, dotted] (EA1.center) -- (EB1.center);
          	   \draw[-, dotted] (EA2.center) -- (EB2.center);
          	   \draw[postaction={decorate}] (EB1.center) -- (Wk);

          	   %% Draw return paths 
          	   \draw[postaction={decorate}] (B2) -- (B1);  
          	   \draw[-] (Bk) -- (EB2.center);
          	   \draw[postaction={decorate}] (EA2.center) -- (B2);  
          	   \draw[postaction={decorate}] ([shift={(-3.75cm,0)}]Base.north east) -- (Bk);   
          	   
          	   \draw[postaction={decorate}] ([shift={(-5cm,2.5cm)}] W1.north) -| (W1.north);     
          	   \draw[postaction={decorate}] (B1.north) |- ([shift={(5cm,2.5cm)}] B1.north);

          	   %% Label each level
          	   \node [rectangle, left=of Box1] {Level 1}; 
          	   \node [rectangle, left=of Box2] {Level 2};   
          	   \node [rectangle, left=of Boxk] {Level $\levels$};    
          	   \node [rectangle, left=of Base] {Base Case};      
                
            \end{tikzpicture}
            
          }  
            \captionof{figure}{A pictorial representation of the recursive framework.} 
            \label{fig:framework2|recursive}
       
        \end{figure}

        A pictorial representation of the execution flow of the recursive framework is depicted in \autoref{fig:framework2|recursive}. 
            
        In order to acquire the \target{} lock, a process starts at the first level as a \normal{} process and waits to acquire the \filter{} lock at level 1.
        It stays on track to become a \fast{} process until an \unsafe{} failure occurs with respect to the \filter{} lock at the first level as a result of which multiple processes may be granted the (\filter{}) lock simultaneously. All of these processes then compete to enter the fast path by navigating through the \splitter. The \splitter{} allows only one process to take the fast path at a time, and the rest are diverted to take the slow path. 
        Note that a \slow{} process is created at the first level only if an \unsafe{} failure occurs with respect to the \filter{} lock at the first level. 
        All \slow{} processes at the first level then move to the second level as \normal{} processes. If no further failure occurs, then no \slow{} process is created at the second level and all processes leave this level one-by-one as \fast{} processes with respect to this level. Thus, only $\bigO{1}$ RMR complexity is \emph{added} to the passages of all the affected processes until the impact of the first failure has subsided. However, if one or more \slow{} processes are created at the second level, then we can infer that a \emph{new} \unsafe{} failure must have occurred with respect to the \filter{} lock at the second level. All these \slow{} processes at the second level then move to the third level as \normal{} processes, and so on and so forth. 
        At each level, a \slow{} process, upon either acquiring the \base{} lock or returning from the adjacent higher level (whichever case applies), becomes a \medium{} process. Irrespective of whether a process is classified as \fast{} or \medium{}, it next waits to acquire the level-specific \arbitrator{} lock. Once granted, it either returns to the adjacent lower level or, if at the initial level, is deemed to have successfully acquired the \target{} lock.
         
        Note that in our algorithm, at least $\fails$ \unsafe{} failures are required at any level to force $\fails$ processes to be ``escalated'' to the next level. Each level except for the last one would add only $\bigO{1}$ RMR complexity to the passages of these process, thus making the \target{} lock \adaptive{} under limited failures. There is no further ``escalation'' of \slow{} processes at the base level and a \bounded{} (possibly \nonadaptive{}) strongly recoverable lock is used to manage all \slow{} processes at that point, thus bounding its RMR complexity under arbitrary number of failures as well. 
        
        As before, in order to release the \target{} lock, a process releases its components locks in the reverse order in which it acquired them.
         
        To prove that our \target{} lock is \sefficient{}, we utilize two properties of our framework, namely, our weakly recoverable lock is \responsive{}, and our \target{} lock, which is a composite lock, satisfies the \locality{} property.

    \subsubsection{Correctness proof}

    Let $\symfilter_i$ and $\symcore_i$ denote the instances of the \filter{} and \core{} locks, respectively, at level $i$ for $i = 1, 2, \ldots, \levels$.

    \begin{theorem}
    \label{theorem:balock_safe}
        For each level $i$ with $1 \leq i \leq \levels$, \salock[i] is a strongly recoverable lock.
    \end{theorem}
    \begin{proof}
        The proof is by backward induction on the level number of \salock{} starting from level $\levels$.
        \begin{itemize}[label=$\square$]
        
        \item \underline{Base case (\salock[$\levels$] is a strongly recoverable lock).} 
        Note that \salock[$\levels$] = \nalock{}. By construction, \nalock{} is a \bounded{} strongly recoverable lock. Thus, \salock[$\levels$] is a strongly recoverable lock.
    
        \item \underline{Induction hypothesis (\salock[$i+1$] is a strongly recoverable lock for some $i$ with $1 \leq i < \levels$).} 
        We prove that \salock[$i$] is also a strongly recoverable lock.
        Note that \salock[$i$] is an instance of our \cadaptive{} lock described in \cref{sec:framework|basic} with 
        \salock[$i+1$] as its \core{} lock. Since \salock[$i+1$] is a strongly recoverable lock, it follows from  \cref{theorem:framework|RME} that \salock[$i$] is also a strongly recoverable lock.
        \end{itemize}
        
        \indent
        Thus, by induction, we can conclude that \salock[$i$] is a strongly recoverable lock for each level $ i = 1, 2, \dots, \levels$.
    \end{proof}

    By construction, \balock{} = \salock[$1$]. Therefore,
    
    \begin{theorem}
    \balock{} is a strongly recoverable lock. 
    \end{theorem}
    
    Using induction similar to the one used in \cref{theorem:balock_safe}, we can show that
    
    \begin{theorem}
    \balock{} satisfies the BR and BE properties.
    \end{theorem}
    
    \subsubsection{Complexity analysis}
     
    To analyze the RMR complexity of a passage, we first prove certain results.

    \begin{theorem}
        \balock{} satisfies the \locality{} property.
    \end{theorem}
    \begin{proof}
        \balock{} uses three types of locks, namely \filter{}, \arbitrator{} and \base{}; only \filter{} lock is weakly recoverable.  There is one instance of the \filter{} lock at each level. By construction, the \Enter{} \segment{s} of any two instances of the \filter{} lock do not overlap. The only \senins{} of the \filter{} lock is the \FAS{} instruction in its \Enter{} \segment{}. Therefore, \balock{} satisfies the \locality{} property.
    \end{proof}
    
    By the construction of our recursive framework, we have
    
    \begin{proposition}
        \label{proposition:coreisnextfilter}
        For each level $i$ and time $t$ with $1 \leq i < \levels$, 
        $\setP{\symcore_{i}}{t} = \setP{\salock{[i+1]}}{t} = \setP{\symfilter_{i+1}}{t}$.
    \end{proposition}

    Note that the set of processes that attempt to acquire the \filter{} lock at any level becomes 
    \emph{progressively smaller} as the level number increases. Furthermore, the number of processes that are escalated to the next level depends on the number of \unsafe{} failures experienced by the \filter{} lock at the current level. This is captured by the next lemma.

    \begin{lemma}
    \label{lemma:level_j_failures}
    Consider a process $p$, time $t$ and level $x$, where $1 \leq x \leq \levels$,
    such that process 
    $p \in \setP{\symfilter_x}{t}$.
    Then there exist $x$ times $t_1, t_2, \ldots, t_{x-1}, t_x$ with $t_1 \leq t_2 \leq \cdots \leq t_{x-1} \leq t_x = t$ such that the following properties hold.
    For each $i$ with $1 \leq i < x$, we have
        \begin{enumerate}[label=(\alph*)]
            \item $\setSuper{p}{t_i} = \setSuper{p}{t}$, 
            \item $\setP{\symfilter_{i}}{t_{i}} \supsetneq \setP{\symfilter_{i+1}}{t_{i+1}}$, and
            \item $\card{\setF{\symfilter_{i}}{t_{i}}} \geq \card{\setP{\symfilter_{i+1}}{t_{i+1}}}$.
        \end{enumerate}
    \end{lemma}
    \begin{proof}
        The proof is by backward induction on $i$ starting from $x-1$. In order to prove our results, we use the following auxiliary properties, which are part of the induction statement. For each $i$ with $1 \leq i < x$, we have,
        \begin{enumerate}[label = (\alph*), start=4]
            \item $\card{\setP{\symfilter_i}{t_i}} > 0$, and
            \item $p \in \setP{\symfilter_i}{t_i}$.
        \end{enumerate}
    
    We are now ready to prove the result. 
    
        \begin{itemize}[label=$\square$]
        
        \item \underline{Base case (properties (a)-(e) hold for $i = x - 1$).}
        By definition, $t_x = t$. By assumption, $p \in \setP{\symfilter_x}{t_x}$. By applying  \cref{proposition:coreisnextfilter}, we obtain that $p \in \setP{\symcore_{x-1}}{t_x}$ thereby implying that $\card{\setP{\symcore_{x-1}}{t_x}} > 0$.
        We can now apply \cref{lemma:failure_count|base} once to infer that there exists time, say $t_{x-1}$ with $t_{x-1} < t_x$, such that the following properties hold.
        \begin{enumerate}[label=(\roman*)]
            
            \item $\setSuper{p}{t_{x-1}} = \setSuper{p}{t_x}$, which, in turn, implies that $\setSuper{p}{t_{x-1}} = \setSuper{p}{t}$ because $t_x = t$ (property (a)).
            
            \item $\setP{\symfilter_{x-1}}{t_{x-1}} \supsetneq \setP{\symcore_{x-1}}{t_x}$, which, in turn, implies that $\setP{\symfilter_{x-1}}{t_{x-1}} \supsetneq \setP{\symfilter_x}{t_x}$ because 
            $\setP{\symcore_{x-1}}{t_x} =  \setP{\symfilter_x}{t_x}$ (property (b)).
            
            \item $\card{\setF{\symfilter_{x-1}}{t_{x-1}}} \geq \card{\setP{\symcore_{x-1}}{t_{x}}}$, which, in turn, implies that $\card{\setF{\symfilter_{x-1}}{t_{x-1}}} \geq \card{\setP{\symfilter_{x}}{t_{x}}}$ (property (c)). 
            
            \item $\card{\setP{\symfilter_{x-1}}{t_{x-1}}} > 0$ because  $\setP{\symfilter_{x-1}}{t_{x-1}} \supsetneq \setP{\symfilter_x}{t_x} \supseteq \{ p \}$ (property (d)).
            
            \item $p \in \setP{\symfilter_{x-1}}{t_{x-1}}$ because $\setP{\symfilter_{x-1}}{t_{x-1}} \supsetneq \setP{\symfilter_x}{t_x} \supseteq \{ p \}$ (property (e)).

        \end{enumerate}

        \item \underline{Induction hypothesis (assume that the properties (a)-(e) hold for some $i$  with $1 < i < x$).}
        We now prove that properties (a)-(e) also hold for $i-1$. 
        Note that, by induction hypothesis, $\card{\setP{\symfilter_i}{t_i}} > 0$. Thus, We can now apply \cref{lemma:failure_count|base} once to infer that there exists time, say $t_{i-1}$ with $t_{i-1} < t_i$, such that the following properties hold.
    
            \begin{enumerate}[label=(\roman*)]
            
            \item $\setSuper{p}{t_{i-1}} = \setSuper{p}{t_i}$, which, in turn, implies that $\setSuper{p}{t_{i-1}} = \setSuper{p}{t}$ (property (a)).
            
            \item $\setP{\symfilter_{i-1}}{t_{i-1}} \supsetneq \setP{\symcore_{i-1}}{t_i}$, which, in turn, implies that $\setP{\symfilter_{i-1}}{t_{i-1}} \supsetneq \setP{\symfilter_i}{t_i}$ because 
            $\setP{\symcore_{i-1}}{t_i} =  \setP{\symfilter_i}{t_i}$ (property (b)).
            
            \item $\card{\setF{\symfilter_{i-1}}{t_{i-1}}} \geq \card{\setP{\symcore_{i-1}}{t_{i}}}$, which, in turn, implies that $\card{\setF{\symfilter_{i-1}}{t_{i-1}}} \geq \card{\setP{\symfilter_i}{t_i}}$ (property (c)). 
            
            \item $\card{\setP{\symfilter_{i-1}}{t_{i-1}}} > 0$ because  $\setP{\symfilter_{i-1}}{t_{i-1}} \supsetneq \setP{\symfilter_i}{t_i} \supseteq \{ p \}$ (property (d)).
            
            \item $p \in \setP{\symfilter_{i-1}}{t_{i-1}}$ because $\setP{\symfilter_{i-1}}{t_{i-1}} \supsetneq \setP{\symfilter_i}{t_i} \supseteq \{ p \}$ (property (e)).

        \end{enumerate}
        
        \end{itemize}

        This establishes the lemma.
    \end{proof}
    
    The next corollary \emph{quantifies} the number of processes that must be present at \emph{each} 
    of the lower levels for some process to be escalated to a certain level.

    \begin{corollary}
        \label{corollary:count_processes_i}
         Consider a process $p$, time $t$ and level $x$, where $1 \leq x \leq \levels$,
    such that process 
    $p \in \setP{\symfilter_x}{t}$.
    Let times $t_1, t_2, \ldots, t_{x-1}, t_x$ be as given by 
    \cref{lemma:level_j_failures}. Then, for each $i$ with $1 \leq i < x$, 
     $\card{\setP{\symfilter_i}{t_i}} \;\geq\; x-i+1$.
    \end{corollary}
    
    The next corollary \emph{quantifies} the number of \unsafe{} failures that must occur with respect to the \filter{} lock at each of the lower levels  for some process to be escalated to a certain level.

    \begin{corollary}
        \label{corollary:count_failures_i}
         Consider a process $p$, time $t$ and level $x$, where $1 \leq x \leq \levels$,
    such that process 
    $p \in \setP{\symfilter_x}{t}$.
    Let times $t_1, t_2, \ldots, t_{x-1}, t_x$ be as given by 
    \cref{lemma:level_j_failures}. Then, for each $i$ with $1 \leq i < x$, 
    $\card{\setF{\symfilter_i}{t_i}} \;\geq\; x-i$.
    \end{corollary}

    For the rest of this section, unless otherwise stated, assume that super-passage of a process and \consequence{} interval of a failure are defined \emph{relative to the \target{} lock}.
    
     \begin{theorem}
     Suppose a process $p$ advances to level $x$ at some time $t$ during its super-passage, where $1 \leq x \leq \levels$. Then, there exist at least $\nicefrac{x (x-1)}{2}$ failures that occurred at or before time $t$ whose \consequence{} interval overlaps with the super-passage of the process $p$.
    \end{theorem}
    \begin{proof}
        Let $t_1, t_2, \dots, t_x$ be the times as given by \cref{lemma:level_j_failures}. 
        Since \balock{} satisfies the \locality{} property, the set of failures that are \unsafe{} with respect to one instance of its \filter{} lock is \emph{disjoint} from the set of failures that are \unsafe{} with respect to another instance of its \filter{} lock. 
        Formally,
        \begin{align*}
            \forall i, j: 1 \leq i, j \leq x \text{ and } i \neq j: \setF{\symfilter_{i}}{t_i} \cap \setF{\symfilter_{j}}{t_j} = \emptyset && \text{(pairwise disjoint property)}
        \end{align*}
        
        Let $\Super$ denote the super-passage of $p$ at time $t$. 
        From the property (a) of \cref{lemma:level_j_failures}, $\Super = \setSuper{p}{t_1} = \setSuper{p}{t_2} = \ldots = \setSuper{p}{t_x}$. In other words, $p$ is executing the same super-passage during the period $[t_1,t_x]$. 
        
        Let $\allF(t)$ denote the set of all failures that occurred at or before time $t$ and whose \consequence{} interval overlaps with the super-passage $\Super$. Note that the \consequence{} interval of any failure with respect to the \target{} lock contains 
        the \consequence{} interval of that failure with respect to any instance of its \filter{} lock. This is because all pending requests for that instance of the \filter{} lock  are also pending requests for the \target{} lock. 
         Thus, $\forall i: 1 \leq i < x: \setF{\symfilter_i}{t_i} \subseteq \allF(t)$. This in turn implies that
         \begin{align*}
         \bigcup_{i=1}^{x-1} \setF{\symfilter_i}{t_i} \; \subseteq  \; \allF(t) && \text{(containment property)}
         \end{align*}
         
         We have
        \begin{align*}
         \card{\allF(t)} & \geq \quad  \card{\bigcup_{i=1}^{x-1} \setF{\symfilter_i}{t_i}} && \text{(using containment property)} \\
                      & =  \quad \sum_{i=1}^{x-1}  \card{\setF{\symfilter_i}{t_i}} && \text{(using pairwise disjoint property)}\\
                      & =  \quad \sum_{i=1}^{x-1} (x-i) && \text{(using \cref{corollary:count_failures_i})}\\
                      & =  \quad (x-1) + \cdots + 2 + 1 && \text{(expanding the sum)} \\
                      & =  \quad \frac{x(x-1)}{2} && \text{(algebra)}
        \end{align*}

        This establishes the result.
    \end{proof}

    The following results then follow from the above two theorems.

    \begin{theorem}[\balock{} is \bounded{} \sadaptive{}]
    The RMR complexity of any passage in a super-passage of \balock{} is given by $\bigO{\min\{\sqrt{\fails+1}, R(\n)\}}$, where $\fails$ denotes the \hazard{} of the super-passage and $R(\n)$ denotes the RMR complexity of the \base{}  \nalock{} for $\n$ processes.
    \end{theorem}

    \begin{corollary}[a \wbounded{} \sadaptive{} lock]
     Assume that we use \JJJ{'s}~\cite{JayJay+:2019:PODC} or \KM{'s} RME algorithm~\cite{KatMor:2020:OPODIS} to implement 
    the \nalock. 
    Then the RMR complexity of any passage in a super-passage of \balock{} is given by $\bigO{\min\{\sqrt{\fails+1}, \nicefrac{\log \n}{\log \log \n}\}}$, where $\fails$ denotes the \hazard{} of the super-passage.
    \end{corollary}

    We now show that \balock{} is \emph{adaptive to contention} as well.
    
    \begin{theorem}
    Suppose a process $p$ advances to level $x$ at some time $t$ during its super-passage, where $1 \leq x \leq \levels$. Then, there exist at least $x-1$ super-passages that are simultaneously \inprogress{} along with the super-passage of the process $p$ at or before time $t$. 
    \end{theorem}
    \begin{proof}
    Let $t_1$ denote the time as postulated in the statement of \cref{lemma:level_j_failures}. Clearly, we can infer that 
    $\Super(p,t_1) = \Super(p,t)$, $p \in \setP{\symfilter_1}{t_1}$ and $\card{\setP{\symfilter_1}{t_1} \setminus \{ p \}} \;\geq\; x-1$.
    \end{proof}
    
    This implies that \balock{} is adaptive to both failures and contention as stated in the next theorem.

    \begin{theorem}[dual adaptivity]
     The RMR complexity of any passage in a super-passage of \balock{} is given by $\bigO{\min\{\pc, \sqrt{\fails+1}, R(\n)\}}$, where $\pc$ denote the point-contention of the super-passage, $\fails$ denotes the \hazard{} of the super-passage and $R(\n)$ denotes the RMR complexity of the \nalock{} for $\n$ processes. 
    \end{theorem}
   
   %%%%%%%%%%%%%%%%%%%%%%%%%%%%%%%%%%%%%%%%%%%%%%%%%

    \section{Fairness}
    \label{sec:fairness}
        
        An important desirable property satisfied by many ME algorithms is \emph{fairness}. Intuitively, fairness ensures that no process is able to monopolize shared resources. To define fairness, the \Enter{} \segment{} is partitioned into two segments---\emph{doorway} followed by \emph{waiting room}. A doorway consists of a \emph{bounded} number of steps that a process executes in the beginning of its \Enter{} \segment, whereas waiting room is the rest of the \Enter{} \segment.
        
        In this work, we use a novel definition for fairness referred to as  \CIFCFS{}.
        To that end, we first define the notion of \singular{} passage as follows. A failure-free passage is said to be \emph{\singular} if its associated super-passage is failure-free (and therefore it is the only passage of its super-passage).
        
        \begin{description}
            \item[\CIFCFS] A history $H$ is said to satisfy \CIFCFS{} if it satisfies the following property.
            Consider a pair of passages $r$ and $s$ in $H$ belonging to processes $p$ and $q$, respectively, such that
            \begin{enumerate*}[label=(\alph*)]
            \item both $r$ and $s$ are \singular{} passages, 
            \item $p$ completes its doorway in $r$ before $q$ begins its doorway in $s$, and
            \item $r$ does not overlap with the \consequence{} interval of any failure.
            \end{enumerate*}
            If $q$ has started the \CS{} \segment{} of $s$,  then $p$ has started the \Exit{} \segment{} of $r$.
        \end{description}
        
        An RME algorithm is said to satisfy \CIFCFS{} if every history generated by the algorithm satisfies \CIFCFS{}. Note that, in the absence of failures, \CIFCFS{} reduces to traditional FCFS (first-come-first-served) property.
        %%
        %% Also, note that if a super-passage of $p$ does not overlap with the \consequence{} interval of any failure, then $p$ itself does not fail during the super-passage and, moreover, the super-passage consists of exactly one passage.
  
        We first argue that both our weakly RME lock, namely \weakMCS{}, 
        and our strongly RME lock, namely \salock{}, satisfy \CIFCFS.
        We then compare our notion of fairness with the one used by Golab and Ramaraju in~\cite{GolRam:2019:DC}.

        Note that \balock{} is a special case of \salock{} and thus would also satisfy \CIFCFS.
        Also, note that we \emph{do not} require the \core{} lock of the framework to be fair.
        
       \subsection{Fairness of our RME Algorithms}
       
        The doorway of \weakMCS{} consists of \crefrange{line:enter:doorway|begin}{line:enter:doorway|end}.
          
          \begin{theorem}
            \label{theorem:MCS_fair}
            \weakMCS{} satisfies the \CIFCFS{} property.
          \end{theorem}
          \begin{proof}  
          Consider a history $H$, and let $r_i$ and $r_j$ be passages of processes $p_i$ and $p_j$ respectively such that:
          \begin{enumerate*}[label=(\alph*)]
          \item $p_i$ completes its doorway in $r_i$ before $p_j$ starts its doorway in $r_j$, %%
          \item both $r_i$ and $r_j$ are \singular{} passages, and
          \item $r_i$ does not overlap with the \consequence{} interval of any failure.
          \end{enumerate*}

          Assume, on the contrary, that there exists a time instant $t$ in $H$ such that, at time $t$, $p_j$ is in the \CS{} \segment{} of $r_j$  but 
          $p_j$ has not started  the \Exit{} \segment{} of $r_i$.
          Note that, by time $t$, both $p_i$ and $p_j$ have completed the doorways for their respective passages. Let $t_i$ and $t_j$ denote the time instants when $p_i$ and $p_j$, respectively, executed their \FAS{} instructions. Further, let $u_i$ and $u_j$ denote the \admissible{} nodes owned by $p_i$ and $p_j$, respectively, at time $t$. 
          Clearly, $t_i < t_j < t$. 
          We first prove the following claim.
          
          \begin{claim}
          \label{claim:MCS_fair}
          As long as $u_i$ is \admissible{}, both $u_i$ and $u_j$  belong to a \emph{single} sub-queue. 
          \end{claim}
          
          The proof of the claim is by contradiction. Assume not. This implies that there exists a process $p_k$ such that $p_k$
          \begin{enumerate*}[label=(\alph*)]
          \item executed an \FAS{} instruction at time $t_k$ with $t_i < t_k < t_j$, but
          \item experienced an \unsafe{} failure $f$ at time $t_f$ with $t_k < t_f < t$. \end{enumerate*}
          Since $t_i < t_k$, $t_k < t_f$ and $t_f < t$, it follows that $t_i < t_f < t$. Clearly, $r_i$ contains both $t_i$ and $t$. This implies that $r_j$ contains $t_f$ and thus overlaps with the \consequence{} interval of $f$. This is a contradiction. Thus the claim holds.
          
          It follows from the claim that, until $u_i$ stays \admissible{}, $u_j$ cannot become the \front{} node of its sub-queue. In other words, $p_j$ cannot enter the \CS{} \segment{} of $r_j$ until $p_i$ has started executing the \Exit{} \segment{} of $r_i$. 
         \end{proof}

         We now prove that the target lock of our framework satisfies the \CIFCFS{} property. To that end, we assume that the \filter{} lock satisfies the \CIFCFS{} property.
         
        \begin{theorem}
            \label{theorem:algo_fair}
            \salock{} satisfies the \CIFCFS{} property.
        \end{theorem}
        \begin{proof}
        Consider a history $H$, and let $r_i$ and $r_j$ be passages of processes $p_i$ and $p_j$  (with respect to the target lock) such that:
        \begin{enumerate*}[label=(\alph*)]
        \item $p_i$ completes its doorway in $r_i$ before $p_j$ starts its doorway in $r_j$, %%
        \item both $r_i$ and $r_j$ are \singular{} passages, and
        \item $r_i$ does not overlap with the \consequence{} interval of any failure.
        \end{enumerate*}
        Assume that $p_j$ in the \CS{} \segment{} of $r_j$. 
        
        Let $r_i(\symfilter) \subseteq r_i$ and $r_j(\symfilter) \subseteq r_j$ denote the passages of $p_i$ and $p_j$, respectively, with respect to the \filter{} lock.
        Due to the arrangements of the locks, we can infer the following. First, 
        if the three conditions of the \CIFCFS{} property hold for $r_i$ and $r_j$, then they also hold for $r_i(\symfilter)$ and $r_i(\symfilter)$. Second, if $p_j$ is the \CS{} \segment{} of $r_j$, then it is also in the \CS{} \segment{} of $r_j(\symfilter)$. Third, if $p_i$ has started the \Exit{} \segment{} of $r_i(\symfilter)$, then it has also started the \Exit{} \segment{} of $r_i$.
        \end{proof}
       
        Since \balock{} is a special case of \salock{}, we have:
        
        \begin{corollary}
        \balock{} satisfies the \CIFCFS{} property.
        \end{corollary}

        \subsection{Comparison with $\mathbf{k}$-FCFS}
    
        \label{subsec:fairness|comparison}
    
        Golab and Ramaraju define the concept of a $k$-failure-concurrent passage, where $k \geq 0$, to capture how a failure may impact the RMR complexity of a passage. The concept is recursively defined.
        Given a history, a passage of a process is said to be 0-failure-concurrent if it either ends with or begins after a crash of the process. It is said to be $k$-failure-concurrent if it is either $(k-1)$-failure-concurrent or its super passage overlaps with another super passage containing a $(k-1)$-failure-concurrent passage.
        Intuitively, the parameter $k$ 
        measures the ``distance'' of a passage from a $0$-failure-concurrent passage in a given history. 
        (Note that a passage is \singular{} if and only if it is not 0-failure-concurrent.)

        We, on the other hand, use the concept of \consequence{} interval of a failure in this work. Given a history, a passage is said to be \emph{\CIFC} if its super-passage overlaps with the \consequence{} interval of one or more failures. 
        
        A natural question is to ask how the concept of $k$-failure-concurrent is related to \CIFC{}. We show that \CIFC{} is ``stronger'' than 2-failure-concurrent and ``incomparable'' with 1-failure-concurrent. 
        
        \begin{theorem}
            Given a passage $r$ in a finite history $H$, if $r$ is \CIFC{}, then it is also 2-failure-concurrent.
        \end{theorem}  
        \begin{proof}
           For convenience, given a passage $r$, we use $\Super(r)$ to denote the super-passage associated with $r$. 
           
           Consider a passage $r_i$ of process $p_i$ that is \CIFC. By definition, $\Super(r_i)$ overlaps with the \consequence{} interval of a failure, say $f$. 
           Let $f$ involve the failure of process $p_j$ while executing the passage $r_j$. 
           We use $\mathbb{P}(f)$ to denote the set of processes that have a super-passage in progress at the time when $f$ occurred. Note that $\mathbb{P}(f) \neq \emptyset$ because $p_j \in \mathbb{P}(f)$. Let $p_k \in \mathbb{P}(f)$ denote the process whose super-passage, which was in progress when $f$ occurred, extends for the \emph{longest} time in $H$.  Finally, let $r_k$ denote the \emph{most recent} passage of $p_k$ that started before $f$ occurred.
           
           Note that $\Super(r_k)$ overlaps with $\Super(r_j)$. This in turn implies that $r_k$ is (at most) 1-failure-concurrent because $r_j$ is 0-failure-concurrent. Also, note that $\Super(r_k)$ contain the \consequence{} interval of $f$ and hence overlaps with $\Super(r_i)$. This implies that $r_i$ is (at most) 2-failure-concurrent.
        \end{proof}
        
        We now provide examples to show that 1-failure-concurrent and \CIFC{} are incomparable properties, \emph{i.e.}, neither implies the other.
        
        \begin{theorem}
            There exists a history $H$ and a passage $r$ in $H$ such that $r$ is  1-failure-concurrent but not \CIFC.    
        \end{theorem}
        \begin{proof}
               \begin{figure}[t]
        \centering
        \begin{tikzpicture}[draw, minimum height=0.5cm]
            \node (p1) at (2,9) {$p_1$};
            \node (p2) at (2,8) {$p_2$};
            
            \node (r1) [rectangle, minimum width=1cm] at (8,9) {$r_1$};
            \node (r1p) [rectangle, minimum width=1cm] at (9,9) {};
            \node (r1r) [rectangle, minimum width=1cm] at (10,9) {$r_2$};
            
            \draw[-] (r1.north west) -- (r1.north east);
            \draw[-] (r1.north west) -- (r1.south west);
            \draw[-] (r1.south west) -- (r1.south east);
            \draw[-, dotted] (r1p.north west) -- (r1p.north east);
            \draw[-, dotted] (r1p.south west) -- (r1p.south east);
            \draw[-] (r1r.north west) -- (r1r.north east);
            \draw[-] (r1r.north east) -- (r1r.south east);
            \draw[-] (r1r.south west) -- (r1r.south east);
            
            \node (r3) [draw = black, rectangle, minimum width=4cm] at (6,8) {$r_3$};

            \node (failure) [above =0.5cm of r1p.north west] {failure};
            \draw[->] (failure.south -| r1p.west) -- (r1p.north west);
            
            \draw[-, dashed] (r1p.north west) -- (8.5,7);
            {\draw[-, dashed] (r1p.north -| r1r.east) -- (10.5,7);
            \draw[<->] (8.5,7) -- (10.5,7);
            \node (CI)[text width=2cm,align=center] at (9.5, 6.5) {\consequence{} interval};}
        \end{tikzpicture}
    \captionof{figure}{Passages $r_1$ and $r_2$ are the only 0-failure-concurrent passages in the history. Passage $r_3$ overlaps $r_1$ and thus is 1-failure-concurrent, but is not \CIFC.}
    \label{fig:1FCnoCI}
    \end{figure}
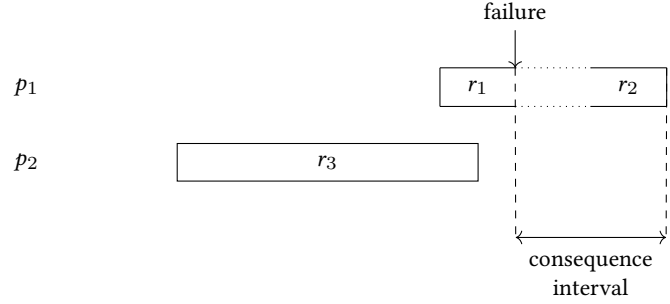
           Consider the history shown in \cref{fig:1FCnoCI}. The passage $r_3$ in the history
           is 1-failure-concurrent but not \CIFC.
        \end{proof}
        
        \begin{theorem}
           There exists a history $H$ and a passage $r$ in $H$ such that $r$ is \CIFC{} but not 1-failure-concurrent.    
        \end{theorem}
        \begin{proof}
               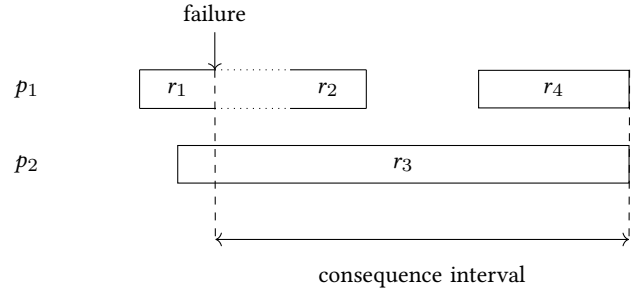
\begin{figure}[t]
        \centering
        \begin{tikzpicture}[draw, minimum height=0.5cm]
            \node (p1) at (2,9) {$p_1$};
            \node (p2) at (2,8) {$p_2$};
            
            \node (r1) [rectangle, minimum width=1cm] at (4,9) {$r_1$};
            \node (r1p) [rectangle, minimum width=1cm] at (5,9) {};
            \node (r1r) [rectangle, minimum width=1cm] at (6,9) {$r_2$};
            
            \draw[-] (r1.north west) -- (r1.north east);
            \draw[-] (r1.north west) -- (r1.south west);
            \draw[-] (r1.south west) -- (r1.south east);
            \draw[-, dotted] (r1p.north west) -- (r1p.north east);
            \draw[-, dotted] (r1p.south west) -- (r1p.south east);
            \draw[-] (r1r.north west) -- (r1r.north east);
            \draw[-] (r1r.north east) -- (r1r.south east);
            \draw[-] (r1r.south west) -- (r1r.south east);
            
            \node (r3) [draw = black, rectangle, minimum width=6cm] at (7,8) {$r_3$};
            
            \node (r4) [draw = black, rectangle, minimum width=2cm] at (9,9) {$r_4$};
            
            \node (failure) [above =0.5cm of r1p.north west] {failure};
            \draw[->] (failure.south -| r1p.west) -- (r1p.north west);
            
            \draw[-, dashed] (r1p.north west) -- (4.5,7);
            {\draw[-, dashed] (r1p.north -| r3.east) -- (10,7);
            \draw[<->] (4.5,7) -- (10,7);
            \node (CI)[text width=10cm,align=center] at (7.25, 6.5) {\consequence{} interval};}
        \end{tikzpicture}
    \captionof{figure}{Passages $r_1$ and $r_2$ are the only 0-failure-concurrent passages in the history. Passage $r_4$ overlaps with the \consequence{} interval of the failure and is 2-failure-concurrent, but not 1-failure-concurrent.}
    \label{fig:CIno1FC}
    \end{figure}
           Consider the history shown in \cref{fig:CIno1FC}. The passage $r_4$ in the history
           is \CIFC{} but not 1-failure-concurrent.
        \end{proof}
        
        Finally, we show that \CIFC{} is a strictly stronger property than 2-failure-concurrent, \emph{i.e.}, the latter does not imply the former.
        
        \begin{theorem}
           There exists a history $H$ and a passage $r$ in $H$ such that $r$ is 2-failure-concurrent but not \CIFC.    
        \end{theorem}
        \begin{proof}
               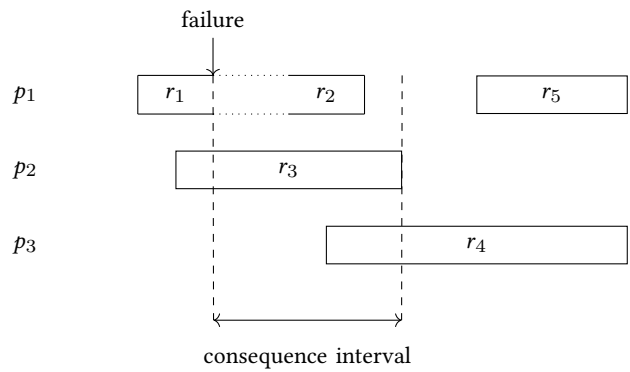
\begin{figure}[t]
        \centering
        \begin{tikzpicture}[draw, minimum height=0.5cm]
            \node (p1) at (2,9) {$p_1$};
            \node (p2) at (2,8) {$p_2$};
            \node (p3) at (2,7) {$p_3$};
            
            \node (r1) [rectangle, minimum width=1cm] at (4,9) {$r_1$};
            \node (r1p) [rectangle, minimum width=1cm] at (5,9) {};
            \node (r1r) [rectangle, minimum width=1cm] at (6,9) {$r_2$};
            
            \draw[-] (r1.north west) -- (r1.north east);
            \draw[-] (r1.north west) -- (r1.south west);
            \draw[-] (r1.south west) -- (r1.south east);
            \draw[-, dotted] (r1p.north west) -- (r1p.north east);
            \draw[-, dotted] (r1p.south west) -- (r1p.south east);
            \draw[-] (r1r.north west) -- (r1r.north east);
            \draw[-] (r1r.north east) -- (r1r.south east);
            \draw[-] (r1r.south west) -- (r1r.south east);
            
            \node (r3) [draw = black, rectangle, minimum width=3cm] at (5.5,8) {$r_3$};
            
            \node (r4) [draw = black, rectangle, minimum width=4cm] at (8,7) {$r_4$};
            
            \node (r5) [draw = black, rectangle, minimum width=2cm] at (9,9) {$r_5$};
            
            \node (failure) [above =0.5cm of r1p.north west] {failure};
            \draw[->] (failure.south -| r1p.west) -- (r1p.north west);
            
            \draw[-, dashed] (r1p.north west) -- (4.5,6);
            {\draw[-, dashed] (r1p.north -| r3.east) -- (7,6);
            \draw[<->] (4.5,6) -- (7,6);
            \node (CI)[text width=4cm,align=center] at (5.75, 5.5) {\consequence{} interval};}
        \end{tikzpicture}
    \captionof{figure}{Passage $r_2$ begins after a failure which makes it 0-failure-concurrent. Passage $r_4$ overlaps with $r_2$, which makes it 1-failure-concurrent. Passage $r_5$ overlaps with $r_4$, which makes it 2-failure-concurrent. However, $r_5$ does not overlap with the \consequence{} interval of any failure.} 
    \label{fig:2FCnoCI}
    \end{figure}
           Consider the history shown in \cref{fig:2FCnoCI}. The passage $r_5$ in the history is 2-failure-concurrent but not \CIFC.
        \end{proof}
        
        Golab and Ramaraju use the notion of $k$-failure concurrency to define a notion of fairness especially suited to RME algorithms, referred to as $k$-FCFS. Intuitively, $k$-FCFS guarantees FCFS among two passages provided none of them is $k$-failure-concurrent.
        The results above also establish the following relationship among different fairness properties
        \begin{enumerate*}[label=(\alph*)]
        \item \CIFCFS{} implies 2-FCFS, and 
        \item \CIFCFS{} and 1-FCFS are incomparable. 
        \end{enumerate*}.
        
        Incidentally, our framework yields an RME algorithm that not only satisfies the \CIFCFS{} property but also satisfies the 1-FCFS property. We have,
        
        \begin{theorem}
        \weakMCS{} satisfies the 1-FCFS property.   
        \end{theorem}
        \begin{proof}
        The proof is analogous to the proof of \cref{theorem:MCS_fair} with the following modification. Let $r_k$ denote the passage of $p_k$ that ends with the failure $f$. Clearly, $r_k$ is 0-failure-concurrent passage and overlaps with $r_i$. This implies that $r_i$ is a 1-failure-concurrent passage---a contradiction. Thus the claim holds in this case as well.
        \end{proof}
        
        \begin{theorem}
         \salock{} satisfies the 1-FCFS property.
        \end{theorem}
        \begin{proof}
        The proof is identical to that of \cref{theorem:algo_fair}.
        \end{proof}
 
        Thus it follows that:
        
        \begin{corollary}
         \balock{} satisfies the 1-FCFS property.
        \end{corollary}

    \section{A Constant-RMR Broadcast Object}
		\label{sec:broadcast|object}
		
	    An important component of our memory reclamation algorithm is a recoverable broadcast object, hereafter denoted by \broadcast{}, that allows a designated process to notify and ``wake up'' one or more processes waiting for it to reach a certain point during its execution. 
	    
	    The \broadcast{} object is similar to the \textsc{Signal} object used by \JJJ{} in~\cite{JayJay+:2019:PODC} to design an RME algorithm with sub-logarithmic RMR complexity. It is also similar to the \textsc{Barrier} object used by Golab and Hendler in~\cite{GolHen:2018:PODC} to design an RME algorithm for system-wide failures with $\bigO{1}$ RMR complexity.

	    The \textsc{Signal} object in~\cite{JayJay+:2019:PODC} can only be used once and different processes must wait on the object sequentially. However, a \broadcast{} object can be used \emph{repeatedly} and multiple processes can wait on the object \emph{concurrently}.
	    
	    The \textsc{Barrier} object  in~\cite{GolHen:2018:PODC} assumes that all processes fail together and may lead to a deadlock if processes fail independently. In that sense, the definition of \broadcast{} object and its implementation can be viewed as a generalization of \textsc{Barrier} object for the independent failure model.

		\subsection{Definition}
		Our broadcast object, in essence, is a recoverable MRSW (Multi-Reader Single-Writer) \emph{counter} object that stores a non-negative value and supports three operations, namely \bSet{}, \bWait{} and \bRead{}, with the following behavior.
			
		\begin{enumerate}
		\item \bSet(~) takes a positive number $x$ as an argument and sets the counter value to $x$ if its current value is smaller than $x$.
		\item \bWait(~) also takes a positive number $x$ as an argument and blocks until the counter value has advanced to at least $x$.
		\item \bRead(~) returns the current counter value.
		\end{enumerate}  
		
		A \broadcast{} object is \emph{owned} by a process and only the owner can change the value of the object. Thus \bSet{} operation can only be invoked by the owner process and \bWait{} operation can only be invoked by a non-owner process.
		
		An operation of a \broadcast{} object is invoked by a process from \emph{within} an RME algorithm. As such, if a process crashes while
		executing the operation, then it \emph{does not} resume execution from where it failed upon recovery. Rather, like in an RME algorithm,  the process restarts from the beginning. In the context of a \broadcast{} object, this translates to the process invoking the operation again at some point in the future. We formalize this behavior by requiring a history of the RME algorithm, \emph{when limited to the steps relevant to the \broadcast{} object}, to satisfy certain properties.

	    As such, in the rest of this  section, we consider a history to only consist of steps taken by processes \emph{while} executing the operations of a \broadcast{} object.
	    We refer to different invocations of operations in a history as \emph{instances}.
	    An instance of an operation \emph{terminates} when the calling process either crashes (while executing the instance) or returns (from the invocation). 
		
		A history $H$ is said to be \emph{\wellformed} with respect to a \broadcast{} object if it satisfies the following conditions:
		\begin{enumerate}[label=(W\arabic*)]
		\item The owner process invokes \bSet{} operation in an \emph{incremental} manner. Specifically, let \bSet($x$) and \bSet($y$) be two consecutive instances of the \bSet{} operation in $H$. Then, $x \leq y \leq x+1$.
		\label{enum:broadcast:set:incremental}
		\item The argument of every \bWait{} operation is ``close'' to the argument of its immediately preceding \bSet{} operation.
		Specifically, let $x$ denote the argument of a \bWait{} operation in $H$, and let $y$ denote the argument of its immediately preceding invocation of \bSet{} operation in $H$. In case no such \bSet{} operation exists, we set $y$ to 0. Then, $x \leq y + 1$.
		\label{enum:broadcast:wait:nearby}
		\end{enumerate}

        Intuitively, a \wellformed{} history helps to guarantee that the \broadcast{} operations are safe.

		An instance of an operation (\bSet{}, \bWait{} or \bRead{}) is referred to as \emph{failure-free} if the process does not crash while executing the instance. 
		An instance of \bWait{} or \bRead{} operation is said to have \emph{\complete[d]} if it has terminated without failing. On the other hand, an instance of \bSet($x$) is said to have \emph{\complete[d]} if it has terminated (possibly with a failure) and an invocation of the \bRead{} operation by the owner immediately afterwards would return the value at least $x$. 
		Note that an instance of \bSet{} operation may have \complete[d] even though it was not failure-free if
		the owner process was able to execute ``enough'' steps before crashing.

	    We now formulate requirements on the history to guarantee that the \broadcast{} operations are live. These requirements are primarily needed for the DSM model because
	    its algorithm is based on helping. To that end, we define the notion of a \legal{}  history, which is based on the notion of \run.
	    
	    Given a history $H$ and a process $p$ that is live in $H$, let $\onlyWait{H}{p}$ denote the sequence of \emph{all} instances of \bWait{} operation invoked by $p$ in $H$. A non-empty sub-sequence $\lambda$ of the sequence $\onlyWait{H}{p}$  forms a \emph{\run} if it satisfies the following conditions:
        \begin{enumerate}[label=(\alph*)]
        \item All instances in $\lambda$ have the same argument.
        \item Two instances are consecutive in $\lambda$ only if they are consecutive in $\onlyWait{H}{p}$.
        \item Every non-last instance in $\lambda$ terminates with a failure (\emph{i.e.}, does not \complete).
        \end{enumerate}

        Note that the last instance in $\lambda$, in case $\lambda$ is finite, may or may not terminate, and, if it does terminate, it may terminate with a failure.
        We are now ready to define the notion of \legal{} history. A  \wellformed{} history $H$ is said to be \emph{\legal} if it satisfies the following conditions:
        \begin{enumerate}[label=(L\arabic*)]
  	    \item If an instance (of an operation) invoked by a process $p$ in $H$ has not terminated, then eventually $p$ either crashes or takes another step in $H$.
	    \label{enum:broadcast:finite:terminated}
        \item Every \run{} in $H$ contains a finite number of instances.
        \label{enum:broadcast:run:finite}
        \item The last instance of every maximal \run{} in $H$ is failure-free.
        \label{enum:broadcast:last:failure-free}
        \end{enumerate}
    	
       The first condition \ref{enum:broadcast:finite:terminated} implies that a history does not abruptly stop in the ``middle'' of an operation.
       The last two conditions \ref{enum:broadcast:run:finite} and \ref{enum:broadcast:last:failure-free} together imply that a process repeatedly invokes an instance of \bWait{} operation with the ``same'' argument
       until it is able to execute a failure-free instance. 
       
       Our algorithm for the \bWait{} operation under the DSM model uses helping as it uses a wakeup-chain to bound the worst-case RMR complexity by $\bigO{1}$. As such, it is guaranteed to  \complete{} only if the history is \legal.

		\subsection{Implementation}
		
		We describe implementations of the \broadcast{} object separately for CC and DSM models since the algorithms to achieve $\bigO{1}$ RMR complexity for all three operations are quite different.

		\subsubsection{CC model}
		It is relatively trivial to implement the \broadcast{} object for the CC model in which all three operations have $\bigO{1}$ RMR complexity in the worst case as shown in \cref{algo:broadcastCC}.

     \addprefix[C.]

     \begin{algorithm}[t]
            \begin{multicols}{2}
            
            \SetKw{Shared}{shared variables}
            \SetKw{Local}{local variables}
            \SetKw{Struct}{struct}
            \SetKw{Integer}{int}
            \SetKw{Boolean}{bool}
            \SetKw{Array}{array}
            \SetKw{Await}{await}
            \SetKw{To}{to}
            \SetKw{Writer}{Designated process}
            \SetKwProg{initialization}{initialization}{\\ begin}{end}
            \SetKwProg{function}{Function}{\\ begin}{end}

            \Shared \\
            \Indp
            \tcp{integer variable to model the counter}
            $\varcount$\;
            \Indm
            \BlankLine
            
            \initialization{}{
                \tcp{initialize counter value}
                $\varcount$ $\leftarrow$ 0\;
            }
            
            \tcc{returns the current counter value}
            integer \function{\bRead(~)}
            {
                \Return $\varcount$\;
                \label{line:broadcastcc:read}
            }
            
            %\columnbreak
            
            \tcc{sets the counter to a desired value; can only be invoked if process $p_i$ is the owner process}
            \function{\bSet( $x$: integer variable )}
            {
                \lIf{($\varcount \geq x$)}{\Return}
                \label{line:broadcastcc:test}
                $\varcount \leftarrow x$\;
                \label{line:broadcastcc:write}
            }
            \BlankLine
            \tcc{wait until the counter has reached a desired value; can only be invoked if process $p_i$ is a non-owner process}
            \function{\bWait( $x$: integer variable )}
            {
               \Await $\varcount[i] \geq x$\;
               \label{line:broadcastcc:spin}
            }
            
            \end{multicols}
            \caption{Pseudocode of process $p_i$ for implementing \broadcast{} object under the CC model.}
            \label{algo:broadcastCC}
     \end{algorithm}

		The algorithm uses a single shared MRSW integer variable $\varcount$.
		In \bSet($x$) operation, the owner process writes $x$ to the $\varcount$ if its current value is smaller than $x$ (\crefrange{line:broadcastcc:test}{line:broadcastcc:write}). In \bWait($x$)  operation, a non-owner process spins until the variable has value of at least $x$ (\cref{line:broadcastcc:spin}). In \bRead{} operation, the process simply returns the current value of $\varcount$ (\cref{line:broadcastcc:read}). 
		Note that only \bWait{} operation has a loop, and \ref{enum:broadcast:wait:nearby} guarantees that a process busy-waiting in the loop incurs only $\bigO{1}$ RMRs.
		
		We refer to the algorithm for the CC model as \broadcastcc.

		\subsubsection{DSM model}
		The solution for the CC model has unbounded RMR complexity for \bWait{} operation in the DSM model.
		This is because, when $\n \geq 3$, at least one non-owner process has to spin on a remote memory location. 
		Another approach is for a non-owner process to spin on a local memory location, while the owner process is responsible for waking up all the waiting processes by updating the memory location of each spinning process. This approach, however, yields $\bigO{\n}$ RMR complexity for \bSet{} operation.
		
		In this section, we describe an efficient algorithm to implement the \broadcast{} object in the DSM model that incurs  only $\bigO{1}$ RMRs for all three operations and uses $\bigO{\n}$ space per \broadcast{} object. The main idea is that, instead of notifying the spinning processes by itself, the owner process creates a \emph{wake-up chain} in its local memory such that each process in the chain is responsible for waking up \emph{at most one} other process. The owner process then only needs to wake up the \emph{first} process in the chain.
	    This technique is similar to the one used by Golab, \emph{et al.} in~\cite{GolHen+:2006:PODC} to derive a leader election algorithm with $\bigO{1}$ RMR complexity under the DSM model. 
		%%
		\begin{comment}
		In their algorithm, Golab, \emph{et al.} create a forest of processes such that only the root of the tree incurs RMRs. After a constant number of RMRs, a new root is selected, thereby bounding the RMR complexity of the algorithm by $\bigO{1}$. 
		\end{comment}
		%%
		However, their algorithm was designed to operate in a failure-free environment, whereas our algorithm is designed to be recoverable.  Further, our \broadcast{} object can be used repeatedly by processes for synchronization with no additional space overhead.

     \addprefix[D.]

     \begin{algorithm}[t]
            \begin{multicols}{2}
            
            \SetKw{Shared}{shared variables}
            \SetKw{Local}{local variables}
            \SetKw{Struct}{struct}
            \SetKw{Integer}{int}
            \SetKw{Boolean}{bool}
            \SetKw{Array}{array}
            \SetKw{Await}{await}
            \SetKw{To}{to}
            \SetKw{Writer}{Designated process}
            \SetKwProg{initialization}{initialization}{\\ begin}{end}
            \SetKwProg{function}{Function}{\\ begin}{end}

            \Shared \\
            \Indp
            \tcp{two integer variables to model the counter}
            $\varbefore$, $\varafter$: integer variable\;
            \tcc{used by a non-owner process to inform the owner process of its intention to wait until the counter has advanced to a certain value; all entries are local to the owner process} 
            $\varannounce$: \Array $[1{\dots}\n]$ of integer variable\;
            \tcc{memory location for each non-owner process to spin on; the $i$-{th} entry is local to process $p_i$} 
            $\vartarget$: \Array $[1{\dots}\n]$ of integer variable\;
            \tcc{used by the owner process to create a wake-up chain; all entries are local to the owner process}
            $\varwakeup$: \Array $[1{\dots}\n]$ of integer variable\;
            \Indm
            
            \BlankLine

            \initialization{}{
                \tcp{initialize both counter values}
                $\varbefore$ $\leftarrow$ 0\;
                $\varafter$ $\leftarrow$ 0\;
                \ForEach{$j \in \{ 1, 2, \dots, \n\}$}
                {
                    \tcc{no process is waiting initially}
                    $\vartarget[j] \leftarrow 0$\;
                    $\varannounce[j] \leftarrow 0$\;
                    $\varwakeup[j] \leftarrow 0$\;
                }
            }

            \tcc{returns the current counter value}
            integer \function{\bRead(~)}
            {
                \Return $\varafter$\;
                \label{line:read:after}
            }
            
            \columnbreak
            
            \tcc{sets the counter to a desired value; can only be invoked if process $p_i$ is the owner process}
            \function{\bSet( $x$: integer variable )}
            {
                \lIf{($\varafter \geq x$)}{\Return}
                \label{line:set:read|after}
                \BlankLine
                $\varbefore \leftarrow x$\tcp*[r]{update the first counter}
                \label{line:set:before} 
                \tcc{create the wake-up chain}
                $\varlast \leftarrow 0$\;
                \label{line:set:wakeup|chain:begin} 
                \For(\tcp*[f]{scan the $\varannounce$ array}){$j \leftarrow$ 1 \To $\n$}
                {
                    \label{line:set:wakeup|chain:loop:begin}
                    \If{($\varannounce[j] = x$)}
                    {
                        \label{line:set:read|announce:first} 
                        \tcc{assign process $p_j$ to wake-up the last waiting process}
                        $\varwakeup[j] \leftarrow \varlast$\;
                        \label{line:set:wakeup|chain:add} 
                        \If{($\varannounce[j] = x$)}
                        {
                            \label{line:set:read|announce:second} 
                            \tcc{affirm that process $p_j$ is still waiting}
                            $\varlast \leftarrow j$\;
                            \label{line:set:verify|waiting} 
                        }
                    }
                }
                \label{line:set:wakeup|chain:loop:end}
                \label{line:set:wakeup|chain:end} 
                \BlankLine
                \If{($\varlast > 0$)}
                {
                    \tcc{release the last process; all other waiting processes will be released one-by-one by the chain}
                    \CAS( $\vartarget[\varlast]$, $x$, 0 )\;
                    \label{line:set:wakeup:CAS} 
                }
                $\varafter \leftarrow x$\tcp*[r]{update the second counter}
                \label{line:set:after} 
            }
            \BlankLine
            \tcc{wait until the counter has reached a desired value; can only be invoked if process $p_i$ is a non-owner process}
            \function{\bWait( $x$: integer variable )}
            {
               $\vartarget[i] \leftarrow x$\tcp*[r]{initialize the location to spin on}
               \label{line:wait:target:initialize}
               $\varannounce[i] \leftarrow x$\tcp*[r]{announce the intention to wait}
               \label{line:wait:announce:initialize}
               \tcc{no need to wait if the owner process has previously invoked \bSet{} operation with argument value $x$}
               \If{($\varbefore \geq x$)}
               {
                  \label{line:wait:read:before}
                  $\vartarget[i] \leftarrow 0$\tcp*[r]{reset the location}
                  \label{line:wait:target:reset}
               }
               \tcc{spin until some process resets the location}
               \Await $\vartarget[i] = 0$\;
               \label{line:wait:spin}

               $\varannounce[i] \leftarrow 0$\tcp*[r]{revoke the announcement}
               \label{line:wait:announce:reset}
               $k \leftarrow \varwakeup[i]$\tcp*[r]{check if any process to wake up}
               \label{line:wait:read:wakeup}
               \If{($k > 0$)}
               {
                   \tcc{wake up the next process in the chain}
                   \CAS( $\vartarget[k]$, $x$, 0 )\;
                   \label{line:wait:wakeup:CAS}
               }
               \label{line:wait:end}
            }
            
            \end{multicols}
            \caption{Pseudocode of process $p_i$ for implementing \broadcast{} object under the DSM model.}
            \label{algo:broadcast}
     \end{algorithm}

		The pseudocode of our algorithm is given in \cref{algo:broadcast}. 
		Our algorithm uses two shared integer variables, namely $\varbefore$ and $\varafter$, and three shared integer array variables, namely $\varannounce$, $\vartarget$ and $\varwakeup$. Each array variable is of size $\n$ with one entry for each process.
		The variables $\varbefore$ and $\varafter$ both store the counter value; they are updated at beginning and end, respectively, of a \bSet{} operation. 
	    The array variables $\varannounce$ and $\varwakeup$ are local to the owner process. However, the array variable $\vartarget$ is distributed among all processes with
	    the $i$-{th} entry local to process $p_i$. 
	    A process $p_i$ uses $\varannounce[i]$ to inform the owner process of its intention to wait until the counter value has advanced to a desired value and
	    $\vartarget[i]$ to busy-wait until it is released (from spinning) by another process (note that $\vartarget[i]$ is local to $p_i$).
		Finally, the owner process uses the $\varwakeup$ array to set up a wake-up chain, where $p_i$, upon waking up, is responsible for notifying the process whose identifier is stored in $\varwakeup[i]$, if any.
		
		\paragraph{\bWait{} operation}
		Let $x$ denote the input argument of the operation. The invoking process, say $p_i$, writes $x$ to $\vartarget[i]$ and $\varannounce[i]$ in order (\crefrange{line:wait:target:initialize}{line:wait:announce:initialize}), thereby informing the owner process of its intention to wait for the counter value to reach $x$. It then reads the value of $\varbefore$ variable (\cref{line:wait:read:before}) to check if the owner process has already invoked \bSet{} operation with argument $x$. If so, it resets $\vartarget[i]$ to zero (\cref{line:wait:target:reset}) so that it does not busy-wait in the next step. It then spins until $\vartarget[i]$ is reset to zero (\cref{line:wait:spin}). Upon quitting the busy-waiting loop, 
		it clears $\varannounce[i]$ (\cref{line:wait:announce:reset})
		and reads $\varwakeup[i]$ (\cref{line:wait:read:wakeup}) to determine the identifier of the next process to be notified, if any. Let $\varwakeup[i] = k$. If $k > 0$, then process $p_i$ notifies process $p_k$ by resetting $\vartarget[k]$ to zero using a \CAS{} instruction (\cref{line:wait:wakeup:CAS}). 
		Note that the $\vartarget$ array may contain multiple chains; our algorithm ensures that 
		all processes in the \emph{same chain} are waiting for the counter to reach the \emph{same value}.

		\paragraph{\bSet{} operation}
		Let $x$ denote the input argument of the operation. The invoking process first writes $x$ to $\varbefore$ variable (\cref{line:set:before}). 
		This ensures that any process that invokes \bWait($x$) hereafter does not block.
		It then scans the $\varannounce$ array looking for processes that may be waiting for the counter value to advance to $x$ and chains all of them together (in reverse order of their identifiers) using the $\varwakeup$ array (\crefrange{line:set:wakeup|chain:begin}{line:set:wakeup|chain:end}). Consider two processes $p_\ell$ and $p_j$ with $\ell < j$  such that $\varannounce[\ell] = x$, $\varannounce[j] = x$ and, for each $k$ with $\ell < k < j$, $\varannounce[k] \neq x$. The owner process then sets $\varwakeup[j] = \ell$ indicating that process $p_j$, upon being released from spinning, is responsible for notifying process $p_\ell$. After writing $\ell$ to $\varwakeup[j]$, the owner process reads $\varannounce[j]$ again (\cref{line:set:read|announce:second}) to ascertain that $p_j$ has not already stopped busy-waiting and thus may have missed reading $\varwakeup[j]$ field; otherwise it may create a situation in which $p_\ell$ spins indefinitely with no process responsible for notifying it. 
		After building this chain, the owner process notifies the first process in the chain by clearing its entry in the $\vartarget$ array (\cref{line:set:wakeup:CAS}), which then leads to a sequence of \emph{cascading} notifications. Finally, it writes $x$ to $\varafter$ variable (\cref{line:set:after}).

		\paragraph{\bRead{} operation.}
		It returns the value of $\varafter$ variable (\cref{line:read:after}). The main use of the operation is to determine the last \bSet{} operation that \complete[d], prior to invoking a new \bSet{} operation. 
		
		\bigskip
		
		All three operations are designed to be idempotent in the sense that executing an operation multiple times with the same argument, possibly partially in some cases, does not lead to any erroneous behavior.

		We refer to the algorithm for the DSM model as \broadcastdsm.

   \subsection{Correctness proofs}
   
    We focus on proving the correctness of \broadcastdsm{} because the correctness proof of \broadcastcc{} is quite straightforward.
     
    In the rest of this section, we use $p_w$ to refer to the owner process. 
    The following proposition follows from the facts that only $p_w$ can write to $\varbefore$ and $p_w$ invokes \bSet{} operation in an incremental manner \ref{enum:broadcast:set:incremental}.
    
    \begin{proposition}
    \label{proposition:before:set}
    Given a \wellformed{} history $H$, if the variable $\varbefore$ has value at least $x$ at time $t$, then there exists an invocation of \bSet($x$) by $p_w$  in $H$ at or before time $t$.
    \end{proposition}

    The next proposition follows from code inspection.
    
    \begin{proposition}
    \label{proposition:owner:CAS:before}
    Given a \wellformed{} history $H$, if process $p_w$ executes the \CAS{} instruction on \cref{line:set:wakeup:CAS} with an expected value of $x$ at some time $t$, then the variable $\varbefore$ has a value that is at least $x$ at time $t$ in $H$.
    \end{proposition}
    
    Next, we prove the same property for a non-owner process.
    
    \begin{lemma}
    \label{lemma:nonowner:CAS:before}
    Given a \wellformed{} history $H$, if a process $p_i$, where $i \neq w$, executes the \CAS{} instruction on \cref{line:wait:wakeup:CAS} with an expected value of $x$ at time $t$, then the variable $\varbefore$ has a value that is at least $x$ at time $t$ in $H$.
    \end{lemma}
    \begin{proof}
    It is sufficient to prove that the statement holds if $p_i$ is the \emph{first} non-owner process in $H$ to invoke the \CAS{} instruction \cref{line:wait:wakeup:CAS} with the expected value of $x$. 
    As the code inspection shows, $p_i$ executes the \CAS{} instruction only after quitting its busy-waiting loop, which can happen only \emph{after} the entry $\vartarget[i]$ has been reset (to zero). The entry can be reset in only three ways. 
    First, $p_i$ resets $\vartarget[i]$ itself at \cref{line:wait:target:reset} because it read the value of $\varbefore$ to be at least $x$ earlier at \cref{line:wait:read:before}. 
    Second, $p_w$ resets $\vartarget[i]$ using the \CAS{} instruction on \cref{line:set:wakeup:CAS}. In this case, it follows from \cref{proposition:owner:CAS:before} that $\varbefore$ has value at least $x$ at the time $p_w$ performed the \CAS{} instruction.
    Third, another non-owner process, say $p_j$ with $j \notin \{ w, i \}$, resets $\vartarget[i]$ using the \CAS{} instruction on \cref{line:wait:wakeup:CAS}. By our assumption that $p_i$ is the first non-owner process to invoke the \CAS{} instruction \cref{line:wait:wakeup:CAS} with an expected value of $x$, the third case cannot occur. 
    \end{proof}

    \begin{theorem}
    \label{theorem:broadcast:safety}
    Given a \wellformed{} history $H$, if an invocation of \bWait($x$) by a process $p_r$, where $r \neq w$, returns at some time $t$ in $H$, then process $p_w$ has invoked \bSet($x$) before time $t$ in $H$.
    \end{theorem}
    \begin{proof}
    As the code inspection shows, $p_r$ quits the busy-waiting loop in the \bWait{} operation after $\vartarget[r]$ has been reset to 0.
    The entry can be reset in only three ways. We show that each of the three cases implies that the value of $\varbefore$ is at least $x$ at the time the entry was reset. The statement then follows from \cref{proposition:before:set}. 
    
    First, $p_r$ resets  $\vartarget[r]$ itself at \cref{line:wait:target:reset} because it read the value of $\varbefore$ to be at least $x$ earlier at \cref{line:wait:read:before}. 
    
    Second, $p_w$ resets $\vartarget[r]$ using a \CAS{} instruction on \cref{line:set:wakeup:CAS}. In this case, it follows from \cref{proposition:owner:CAS:before} that the value of $\varbefore$ is  at least $x$ at the time the entry was set.
    
    Third, another non-owner process, say $p_s$, where $s \notin \{ w, r \}$, resets $\vartarget[r]$ using a \CAS{} instruction on \cref{line:wait:wakeup:CAS}. In this case, it follows from \cref{lemma:nonowner:CAS:before} that the value of $\varbefore$ is at least $x$ at the time the entry was reset. The result then follows from \cref{proposition:before:set}.
    \end{proof}

     \begin{comment}
  
     If an instance of an operation did not \complete, then we expect the calling process to invoke another instance of the same operation with the same argument again. This is particularly important for the \bWait{} operation because it uses helping chain to wake-up all processes. If a process does not re-invoke the \bWait{} operation  upon failing, then some process may be blocked forever. 
     This requirement is formalized using the notion of a \legal{} history, which is defined using the notion of a \run.
	      
     \end{comment}
	
   	 In \broadcastdsm, \bSet{} operation \complete[s] once it has successfully executed 
   	 \cref{line:set:after}.

    \begin{theorem}
    \label{theorem:broadcast:liveness}
    Let $H$ be a \legal{} history in which at least one instance of \bSet($x$) \complete[s]. Then every instance of \bWait($x$) eventually terminates.  
    \end{theorem}
    \begin{proof}
    It suffices to only consider failure-free instances of \bWait($x$). 
    Assume, on the contrary, that there exists at least one process in $H$ that invokes \bWait($x$) and does not crash but the invocation never returns. 
    Among all such processes, pick the one with the \emph{largest} identifier, say $p_{\ell}$, and let $I_{\ell}$ denote an instance of \bWait($x$) in $H$ invoked by $p_{\ell}$ that never terminates. This also implies that $p_{\ell}$ eventually advances to the busy-waiting loop during $I_{\ell}$ \ref{enum:broadcast:finite:terminated} and gets stuck in the loop.
    Let $I_w$ denote the \emph{first} instance of \bSet($x$) in $H$ that $p_w$ is able to \complete.
    It follows from code inspection that $p_w$ is able to write $x$ to variable $\varafter$ (\cref{line:set:after}) during $I_w$. Further, any subsequent invocation of \bSet($x$) by $p_w$ in $H$ simply returns after ascertaining that variable $\varafter$ has value at least $x$, without creating any ``new'' wake-up chain.
    
    Before proceeding further with the proof, we define some notation. Suppose a process $p$ is executing an instance $I$ of an operation $op$, where $op \in \{ \bSet{}, \bWait{} \}$. Let $m$ denote the number of a line belonging to $op$ in the pseudocode shown in \cref{algo:broadcast}. 
    If the line is not part of the for-loop in \bSet{}, then we use $\atTime{I}{m}$ to denote the time when $p$ finished executing the line numbered $m$ during $I$. Otherwise, we use $\atTimeFor{I}{m}{k}$ to denote the time when $p$ finished executing the line for the iteration (of the for-loop) with $j$ set to $k$ during $I$. Finally, given a variable $var$, we use $var^t$ to denote the value of $var$ at time $t$.
    
    For ease of exposition, unless otherwise stated, steps of processes $p_w$ and $p_{\ell}$ are assumed to belong to instances $I_w$ and $I_{\ell}$, respectively.

    Since $p_{\ell}$ never advances beyond the busy-waiting loop, $\vartarget[\ell]$ and $\varannounce[\ell]$ are never reset. Formally,
    \begin{align}
    \label{eq:broadcast:fact:1}
    \forall t : t \geq \atTime{I_{\ell}}{\ref{line:wait:read:before}} : (\varannounce^t[\ell] = x) \land (\vartarget^t[\ell] = x)
    \end{align}
    Clearly, $p_{\ell}$ reads the value of $\varbefore$  \emph{before} $p_w$ writes value $x$ to $\varbefore$. Formally,
    \begin{align}
    \label{eq:broadcast:fact:2}
    \atTime{I_{\ell}}{\ref{line:wait:read:before}} \ < \ \atTime{I_w}{\ref{line:set:before}} 
    \end{align}
    We use $\mathcal{C} = \{ c_1, c_2, \ldots, c_h \}$ to denote the set of process indices in the wake-up chain created by $p_w$ (in the reverse order of their identifiers). We  prove that $p_{\ell}$ must be part of this wake-up chain, \emph{i.e.}, $\ell \in \mathcal{C}$.
    Using \eqref{eq:broadcast:fact:1} and \eqref{eq:broadcast:fact:2}, we can infer that  
    \begin{align}
    \label{eq:broadcast:fact:3}
    \forall t : t \geq \atTime{I_w}{\ref{line:set:before}} : (\varannounce^t[\ell] = x) \land (\vartarget^t[\ell] = x)
    \end{align}   
    This implies that, when $p_w$ reads the value of $\varannounce[\ell]$ twice at \cref{line:set:read|announce:first,line:set:read|announce:second}, it finds it to be $x$ both times. Thus $\ell \in C$.
    
    It is now sufficient to prove that some process executes \CAS{}($\vartarget[\ell]$, $x$, 0) after  time $\atTime{I_w}{\ref{line:set:before}}$, thereby resetting $\vartarget[\ell]$ to 0 and enabling $p_{\ell}$ to quit its busy-waiting loop. The remainder of the code after the busy-waiting loop is wait-free, thereby implying that $I_{\ell}$ eventually returns. This will contradict our original assumption that $I_{\ell}$ never returns. 
    There are two cases to consider.
    
    \begin{description}
    \item[Case 1 ($\ell = c_1$)]: It follows from code inspection that $p_w$ proceeds to reset $\vartarget[c_1]$ to 0 using a \CAS{} instruction, which is guaranteed to succeed due to \eqref{eq:broadcast:fact:3}.
    \item[Case 2 ($\ell = c_j$, where $j > 1$)]: 
    Since $p_{c_{j-1}}$ is part of the wake-up chain created by $p_w$, when $p_w$ reads $\varannounce[c_{j-1}]$ for the second time, $p_w$ finds that $\varannounce[c_{j-1}]$ has value $x$. 
    This implies that $I_w$ ``overlaps'' with one or more \run{s} of \bWait($x$) invoked by $p_{c_{j-1}}$. Let $I_{c_{j-1}}$ denote the \emph{last} instance of the \emph{last} \run{} of \bWait($x$) invoked by $p_{c_{j-1}}$ that ``overlaps'' with $I_w$. Note that $I_{c_{j-1}}$ is guaranteed to exist because:
    \begin{enumerate*}[label=(\roman*)]
    \item $I_w$ is finite (by choice of $I_w$) and hence there is a last \run, and 
    \item every \run{} in $H$ is of finite length \ref{enum:broadcast:run:finite} and hence the last \run{} has a last instance.
    \end{enumerate*}
    Furthermore, $I_{c_{j-1}}$  is failure-free since the last instance of every maximal \run{} in $H$ is failure-free \ref{enum:broadcast:last:failure-free}.
    By our choice of $p_{\ell}$, $I_{c_{j-1}}, $ is finite (\emph{i.e.}, \complete[s]). Again, for ease of exposition, unless otherwise stated, steps of process $p_{c_{j-1}}$ are assumed to belong to instance $I_{c_{j-1}}$.

    By our choice of $I_{c_{j-1}}$, once $p_{c_{j-1}}$ resets $\varannounce[c_{j-1}]$ to 0, it is not set to $x$ again at least until $I_w$ terminates.  Formally,
    \begin{align}
    \label{eq:broadcast:fact:4}
    \forall t : \atTime{I_{c_{j-1}}}{\ref{line:wait:announce:reset}} \leq t \leq \atTime{I_w}{\ref{line:set:after}}: \varannounce^t[c_{j-1}] \neq x 
    \end{align}

    It follows from code inspection that $p_w$ writes $c_j$ to $\varwakeup[c_{j-1}]$ before it reads $\varannounce[c_{j-1}]$ for the second time. Formally,
    \begin{align}
    \label{eq:broadcast:fact:5}
    \atTimeFor{I_w}{\ref{line:set:wakeup|chain:add}}{c_{j-1}} \ < \ 
    \atTimeFor{I_w}{\ref{line:set:read|announce:second}}{c_{j-1}}
    \end{align}
    
    Since $p_{c_{j-1}}$ is part of the wake-up chain created by $p_w$, when $p_w$ reads $\varannounce[c_{j-1}]$ for the second time, $p_w$ finds that $\varannounce[c_{j-1}]$ has value $x$. 
    Thus, using \eqref{eq:broadcast:fact:4}, we can infer that
    \begin{align}
    \label{eq:broadcast:fact:6}
    \atTimeFor{I_w}{\ref{line:set:read|announce:second}}{c_{j-1}} \ <  \ 
    \atTime{I_{c_{j-1}}}{\ref{line:wait:announce:reset}} 
    \end{align}
    
    It also follows from code inspection that $p_{c_{j-1}}$ resets $\varannounce[c_{j-1}]$ to 0 before it reads the value of $\varwakeup[c_{j-1}]$. Formally,
    \begin{align}
    \label{eq:broadcast:fact:7}
    \atTime{I_{c_{j-1}}}{\ref{line:wait:announce:reset}} \ < \    
    \atTime{I_{c_{j-1}}}{\ref{line:wait:read:wakeup}} 
    \end{align}
   
    Combining \eqref{eq:broadcast:fact:5}, \eqref{eq:broadcast:fact:6} and \eqref{eq:broadcast:fact:7}, we obtain
    that $p_w$ writes $c_j$ to $\varwakeup[c_{j-1}]$ before $p_{c_{j-1}}$ reads from $\varwakeup[c_{j-1}]$. Formally,
    \begin{align}
    \label{eq:broadcast:fact:8}
    \atTimeFor{I_w}{\ref{line:set:wakeup|chain:add}}{c_{j-1}} \ < \    
    \atTime{I_{c_{j-1}}}{\ref{line:wait:read:wakeup}} 
    \end{align}
    
    When $p_w$ reads $\varannounce[c_{j-1}]$ for the second time, it finds it to be $x$. Thus, using \eqref{enum:broadcast:set:incremental}, we can infer that
    \begin{align}
    \forall t : \atTimeFor{I_w}{\ref{line:set:read|announce:second}}{c_{j-1}} \leq t \leq \atTime{I_{c_{j-1}}}{\ref{line:wait:read:wakeup}} : \varannounce^t[c_{j-1}] \in \{ x, 0 \}
    \end{align}
    
    Since any subsequent invocation of \bSet($x$) by $p_w$ (after $I_w$) does not create any new wake-up chain, it implies that $p_w$ does not write to $\varwakeup[c_{j-1}]$ between $\atTimeFor{I_w}{\ref{line:set:wakeup|chain:add}}{c_{j-1}}$ and  $\atTime{I_c}{\ref{line:wait:read:wakeup}}$. %%
    As a result, when $p_{c_{j-1}}$ reads the value of $\varwakeup[c_{j-1}]$, it finds that it is set to $c_j~(= \ell)$. It follows from code inspection that $p_{c_{j-1}}$ proceeds to execute \CAS{}($\vartarget[\ell]$, $x$, 0), which is guaranteed to succeed due to \eqref{eq:broadcast:fact:1}.
    \end{description}
    In both cases, we arrive at a contradiction, thereby proving the statement.
    \end{proof}
    
    The following theorem follows directly from code inspection since \bSet{} and \bRead{} operations are wait-free.
    
    \begin{theorem}
    \label{theorem:broadcast:rest}
        Every instance of \bSet{} and \bRead{} operation in a \legal{} history eventually terminates (possibly with a failure). 
    \end{theorem}
    
    \begin{theorem}
    \label{theorem:broadcast:performance}
        Each instance of \bRead{}, \bSet{} or \bWait{} operation has worst-case RMR complexity of $\bigO{1}$.
    \end{theorem}
    \begin{proof}
        It is trivial to see that \bRead{} incurs $\bigO{1}$ RMRs. The \bSet{} operation contains a loop from \cref{line:set:wakeup|chain:begin} to \cref{line:set:wakeup|chain:end} where process $p_w$ accesses variables $\varannounce$ and $\varwakeup$. Both these variables are local to process $p_w$ and thus the \bSet{} operation incurs $\bigO{1}$ RMRs. Finally, the \bWait{} operation also contains only one loop at \cref{line:wait:spin} where only variable $\vartarget[i]$ is repeatedly accessed by process $p_i$. The variable $\vartarget[i]$ is local to process $p_i$ and therefore the \bWait{} operation also incurs only $\bigO{1}$ RMRs. Thus, the theorem holds. 
    \end{proof}
    
    It can be easily verified that \broadcastcc{} satisfies  \cref{theorem:broadcast:safety,theorem:broadcast:liveness,theorem:broadcast:rest,theorem:broadcast:performance} as well.

    \section{Bounding Space Complexity of the Framework}
    \label{sec:memory}
    
        Note that our framework uses three types of locks---a weakly recoverable \filter{} lock, a strongly recoverable $\n$-process \base{} lock and a strongly recoverable dual-port \arbitrator{} lock. 
        The first lock can be implemented using the weakly RME algorithm described in \cref{sec:weak_recoverability}. The second lock can be implemented using sub-logarithmic RME algorithm proposed by \KM\footnote{The algorithm has the added feature of abortability, which is not used in our framework.} in~\cite{KatMor:2020:OPODIS}. Finally, the third lock can be implemented using Yang and Anderson's algorithm augmented to handle failures~\cite{GolRam:2019:DC}. The RME algorithm by \KM{} has space complexity of $\bigO{\nicefrac{\n\log^2\n}{\log\log\n}}$ under both CC and DSM models. The augmented Yang and Anderson's algorithm has $O(\n)$ space complexity under both CC and DSM models. However, the weakly RME algorithm described in \cref{sec:weak_recoverability} (\weakMCS) has unbounded space complexity because, upon generating a request, it allocates a queue node at run time (additional in case of \unsafe{} failures). It is non-trivial to determine when the memory of these nodes can be reclaimed without causing the algorithm to misbehave due to dangling pointers while, at the same time, maintaining its $\bigO{1}$ RMR complexity.
        
        In this section, we describe an algorithm for memory reclamation that can be used to bound the space complexity of augmented \weakMCS{} by  $\bigO{\n^2}$ under both CC and DSM models.
        Our memory reclamation algorithm uses ideas from two well known approaches for memory reclamation, namely \emph{epoch-based}~\cite{Fra:2004:PhD} and \emph{quiescent-state-based}~\cite{ArcCao+:2003:ATC,McKSli:1998:ICPDCS}, with the added benefits that it 
        \begin{enumerate*}[label=(\alph*)]
        \item has bounded space complexity and 
        \item is recoverable. 
        \end{enumerate*}

        The quiescent-state-based memory reclamation algorithms assume that processes exhibit the following behavior:
        \begin{enumerate}[label=(Q\arabic*)]
        \item Every process \emph{alternates} between quiescent and non-quiescent states during its execution. 
        \label{enum:quiescent:1}
        \item A process can access a shared object \emph{only} when it is in a non-quiescent state. 
        \item A shared object has to be \emph{retired} before its memory can be reclaimed.  
        \item While in a non-quiescent state, a process cannot access any shared object that was retired \emph{prior} to it entering its non-quiescent state.
        \label{enum:quiescent:4}
        \end{enumerate}
        This behavior can be leveraged to obtain the following general memory reclamation scheme: upon retiring a shared object, a process can reclaim the memory of the object safely after every process in the system has been in its quiescent state at least once since then.
        We use this main idea to design a memory reclamation algorithm that has the desired RMR and space complexities.

        If processes do not fail, then quiescent and non-quiescent states of a process can be taken to be \NCS{} \segment{} and passage, respectively, and it can be verified that properties \ref{enum:quiescent:1}--\ref{enum:quiescent:4} hold. If processes can fail, then one possible approach is to consider a process to be in non-quiescent state when it is executing its super-passage. Although the properties \ref{enum:quiescent:1}--\ref{enum:quiescent:4} hold, however, this straightforward extension does not yield a memory reclamation algorithm with bounded space complexity, at least directly. This is because, every time a process experiences an \unsafe{} failure, it needs a ``fresh'' queue node. As a result, a process may ``consume'' an arbitrarily large number of queue nodes during a single super-passage.
        
        To obtain a memory reclamation algorithm with bounded space complexity, we model a super-passage of \weakMCS{} as consisting of a sequence of \emph{\attempt{s}}. Loosely speaking, an \attempt{} begins when a process requests 
        allocation of a new node at \cref{line:MCS:enter:newnode} (using the \FnNewNode{} function) and ends when it retires\footnote{Note that the notion of ``retire'' is distinct from that of ``\relieve''~(defined in \cref{sec:MCS:proof}). Also, note that a node is \relieve{d} before it is retired.} the allocated node at \cref{line:exit:retire} (using the \FnRetireNode{} function).
        We use two monotonically non-decreasing counters for every process to \emph{demarcate} the beginning and end of its \attempt{}. The counters for process $p_i$ are denoted by $\varstart[i]$ and $\varfinish[i]$; the latter is a \broadcast{} object described in \cref{sec:broadcast|object}. They represent the number of \attempt{s} a process has started (respectively, finished) across all super-passages so far. 
        A process increments its start counter just before it enters the doorway of \weakMCS, and increments its finish counter at the completion of the \FnRetireNode{} function.
        
        If a process is not executing an \attempt{}, it is said to be in a \emph{quiescent} state, at which point its two counters will have the same value. At all other times, the finish counter of each process lags behind its start counter by exactly one. 
        As desired, when a process is executing the doorway or waiting room of \weakMCS{} as part of an \attempt{}, it is in non-quiescent state.
        
        Note that ``\attempt'' and ``passage'' are related but different concepts. They provide two different ways to model the execution of a process \emph{within its super-passage}.  The boundaries of \attempt{s} and passages do \emph{not} align. A failure ends a passage but not an \attempt. A process may fail multiple times during an \attempt.
        As such, an \attempt{} of a process can overlap with multiple passages of that process, but a passage of a process can overlap with most two \attempt{s} of that process.
        
        \begin{comment}
        
        This because an \attempt{} begins in an \Enter{} \segment{} (when the start counter is incremented), whereas a passage begins with a \Recover{} \segment{}. Further, an \attempt{} ends in either the \Recover{} or the \Exit{} \segment{} (when the finish counter is incremented), whereas a passage ends when the \Exit{} \segment{} completes or the process crashes. 
      
        It can be verified that an \attempt{} ends in the \Recover{} \segment{} if (and only if) the process experiences an \unsafe{} failure (\emph{i.e.}, crashes while executing the \FAS{} instruction in the doorway). Upon recovery, the process retires the node allocated during this \attempt{} and starts a new \attempt{} with a fresh node. 
        %%
        Also, an \attempt{} of a process can overlap with multiple passages of that process. But a passage of a process can overlap with most two \attempt{s} of that process.
        
        \end{comment}
        
        \begin{definition}[\useful{} \attempt]
        An \attempt{} of a process is said to be \emph{\useful} if the process takes at least one step of its \CS{} \segment{} during the \attempt; otherwise, it is said to be \useless{}.
        \end{definition}

        It can be shown that an \attempt{} of a process becomes \useless{} if and only if the process crashes while executing the \FAS{} instruction at \cref{line:enter:FAS}.
        Every completed super-passage in an infinite fair history consists of \emph{at least one} \useful{} \attempt{}. In particular, the last \attempt{} of a completed super-passage is \useful.
        A node allocated during an \attempt{} is said to be have been \emph{retired} if the process has completed the execution of the \FnRetireNode{} function during that \attempt{} at least once without crashing.
        %%
		\begin{comment}
		%%
		%% Do we need this?
		%%
		For ease of exposition, assume that a retired node is never reused; otherwise, the properties described next do not hold in the strict sense. In practice, however, a retired node can be reused once the algorithm described in this section has reclaimed the memory of the node.
		%SAHIL: Retired and reclaimed are not defined
		\end{comment}
		%%
		It can be easily verified that:

        \begin{proposition}
        \label{proposition:quiescent:attempt}
        \weakMCS{} satisfies \ref{enum:quiescent:1}--\ref{enum:quiescent:4}.
        \end{proposition}

        \begin{proposition}
        \label{proposition:attempt:allocate}
        A process consumes at most one queue node during an \attempt.
        \end{proposition}
 
		Note that, even after a node has retired, another process may still hold a reference to that node and may dereference it in the future. 
		Our memory reclamation algorithm relies on the notion of a \emph{grace period} to determine when is it safe to reclaim the memory of a node after it has been retired~\cite{McKSli:1998:ICPDCS,ArcCao+:2003:ATC,HarMcK+:2007:JPDC}. 
		%%
		%%\begin{definition}[grace period]
		%%
		A time interval $[t,t']$ is said to be a \emph{grace period}  if, after time $t'$, no process holds a reference to any node that was retired at or before time $t$. 
		%%
		%%\end{definition}
		%%
	    We define a related notion of \allowance to highlight the grace period \emph{with respect to a retired node} as follows.
		
	    \begin{definition}[\allowance{}]
	    Given a history $H$ and a node $x$, an interval $[t,t']$ in $H$ is said to be the \emph{\allowance{}} with respect to $x$ if $x$ was retired at time $t$ and no process holds any reference to $x$ after time $t'$.
	    \end{definition}
	    
	    It follows from \cref{proposition:quiescent:attempt} that:
	    
	    \begin{lemma}
	    The \allowance{} with respect to a node retired at time $t$ expires once every process has been in a quiescent state \emph{at least once} after time $t$.
	    \end{lemma}
	    
		Typically, in existing epoch based memory reclamation algorithms, a process uses a \emph{non-blocking} method to \emph{detect} that the \allowance{} of a node it retired earlier has expired. The length of \allowance{} depends on the execution speeds of other processes. As such, it may last arbitrarily long, during which period the process may retire many more (possibly unbounded number of) nodes. This makes it hard to achieve bounded space complexity. 
		In this work, we use a \emph{blocking} approach to detect that the \allowance{} of a retired node has expired. This is feasible in our case because the underlying application, namely mutual exclusion, is inherently blocking (provided that any additional blocking does not create a deadlock). One of the main components of our memory reclamation algorithm is a routine to detect expiration of the \allowance{} with respect to a retired node (or, more precisely, a set of retired nodes), denoted by \emph{\GPTD{} routine}, using a blocking synchronization that yields a bounded space complexity. In particular, the routine is designed to satisfy the following property:

		\begin{lemma}
		\label{lem:gp:routine}
	    Suppose an instance of \GPTD{} routine began at time $t$ and completed at time $t'$. Then, the \allowance{} of every node retired at or before time $t$ has expired at or after time $t'$.
	    \end{lemma}
	    
	    A simple approach to implement the \GPTD{} routine is as follows. A process, while in its quiescent state, waits for the $\varfinish$ counter of every process to ``catch up'' to its $\varstart$ counter, one-by-one. After the \GPTD{} routine completes,  it follows from \cref{lem:gp:routine} that the process can safely reclaim the memory of any node it retired in the previous \attempt. Due to its blocking nature, a process executes the \GPTD{} routine
		in the \Enter{} \segment{} before starting a new \attempt{} in order to maintain BR and BE properties. Executing the routine in a quiescent state cannot create any deadlock since, both $\varstart$ and $\varfinish$ counters have the same value while a process is in quiescent state,and the process does not own any node in any sub-queue~\cref{sec:MCS:proof}. Thus, no process can be busy-waiting on another process executing the \GPTD{} routine.
		This approach, however, increases the RMR complexity of each passage of \weakMCS{} to $\bigO{\n}$.
		
		\subsection{Achieving constant RMR complexity}
		A possible approach to reduce RMR complexity is to \emph{amortize} the overhead of executing a \GPTD{} routine over $\Omega(\n)$ \attempt{s}. For example, a process can execute the \GPTD{} routine after every $\n$ \attempt{s}. Once the \GPTD{} routine completes, all nodes retired prior to the $\n$ \attempt{s} can be safely reclaimed. This, however, requires another set of $\n$ nodes that can be used to service requests for allocating queue nodes during these $\n$ \attempt{s}. Thus, to use this optimization, a process has to maintain \emph{two} pools of nodes, each containing $\n$ nodes. While nodes in one pool are waiting for their \allowance{} to expire, nodes in the other pool can be used to serve requests for node allocation. We refer to the two pools as \emph{active} and \emph{backup} with obvious meaning. The role of the two pools is then \emph{switched} after every $\n$ \attempt{s}. With this optimization, each passage of \weakMCS{} has $\bigO{1}$ RMR complexity in the \emph{amortized case} but $\bigO{\n}$ RMR complexity in the \emph{worst case}.
		
		We now describe a way to keep the worst-case RMR complexity of each passage at $\bigO{1}$. The main idea is to execute the \GPTD{} routine \emph{incrementally} over $\Theta(\n)$ \attempt{s} in such a way that it
		\begin{enumerate*}[label=(\alph*)]
		\item is recoverable,
		\item adds only $\bigO{1}$ RMRs to each passage of \weakMCS{},
		\item uses only $\bigO{\n^2}$ space, and
		\item maintains the \CIFCFS{} property of  \weakMCS{}.
		\end{enumerate*}
		
		To maintain the fairness guarantee, we use a third counter that is incremented after a process has completed the doorway of the original \weakMCS{} without crashing during a passage, referred to as \emph{\checkpoint} counter. The \checkpoint{} counter for process $p_i$ is denoted by $\varcheckpoint[i]$ and is basically a \broadcast{} object described in \cref{sec:broadcast|object}.
		At any given time, the counters satisfy the following invariants for each process $p_i$:
	    \begin{gather}
	    \label{eq:counters:invariant:1}
	    \varstart[i] - 1 \; \leq \; \varfinish[i] \; \leq \; \varcheckpoint[i] \; \leq \; \varstart[i] 
	    \end{gather}
	    In case an \attempt{} is rendered \useless{} due to an \unsafe{} failure, the \checkpoint{} counter is incremented before the finish counter even though the process crashed while executing the doorway in order to maintain the above invariant. 
	    
	    \paragraph{\underline{Phases and strides}}
	    To design a recoverable \GPTD{} routine that can be executed incrementally, we divide the execution of the \GPTD{} routine into \emph{four} phases as follows:
		\begin{itemize}
		\item \emph{Phase 1 (\snapshot{} phase):} the process reads and records the value of the start counter of each process.
		\item \emph{Phase 2 (\rendezvous{} phase):} the process waits for the \checkpoint{} counter of each process to catch up to its start counter.
		\item \emph{Phase 3 (\barrier{} phase):} the process waits for the finish counter of each process to catch up to its start counter.
		\item \emph{Phase 4 (\switch{} phase):} the process switches the role of the two pools.
		\end{itemize}
		
		The second phase is required to achieve the desired fairness guarantee.
		To enable incremental execution of the \GPTD{} routing, we further divide each phase into multiple \emph{strides\footnote{A stride means a long step, which is an apt description of what it denotes in this context.}}. Intuitively, a stride constitutes a ``unit of execution'' of the \GPTD{} routine and incurs $\bigO{1}$ RMRs under both CC and DSM models. The first three phases consist of $\n$ strides each and the fourth phase consist of two strides. Thus the routine as a whole consists of $3\n + 2$ strides. A \emph{single} stride of the \GPTD{} routine is executed by invoking the function \FnAdvanceOne. Each process maintains a stride counter to keep track of the numbers of strides of the \emph{current instance} of the \GPTD{} routine it has executed so far; the counter is reset during the \switch{} phase as the designation of the two pools is flipped. We use $\varstep[i]$ to denote the stride counter of process $p_i$. 
		The pseudocode of the function \FnAdvanceOne{} is shown in \cref{algo:fair|mem_rec}. 
		The function is written in a way such that executing it multiple times with the same value of the stride counter does not cause any undesirable behavior. 
		Further, we consider repeated invocations of \FnAdvanceOne{} function when the stride counter of the invoking process has the same value until the value of the stride counter changes as essentially the same stride.

		\paragraph{\underline{When to execute a stride?}}
		The main idea is to execute one stride of the \GPTD{} routine ``between'' two consecutive \attempt{s}. Specifically, our memory reclamation algorithm satisfies the following property:
		
		\begin{proposition}
		\label{proposition:attempts:stride}
		Suppose a process begins two consecutive \attempt{s} at times $t$ and $t'$ with $t < t'$. Then, it executes at least one stride of the \GPTD{} routine without failing between times $t$ and $t'$. 
		\end{proposition}
		
		To ensure that the currently active pool does not run out of nodes at least until all strides of the \GPTD{} routine have been executed successfully, each pool now consists of $3\n + 2$ nodes. When combined with \cref{proposition:attempts:stride}, this is necessary and sufficient because the \GPTD{} routine consists of $3\n+2$ strides and a process can consume at most one node in an \attempt.

		If fairness can be foregone, then a simple memory reclamation algorithm that meets all other desirable requirements (except for fairness) works as follows. A process executes one stride of the \GPTD{} routine in its \Enter{} \segment{} if it is in quiescent state. Executing a stride of the \GPTD{} routine in quiescent state, specifically after finishing an \attempt{} but before starting a new \attempt, helps to easily avoid a deadlock.

		\begin{comment}
		
		We refer to the resulting weakly recoverable lock as \weakMCSUnfair. Clearly, we have
		
		\begin{theorem}
		\weakMCSUnfair{} satisfies ME, SF, BCSR, BE and BR properties. Further, it has $\bigO{1}$ (worst-case) RMR complexity and $\bigO{\n^2}$ space complexity.             
		\end{theorem}
		
		The BCSR property requires some explanation. 
		If a process crashes inside its \CS{} \segment, then it implies that the process is currently executing a \useful{} \attempt{}. Thus, upon restarting, it will \emph{not} execute a stride of the \GPTD{} routine since it will not be in quiescent state.
		
		\end{comment}

		\subsection{Achieving fairness}
		
	    A fairness property is typically defined with respect to a doorway, which is a \emph{wait-free} piece of code that a process executes at the beginning of an \Enter{} \segment. As described earlier, a process executes a stride of the \GPTD{} routine at the \emph{beginning} of its \Enter{} \segment{} when in a quiescent state before starting a new \attempt. Since a stride may involve blocking (\emph{i.e.}, busy waiting in a loop), it makes it infeasible to define the notion of doorway in the algorithm described so far. 
	    
	    To remedy this shortcoming, we proceed as follows. If an \attempt{} of a process is \useful, then the process executes a stride of the \GPTD{} routine immediately after completing the doorway of \weakMCS{} rather than after completing that \attempt{}. We refer to this stride as \emph{regular stride}. However, if an \attempt{} of a process is \useless, then the process executes a stride of the \GPTD{} routine immediately after completing that \attempt{} as before (in quiescent state). We refer to this stride as \emph{penalty stride}. With this change, it can be verified that \cref{proposition:attempts:stride} still holds.
	    Note that a process executes both types of strides during its \Enter{} \segment. Specifically, a process executes \emph{at most one} regular and \emph{at most one} penalty stride during its \Enter{} \segment{} so that the (worst-case) RMR complexity of a passage remains $\bigO{1}$. 
	    
	    To ensure the BCSR property, we guarantee that a process executes a regular stride during an \attempt{} exactly once. In other words, once it has completed a stride of the \GPTD{} routine, it does not execute another stride during that \attempt{} even it were to fail and start a new passage (but same \attempt); otherwise the process may not be able to reenter its \CS{} \segment{} within a bounded number of its own steps. To that end, whenever it starts a new \attempt, it records the current value of its stride counter and executes a stride later only if the recorded and current values of its stride counter match. 
	    This also helps to avoid a deadlock by guaranteeing that a process that owns a node in a sub-queue never waits on a process that appended its node to the same sub-queue \emph{after} its own. Specifically, we prove later that if a process $p_i$ is waiting on a process $p_j$ (either in the \barrier{} phase of the \GPTD{} routine or in the waiting room of \weakMCS{}) and \attempt{s} of both processes are \useful, then $p_i$ executed its \FAS{} instruction after $p_j$.

     \addprefix[M.]

     \begin{algorithm}[ht!]
            \begin{multicols}{2}

            \SetKw{Shared}{shared variables}
            \SetKw{Local}{local variables}
            \SetKw{Struct}{struct}
            \SetKw{Integer}{int}
            \SetKw{Boolean}{bool}
            \SetKw{Array}{array}
            \SetKw{Await}{await}
            \SetKw{LAnd}{and}
            \SetKwProg{initialization}{initialization}{\\ begin}{end}
            \SetKwProg{function}{Function}{\\ begin}{end}
            
            %% \DontPrintSemicolon
            
            \Shared \\
            \Indp
            $\symfilter{}:$ an instance of \weakMCS\;
            \tcc{the remaining variables are arrays whose $i$-{th} entry is local to process $p_i$}
            \tcc{three counters to keep track of the progress of each process during an attempt}
            \tcc{the number of attempts that each process has started}
            $\varstart$: \Array $[1{\dots}\n]$ of integer variable\;
            \tcc{the number of doorways that each process has completed}
            $\varcheckpoint$: \Array $[1{\dots}\n]$ of \broadcast{} object\;
            \tcc{the number of attempts that each process has completed}
            $\varfinish$: \Array $[1{\dots}\n]$ of \broadcast{} object\;
            \BlankLine\BlankLine
            %%
            %% pool
            %%
            \tcc{two pools of queue nodes (current and backup) for each process}
            $\varpool$: \parbox[t]{2in}{\Array $[1{\dots}\n][0,1][1{\dots}3\n+2]$ of two pools \\ of $3\n+2$ QNode;} \\
            \tcc{the current pool of each process}
            $\varcurrentpool$: \Array $[1{\dots}\n]$ of integer variable\;
            \tcc{the backup pool of each process}
            $\varbackuppool$: \Array $[1{\dots}\n]$ of integer variable\;
            \BlankLine\BlankLine
            %%
            %%
            %% snapshot and related
            %%
            \tcc{stores a snapshot of how many attempts other processes have started as observed by each process}
            $\varsnapshot$: \Array $[1{\dots}\n][1{\dots}\n]$ of integer variable\;
            \tcc{counts the number of steps pertaining to memory reclamation that each process has executed since its last pool switch}
            $\varstep$: \Array $[1{\dots}\n]$ of integer variable\;
			\tcc{flag for each process to indicate if that process should be penalized for crashing}
			$\varpenalty$: \Array $[1{\dots}\n]$ of boolean variable\;
			\tcc{used to determine if a process has executed a step of memory reclamation without failing}
			$\varlaststep$: \Array $[1{\dots}\n]$ of integer variable\;
	        \Indm
            
            \PrintSemicolon
                 
            \columnbreak
            
            \initialization{}{
                \ForEach{$j \in \{ 1, 2, \dots, \n\}$}
                {
                    $\varcurrentpool[j] \leftarrow 0$\;
                    $\varbackuppool[j] \leftarrow 1 - \varcurrentpool[j]$\;
                    $\varstart[j] \leftarrow 0$\;
                    $\varstep[j] \leftarrow 1$\;
    				$\varlaststep[j] \leftarrow \varstep[j]$\;
    				$\varpenalty[j] \leftarrow \false$\;
    				\BlankLine
    				\ForEach{$k \in \{0,1\}$, $\ell \in \{1, \ldots, 3\n+2\}$}
                    {
                       $\varpool[j][k][\ell] \leftarrow$ new QNode\;
                    }
                }
            }
            \BlankLine
            \function{\FnAdvanceOne(~)}
            {
                \tcc{blocking method that executes one stride of memory reclamation}
                $\varindex \leftarrow \varstep[i] \pmod \n$\;
                \BlankLine
            	\uIf{($\varstep[i] \in [1,\n]$)}
            	{
					\label{line:fair|mem_rec:step:snapshot:begin}
            		\tcc{\snapshot{} phase: take a snapshot of the start counters of all processes}
            		%% $\varindex = \varstep[i]$\;
            		$\varsnapshot[i][\varindex] \leftarrow \varstart[\varindex]$\;
            		%% $\varstep[i]$++\;
					\label{line:fair|mem_rec:step:snapshot:end}
					\BlankLine
            	}
            	%% \uElseIf{($\varstep[i] > \n$) \LAnd ($\varstep[i] <= 2\n$)}
            	\uElseIf{($\varstep[i] \in [\n+1,2\n]$) \LAnd ($\varindex \neq i$)}
            	{
					\label{line:fair|mem_rec:step:checkpoint:begin}
            		\tcc{\rendezvous{} phase: wait for other processes to complete their doorway}
            		%% $\varindex = \varstep[i] - \n$\;
            		%% \If{($\varindex \neq i$)}  
            		{
            		    
            			$\varcheckpoint[\varindex]$.\bWait{( $\varsnapshot[i][\varindex]$ )}\;
            		}
            		%% $\varstep[i]$++\;
					\label{line:fair|mem_rec:step:checkpoint:end}
					\BlankLine
            	}
            	\uElseIf{($\varstep[i] \in [2\n+1,3\n]$) \LAnd ($\varindex \neq i$)}
            	{
					\label{line:fair|mem_rec:step:barrier:begin}
            		\tcc{\barrier{} phase: wait for other processes to complete their attempt}
            		%% $\varindex = \varstep[i] - 2\n$\;
            		%% \If{($\varindex \neq i$)}
            		{
            			$\varfinish[\varindex]$.\bWait{( $\varsnapshot[i][\varindex]$ )}\;
            		}
            		%% $\varstep[i]$++\;
					\label{line:fair|mem_rec:step:barrier:end}
					\BlankLine
            	}
            	\ElseIf{($\varstep[i] \in [3\n+1,3\n+2]$)} {
            		\tcc{\switch{} phase: switch current and backup pools}
                	\uIf{($\varstep[i] = 3\n + 1$)}
                	{
    					\label{line:fair|mem_rec:step:switch:begin}
    					\tcc{switch the backup pool}
                		$\varbackuppool[i] \leftarrow \varcurrentpool[i]$\;
    					\label{line:fair|mem_rec:step:switch:backup}
                		%% $\varstep[i]$++\;
                	}
                	\Else
                	{
                		\tcc{switch the current pool}
                		$\varcurrentpool[i] \leftarrow 1 - \varbackuppool[i]$\;
    					\label{line:fair|mem_rec:step:switch:current}
                		%% \tcc{reset step}
                		%% $\varstep[i] \leftarrow 1$\;
    					\label{line:fair|mem_rec:step:switch:end}
                	}
                }
                \BlankLine
                \lIf{($\varstep[i] < 3\n + 2$)}
                {%%
                   $\varstep[i] := \varstep[i] + 1$%%
                   \label{line:fair|mem_rec:step:advance}%%
                } \lElse
                {%%
                   $\varstep[i] \leftarrow 1$%%
                   \label{line:fair|mem_rec:step:reset}%%
                }
            }

            \end{multicols}
            \caption{Pseudocode of process $p_i$ for bounding space complexity of \weakMCS{} described in \autoref{algo:weak|MCS}.}
            \label{algo:fair|mem_rec}
     \end{algorithm}

     \begin{algorithm}[ht!]
            \begin{multicols}{2}
    
            \SetKw{Shared}{shared variables}
            \SetKw{Local}{local variables}
            \SetKw{Struct}{struct}
            \SetKw{Integer}{int}
            \SetKw{Boolean}{bool}
            \SetKw{Array}{array}
            \SetKw{Await}{await}
            \SetKw{LAnd}{and}
            \SetKwProg{initialization}{initialization}{\\ begin}{end}
            \SetKwProg{function}{Function}{\\ begin}{end}

			\function{\FnEnter(~)}
		    {
		    
		        \If(\tcp*[f]{the penalty flag is set}){($\varpenalty[i] = \true$)}
		        {
		            \label{line:fair|mem_rec:enter:penalty:begin}
		            \tcc{execute a stride of the \GPTD{} routine if not done already}
		            \If{($\varstep[i] = \varlaststep[i]$)}
		            {
		                \FnAdvanceOne(~)\tcp*[r]{penalty stride}
		                \label{line:fair|mem_rec:enter:penalty:step}
		            }
		            $\varpenalty[i] \leftarrow \false$\tcp*[r]{reset the penalty flag}
		        }
		        \label{line:fair|mem_rec:enter:penalty:end}
		        \BlankLine
		        \If{($\varstart[i] = \varfinish[i].\bRead{(~)}$)}
            	{
            	    \label{line:fair|mem_rec:enter:attempt|new:begin}
            	    \tcc{start a new attempt}
            	    \tcc{record the stride counter}
            	    $\varlaststep[i] \leftarrow \varstep[i]$\;
            	    \label{line:fair|mem_rec:enter:attempt|new:snapshot}
            	    \tcc{increment the start counter}
            	    $\varstart[i] := \varstart[i] + 1$\;
            	    \label{line:fair|mem_rec:enter:attempt|new:increment}
            	}
            	\label{line:fair|mem_rec:enter:attempt|new:end}
		        \BlankLine
            	Execute the doorway of $\symfilter$  (\crefrange{line:enter:doorway|begin}{line:enter:doorway|end} in \cref{algo:weak|MCS})
            	\label{line:fair|mem_rec:enter:doorway:base}
		        \BlankLine
		        %%
            	%% \If{($\varstart[i] \neq \varcheckpoint[i].\bRead{(~)}$)}
            	%% {
            	%% \label{line:fair|mem_rec:enter:doorway:test}
            	%%
				\tcc{announce completion of the doorway}
            	$\varcheckpoint[i]$.\bSet{( $\varstart[i]$ )}\;
				\label{line:fair|mem_rec:enter:doorway:increment}
				%%
            	%% }
                \label{line:fair|mem_rec:enter:doorway:end}
		        \BlankLine
		        \tcc{execute a stride of the \GPTD{} routine if not done already}
            	\If{($\varstep[i] = \varlaststep[i]$)}
	            {
	               \label{line:fair|mem_rec:enter:regular:begin}
		           \FnAdvanceOne(~)\tcp*[r]{regular stride}
		           \label{line:fair|mem_rec:enter:regular:step}
	            }
	            \label{line:fair|mem_rec:enter:regular:end}
		        \BlankLine
            	Execute the waiting room of $\symfilter$  (\crefrange{line:enter:waitingroom|begin}{line:enter:waitingroom|end} in \cref{algo:weak|MCS})
                \label{line:fair|mem_rec:enter:waitingroom:base}
            	\label{line:fair|mem_rec:enter:end}
			}
            \columnbreak
            \function{\FnNewNode(~)}
            {
             
                \tcc{return a reference to the node pointed to by the stride counter in the active pool}
                \Return $\varpool[i][\varcurrentpool[i]][\varstep[i]]$\;
                \label{line:fair|mem_rec:newnode}
               
            }
            \label{line:fair|mem_rec:newnode:end}
            \BlankLine
            \function{\FnRetireNode(~)}
            {
            
                \tcc{check if the \attempt{} is \inadmissible}
                \If{($\varstart[i] \neq \varcheckpoint[i].\bRead{(~)}$)}
            	{
            	    \label{line:fair|mem_rec:retirenode:inadmissible:begin}
					\tcc{attempt is being aborted due to a crash; set the penalty flag, record the stride counter and fix the \checkpoint{} counter}
            	    $\varpenalty[i] \leftarrow \true$\;
            	    $\varlaststep[i] \leftarrow \varstep[i]$\;
            		$\varcheckpoint[i]$.\bSet{( $\varstart[i]$ )}\;
            	}
            	\label{line:fair|mem_rec:retirenode:inadmissible:end}
            	\BlankLine
            	\tcc{increment the finish counter if needed}
            	%% \If{($\varstart[i] \neq \varfinish[i].\bRead{(~)}$)}
            	%% {   	
            	%% \label{line:fair|mem_rec:retirenode:finish:test}
            	%%
				$\varfinish[i]$.\bSet{( $\varstart[i]$ )};
				%%
            	%% }
            	\label{line:fair|mem_rec:retirenode:finish:increment}
            }
          
        \end{multicols}
        \caption{Pseudocode of process $p_i$ for bounding space complexity of \weakMCS{} described in \autoref{algo:weak|MCS} (continued).}
        \label{algo:fair|mem_rec|2}
     \end{algorithm}

	    We refer to the augmented weakly recoverable lock with bounded space complexity as \weakMCSFair, and
	    its pseudocode is given in \cref{algo:fair|mem_rec,algo:fair|mem_rec|2}. To avoid repetition,  
	    we only provide the pseudocode for \Enter{} \segment{}, \FnNewNode{} function and \FnRetireNode{} function. The pseudocode for \Recover{} and \Exit{} \segment{s} is same as that for \weakMCS{} and has been omitted.
	    
	    First, consider the \Enter{} \segment. A process first executes a penalty stride, if applicable (\crefrange{line:fair|mem_rec:enter:penalty:begin}{line:fair|mem_rec:enter:penalty:end}). It then begins a new \attempt{}, if in quiescent state, and also records the current value of the stride counter  (\crefrange{line:fair|mem_rec:enter:attempt|new:begin}{line:fair|mem_rec:enter:attempt|new:end}). It next executes the doorway of \weakMCS{} (\cref{line:fair|mem_rec:enter:doorway:base}) and increments the \checkpoint{} counter if lagging (\cref{line:fair|mem_rec:enter:doorway:increment}). It then executes a regular stride, if needed (\crefrange{line:fair|mem_rec:enter:regular:begin}{line:fair|mem_rec:enter:regular:end}). Finally, it executes the waiting room of \weakMCS{} (\cref{line:fair|mem_rec:enter:waitingroom:base}).
	    
	    Next, consider the \FnNewNode{} function. We use stride counter as an \emph{index} into the active pool. So an invocation of \FnNewNode{} simply returns the reference to the node stored at the corresponding location in the pool (\cref{line:fair|mem_rec:newnode:end}).
	    
	    Finally, consider the \FnRetireNode{} function. A process checks if the current \attempt{} is \useless{}, and, if so, sets the penalty flag and records the current value of the stride counter (\cref{line:fair|mem_rec:retirenode:inadmissible:begin}). It then increments the finish counter if lagging (\cref{line:fair|mem_rec:retirenode:finish:increment}).

	    We now define the doorway of \weakMCSFair{} as follows. Let $B$ denote the maximum number of steps a process executes in the \Enter{} \segment{} in order to reach \cref{line:fair|mem_rec:enter:doorway:end} provided that the passage is \singular. 
	    Recall that, in a \singular{} passage, the process does not execute a penalty stride.
	    In the absence of penalty stride, the code in the \Enter{} \segment{}  contains one only bounded loop, namely in the \bSet{} operation (\crefrange{line:set:wakeup|chain:loop:begin}{line:set:wakeup|chain:loop:end}).
	    The doorway of \weakMCSFair{} then consists of the first $B$ steps of a process in the \Enter{} \segment{}.

    \subsection{Correctness proofs}
      
      Note that \weakMCSFair{} uses $2\n$ \broadcast{} objects for synchronization among processes given by $\varcheckpoint[i]$ and $\varfinish[i]$ for each $i = 1, 2, \ldots, \n$.
      As such, the correctness of \weakMCS{} depends on the broadcast objects behaving as expected (satisfying safety and liveness properties). In particular, the conditions mentioned in the statements of \cref{theorem:broadcast:safety,theorem:broadcast:liveness} should hold as and when needed. For example, \cref{theorem:broadcast:safety} requires that the (sub)history with respect to a \broadcast{} object should be \wellformed, while \cref{theorem:broadcast:liveness} requires that the (sub)history should be \legal{} as well. 
       
      The correctness proof consists of two parts. We first argue that the conditions required for a \broadcast{} object to behave correctly actually hold. We then argue that \weakMCSFair{} behaves correctly.

      \subsubsection{\broadcast{} objects behave correctly}
      For each $i = 1, 2, \ldots, \n$, process $p_i$ owns the two \broadcast{} objects $\varcheckpoint[i]$ and $\varfinish[i]$. As required, only process $p_i$ invokes \bSet{} operation on $\varcheckpoint[i]$ and $\varfinish[i]$, and, only process $p_j$, where $j \in \{ 1, 2, \ldots, \n \} \setminus \{ i \}$, invokes \bWait{} operation on $\varcheckpoint[i]$ and $\varfinish[i]$. For convenience, 
      \[
        \mathcal{B} = \big\{ \varcheckpoint[i] \mid 1 \leq i \leq \n  \big\} \ \cup \ \big\{ \varfinish[i] \ | \mid 1 \leq i \leq \n  \big\}
      \]
       
       Given a history $H$ and a \broadcast{} object $b \in \mathcal{B}$, we use $H \mid b$ to denote the sub-history that consists of only those steps of $H$ that processes take while executing an instance of one of the three operations of $b$.
       
       We first establish that the assumption postulated in \cref{theorem:broadcast:safety} holds.

		\begin{theorem}
		\label{theorem:memrec:wellformed}
		Let $H$ denote a history  of \weakMCSFair{}. Then, for every \broadcast{} object $b \in \mathcal{B}$, $H \mid b$ is \wellformed{} with respect to $b$.
		\end{theorem}
		\begin{proof}
		Consider a \broadcast{} object $b$ owned by process $p_i$.
		
		The property \ref{enum:broadcast:set:incremental} holds because
		\begin{enumerate*}[label=(\alph*)]
		\item the argument passed to any instance of \bSet{} operation on $\varcheckpoint[i]$ or $\varfinish[i]$ counter is the current value of $\varstart[i]$ counter, and
		\item $\varstart[i]$ counter has monotonically non-decreasing value.
		\end{enumerate*}
		
		The property \ref{enum:broadcast:wait:nearby} holds
		\begin{enumerate*}[label=(\alph*)]
		\item the argument passed to any instance of \bWait{} operation on $\varcheckpoint[i]$ or $\varfinish[i]$ counter is the value of $\varstart[i]$ counter read in the past,
		\item $\varstart[i]$ counter has monotonically non-decreasing value, and
		\item the three counters of a process ($\varstart[i]$, $\varcheckpoint[i]$ and $\varfinish[i]$) satisfy \eqref{eq:counters:invariant:1}.
		\end{enumerate*}
		\end{proof}
		
		We next establish that the assumptions postulated in \cref{theorem:broadcast:liveness} also hold.  To that end, we need to prove  that, \emph{under the assumptions made in the definition of the SF property},
		\begin{enumerate*}[label=(\roman*)]
		\item the (sub)history with respect to a \broadcast{} object is \legal, 
		\item a \checkpoint{} counter value eventually catches up to the corresponding start counter value and, 
		\item a finish counter value eventually catches up to the corresponding start counter value.
        \end{enumerate*}
		We prove each of the three conditions one-by-one.
		
		A line $\ell$ of a function $f$ is said to be \emph{\inevitable} if it satisfies the following property: 
        if a process fails \emph{while} executing the line $\ell$, then, in every invocation of the function $f$ by the (same) process thereafter, the line $\ell$ \emph{lies} on \emph{all} possible execution paths that can be taken by the process inside the function $f$, until the execution of the line has been \complete[d].
        Typically, a line is considered to have \complete[d] if the execution of the line was failure-free, except in some cases as follows.
        We say that \cref{line:fair|mem_rec:enter:penalty:step} or  \cref{line:fair|mem_rec:enter:regular:step} of the \FnEnter{} function has 
        been \complete[d] if the value of the stride counter is modified inside the function.

        \begin{proposition} 
        \label{proposition:inevitable:1}
        \Cref{line:fair|mem_rec:enter:penalty:step,line:fair|mem_rec:enter:regular:step} of the \FnEnter{} function are \inevitable.
        \end{proposition}
        
        \begin{proposition} 
        \label{proposition:inevitable:2}
        \Cref{line:fair|mem_rec:step:barrier:end,line:fair|mem_rec:step:checkpoint:end} of the \FnAdvanceOne{} function are \inevitable.
        \end{proposition}
      
        We are now ready to prove the desired results.
		
		\begin{theorem}
		Assume that every process fails only a finite number of times in each of its super-passage. Let $H$ denote an infinite fair history  of \weakMCSFair{}. Then,  for every \broadcast{} object $b \in \mathcal{B}$, $H \mid b$ is \legal{} with respect to $b$.       
		\end{theorem}
		\begin{proof}
		Consider a \broadcast{} object $b$ owned by process $p_i$.
		
		The property \ref{enum:broadcast:finite:terminated} holds because, in an infinite fair history, a process can stop taking steps only \emph{after} executing a failure-free passage.
		
		The properties \ref{enum:broadcast:run:finite} and  \ref{enum:broadcast:last:failure-free} hold because, as implied by \cref{proposition:inevitable:1,proposition:inevitable:2}, if a process fails during an instance of \bWait($x$) operation, then it is guaranteed to repeatedly invoke instances of \bWait($x$) until the stride counter has changed. Moreover, a process can fail only a finite number of times during its super-passage, thereby implying that any \run{} of \bWait{} operation in $H$ is of finite length and ends with a failure-free instance. 
		\end{proof}

		\begin{theorem}
		\label{theorem:checkpoint:start}
		Assume that every process fails only a finite number of times in each of its super-passage. Let $H$ denote an infinite fair history  of \weakMCSFair{} and consider a process $p_i$. Then, 
	    for every positive value $x$ assumed by $\varstart[i]$ counter in $H$,  there exists at least one instance of \bSet($x$) operation on $\varcheckpoint[i]$ in $H$ that \complete[s].
		\end{theorem}	
		\begin{proof}
		After a process \emph{begins} an \attempt, it has to execute only a bounded number of steps to increment its checkpoint counter unless it fails. However, by assumption a process can fail only a finite number of times during each of its super-passage and hence during each of its \attempt.
		\end{proof}

    	The proof for the last assumption (the finish counter value eventually catches up to the corresponding start counter value) is more involved since a process may have to wait \emph{after} incrementing the \checkpoint{} counter value but \emph{before} incrementing the finish counter value.
    	We classify a \emph{wait-for dependency} into three types: 
    	\begin{enumerate}[label=(T\arabic*)]
    	\item Waiting for a \checkpoint{} counter value to catch up to the associated start counter value (\cref{line:fair|mem_rec:step:checkpoint:end} executed as part of penalty stride or regular stride).
    	\label{enum:waiting:1}
    	\item Waiting for a finish counter value to catch up to the associated start counter value (\cref{line:fair|mem_rec:step:barrier:end} executed as part of penalty stride or regular stride).
    	\label{enum:waiting:2}
    	\item Waiting inside the waiting-room as part of \weakMCS{} (\cref{line:fair|mem_rec:enter:waitingroom:base}).
    	\label{enum:waiting:3}
    	\end{enumerate}

        \Cref{theorem:checkpoint:start} implies that waiting of type \cref{enum:waiting:1} is finite.
        So, we focus on the other two types of waiting and prove that they are finite as well.
        Note that, if an \attempt{} of a process is rendered \useless, then the process eventually abandons the queue node it allocated at the beginning of the \attempt{} and increments its finish counter for that \attempt. Therefore, we only focus on \useful{} \attempt{s} and argue that every \useful{} \attempt{} eventually completes.  To that end,  we are only concerned with the situation in which a process $p$ is waiting on another process $q$ such that:
	    \begin{enumerate}[label=(A\arabic*)]
	    \item Both $p$ and $q$ are currently executing a \useful{} \attempt.
	    \label{enum:deadlock:admissible}
	    \item Both $p$ and $q$ have appended their respective nodes in the queue successfully using an \FAS{} instruction.
	    \label{enum:deadlock:appended}
	    \end{enumerate}
        
        The next lemma proves a crucial property common to both types of waiting.
        
        \begin{lemma}
        \label{lemma:waiting:FAS:before}
        If a process $p$ is waiting on another process $q$ such that \ref{enum:deadlock:admissible} and \ref{enum:deadlock:appended} hold, then $p$ executed its most recent \FAS{} instruction \emph{after} that of $q$. 
        \end{lemma}
        \begin{proof}
        Clearly, the statement holds if $p$ is waiting for $q$ in the waiting room of \weakMCS{}. Thus, it is sufficient to focus on the case in which $p$ is waiting for $q$ in the \barrier{} phase of a penalty or regular stride. Let the \useful{} \attempt{s} of $p$ and $q$ be denoted by $A_p$ and $A_q$, respectively.
        We define four instants of time as follows:
           
           \begin{itemize}[label=$\triangleright$]
              \item $t_q$ is the time when $q$ executed the \FAS{} instruction for $A_q$
              \item  $t_c$ is the time when $p$ completed its stride for the first time during which it waited for the \checkpoint{} counter of $q$ to catch up to its start counter for $A_q$
              \item $t_a$ is the time when $p$ started the \attempt{} $A_p$
              \item $t_p$ is the time when $p$ executed the \FAS{} instruction for $A_p$
           \end{itemize}
           
           We show that $t_q < t_c < t_a < t_p$, thereby proving that $t_q < t_p$.
           
           \begin{itemize}[label=$\triangleright$]
              \item $t_q < t_c$ because a process executes the \FAS{} instruction \emph{before} incrementing the \checkpoint{} counter for a \admissible{} \attempt{}
              \item $t_c < t_a$ because a process executes only \emph{one} regular stride of \GPTD{} routine during a \useful{} \attempt{} (and, in the regular stride for $A_p$, $p$ is waiting for the finish counter of $q$ to catch up to its start counter for $A_q$)
              \item $t_a < t_p$ because a process executes the \FAS{} instruction \emph{after} starting an \attempt{}
           \end{itemize} 
           
           This proves that the statement holds.
        \end{proof}

        \begin{comment}

        The above lemma implies that, in any chain of waiting processes such that \ref{enum:deadlock:admissible} and \ref{enum:deadlock:appended} hold along every edge in the chain, the first process in the chain executed its most recent \FAS{} instruction after that of the last process. 
        %%
        This, in turn, guarantees that processes cannot be involved in a circular wait. Moreover, from \cref{theorem:memrec:legal} and \cref{theorem:broadcast:liveness}, and using the SF property of the \weakMCS{}, we know that every waiting process will eventually get unblocked. Therefore, we have:
      
        \end{comment}        
                
        Using the above lemma, we are now able to show the following. 
        
       \begin{theorem}
	   \label{theorem:attempt:finite}
	   Assume that every process fails only a finite number of times in each of its super-passage. Let $H$ denote an infinite fair history  of \weakMCSFair{}. Then, every \attempt{} in $H$ eventually completes.
	   \end{theorem}
       \begin{proof}
       Let $\mathcal{I}(t)$ denote the set of \attempt{s} in $H$ that execute their \FAS{} instructions at or before time $t$ and have not completed by time $t$.

       Consider an arbitrary \attempt, say $A$ in $H$. Clearly, a \useless{} \attempt{} has no busy-waiting loop and, thus, is guaranteed to complete. Therefore, assume that $A$ is a \useful{} \attempt. Let $p$ denote the process to which $A$ belongs, and let $t_A$ denote the time when $p$ performs the \FAS{} instruction during $A$ to append its node to the queue. 
       
       Consider $\mathcal{I}(t_A)$. We can order all \attempt{s} in $\mathcal{I}(t_A)$ based on the sequence in which their \FAS{} instructions are executed.  \Cref{lemma:waiting:FAS:before} implies that, after time $t$, an \attempt{} in $\mathcal{I}(t_A)$ can only busy-wait on another \attempt{} in $\mathcal{I}(t_A)$. We can prove using induction (on the order in which \FAS{} instructions are performed) that every \attempt{} in $\mathcal{I}(t_0)$ eventually completes.
      
       Since $A$ was chosen arbitrarily, we can infer that every \attempt{} in $H$ eventually completes.
       \end{proof}

       Clearly, it follows that:
       
	   \begin{corollary}
	   \label{corollary:finish:start}
	   Assume that every process fails only a finite number of times in each of its super-passage. Let $H$ denote an infinite fair history  of \weakMCSFair{} and consider a process $p_i$. Then, 
	   for every positive value $x$ assumed by $\varstart[i]$ counter in $H$,  there exists at least one instance of \bSet($x$) operation on $\varfinish[i]$ in $H$ that \complete[s].
	   \end{corollary}	
       
       In other words, all three assumptions required for \cref{theorem:broadcast:liveness} to apply hold.

		\subsubsection{\weakMCSFair{} behaves correctly}
		We only focus on SF and BCSR properties of the RME problem because the proofs for other properties (ME, BE, BR and \CIFCFS) are almost identical to those for \weakMCS.
		
	    To establish the SF property, we first prove the following lemma about a penalty stride.

       \begin{lemma}
       \label{lemma:fair|mem_rec:penalty:finite}
	   Assume that every process fails only a finite number of times in each of its super-passage. Let $H$ denote an infinite fair history  of \weakMCSFair{} and consider a process $p_i$. Then every penalty stride executed by a process in $H$ eventually terminates.
	   \end{lemma}
	   \begin{proof}
	   A process executing a penalty stride may either wait in the \rendezvous{} phase or the \barrier{} phase of the \GPTD{} routine. The first type of wait is finite due to \cref{theorem:checkpoint:start}, while the second type due to \cref{corollary:finish:start}. 
	   \end{proof}

         Thus we have:
         
		\begin{lemma}
		\weakMCSFair{} satisfies the SF property.
		\end{lemma}
		\begin{proof}
		When executing a passage, a process  may wait at three different points:
		\begin{enumerate*}[label=(\alph*)]
		\item while performing a penalty stride (\cref{line:fair|mem_rec:enter:penalty:step}),
		\item while performing a regular stride (\cref{line:fair|mem_rec:enter:regular:step}), or
		\item while executing the waiting room of \weakMCS{} (\cref{line:fair|mem_rec:enter:waitingroom:base})
		\end{enumerate*}
		\Cref{lemma:fair|mem_rec:penalty:finite} establishes that the first type of waiting is finite. \Cref{theorem:attempt:finite} establishes that the other two types of waiting are also finite.
		\end{proof}
		
		\begin{lemma}
		\weakMCSFair{} satisfies the BCSR property.
		\end{lemma}
		\begin{proof}
		   If a process crashes inside its critical section, then it implies that the process is currently executing a \useful{} \attempt. We show that, upon restarting, it does not execute any stride of the \GPTD{} routine---penalty or regular.
		   
		   It does not execute a penalty stride because it is not in a quiescent state. It does not execute a regular stride because it would have already executed one prior to entering the critical section for the first time during this \attempt{}. By design, it does not execute another regular stride until it completes the current \attempt{} by executing the \Exit{} \segment.
		\end{proof}
		
		\begin{lemma}
		\weakMCSFair{} satisfies the \CIFCFS{} property.       
		\end{lemma}
		
		It now follows that:
		
	    \begin{theorem}
		\weakMCSFair{} satisfies ME, SF, BCSR, BE, BR and \CIFCFS{} properties. Further, it has $\bigO{1}$ (worst-case) RMR complexity and $\bigO{\n^2}$ space complexity.             
		\end{theorem}

	 \begin{theorem}
     The space complexity of \balock{} is given by $\bigO{\nicefrac{n^2 \log \n}{\log \log \n} + S(\n)}$, where $S(\n)$ denotes the space complexity of the \base{}  \nalock{} for $\n$ processes.
    \end{theorem}

    \begin{corollary}
     Assume that we use \KM{'s} RME algorithm~\cite{KatMor:2020:OPODIS} to implement 
    the \nalock. 
    Then the space complexity of \balock{} is given by $\bigO{\nicefrac{n^2 \log \n}{\log \log \n}}$. 
    \end{corollary}

   \section{Related Work}
   \label{sec:related}
    
    Bohannon \emph{et al.}~\cite{BohLie+:1995:SPDP,BohLie+:1996:SPDP} were the first ones to investigate the RME problem. However, their system model is different from the one assumed in this work. Specifically, in their system model, at least one process is reliable while other processes may be unreliable. Once an unreliable process fails, it never restarts. The reliable process is responsible for continuously monitoring the health of all other processes, and, upon detecting
    that an unreliable process has failed during its passage, it performs recovery by ``fixing'' the lock. The two RME algorithms differ in the way they implement the lock; the one in~\cite{BohLie+:1995:SPDP} uses test-and-set instruction whereas the one in~\cite{BohLie+:1996:SPDP} uses MCS queue-based algorithm.
    
    Golab and Ramaraju formally defined the RME problem in~\cite{GolRam:2016:PODC}. 
    %%
    %% Their formalization, especially the system model, has served as the basis for the 
    %% subsequent work in this area, including this work. 
    %%
    We use the same system model as in their work.
    In~\cite{GolRam:2016:PODC}, Golab and Ramaraju also presented four different RME algorithms---a 2-process RME algorithm  and three $\n$-process RME algorithms. The first algorithm is based on Yang and Anderson's lock \cite{YanAnd:1995:DC}, and is used as a building block to design an $n$-process RME algorithm. Both RME algorithms use only read, write and comparison-based primitives. The worst-case RMR complexity of the 2-process algorithm is $\bigO{1}$ whereas that of the resultant $\n$-process algorithm is $\bigO{\log \n}$. Both RME algorithms have optimal RMR complexity because, as 
    shown in~\cite{AttHen+:2008:STOC, AndKim:2002:DC, YanAnd:1995:DC}, any mutual exclusion algorithm that uses only read, write and comparison-based primitives has worst-case RMR complexity of $\bigOmega{\log \n}$.  The remaining two algorithms are \fadaptive{} (with $f(x) = x$) and \cadaptive{} (with $g(x) = x$), respectively (where $f$ and $g$ are as per the definitions of adaptivity and boundedness respectively from \autoref{sec:model|problem}).
    
     Later, Golab and Hendler \cite{GolHen:2017:PODC} proposed an RME algorithm with sub-logarithmic RMR complexity of $\bigO{\nicefrac{\log \n}{\log \log \n}}$ under the CC model using MCS queue based lock~\cite{MelSco:1991:TrCS} as a building block. Note that MCS uses \FAS{} instruction, which is \emph{not} a comparison-based RMW instruction, and thus the result does not violate the previously mentioned lower bound. Their algorithm does not satisfy the bounded exit property. Moreover, it has been shown to be vulnerable to starvation~\cite{JayJay+:2019:PODC}.

    Ramaraju showed in~\cite{Ram:2015:Thesis} that it is possible to design an RME algorithm with $\bigO{1}$ RMR complexity provided the hardware provides a special RMW instruction to swap the contents of two arbitrary locations in memory atomically. Unfortunately,  at present, no hardware supports such an instruction to our knowledge.
    
    In~\cite{JayJos:2017:DISC}, Jayanti and Joshi presented an RME algorithm with $\bigO{\log{n}}$ RMR complexity.  Their algorithm satisfies bounded (wait-free) exit and FCFS (first-come-first-served) properties.
    
    In~\cite{JayJay+:2019:PODC}, \JJJ{}  proposed an RME algorithm 
    that has sub-logarithmic RMR complexity of $\bigO{\nicefrac{\log n}{\log \log n}}$. To our knowledge, this is the best known RME algorithm  as far as the worst-case RMR complexity is concerned that also satisfies bounded recovery and bounded exit properties.
    
   Using a weaker version of starvation freedom, Chan and Woelfel \cite{ChaWoe:2020:PODC} present a novel solution to the RME problem that incurs a constant number of RMRs in the amortized case, but its worst case RMR complexity may be unbounded.
   
   %However, the worst case RMR complexity of a passage in their algorithm depends on the number of failures, which may grow unboundedly. Moreover, as acknowledged by the authors, their algorithm does not satisfy the bounded (stronger) starvation freedom property.  In other words, it is possible for a (slow) process to be starved indefinitely even though every process only crashes finitely many times during its super passage. Lastly, their algorithm uses an infinite array, which makes it unsuitable even for those languages that implement their own garbage collector (\emph{e.g.}, Java). It is not clear how it can be modified to use only an array of bounded size while maintaining constant RMR amortized complexity.
    
    In~\cite{GolHen:2018:PODC}, Golab and Hendler proposed an RME algorithm under the assumption of system-wide failure (all processes fail and restart) with $\bigO{1}$ RMR complexity.

    A useful extension to the RME problem is when a process may decide to \emph{abort} its request for critical section; this extension is referred to as the \emph{abortable RME problem}. Recently, 
    Katzan and Morrison~\cite{KatMor:2020:OPODIS} and Jayanti and Joshi~\cite{JayJos:2019:NETSYS} have proposed efficient algorithms for solving the problem under both CC and DSM models. The algorithm by Jayanti and Joshi uses $f$-arrays and has $\bigO{\log \n}$ RMR complexity~\cite{JayJos:2019:NETSYS}. On the other hand, the algorithm by Katzan and Morrison uses a $k$-port abortable RME algorithm as a building block to design a sub-logarithmic 
    abortable RME algorithm with RMR complexity of $\bigO{\nicefrac{\log \n}{\log \log \n}}$~\cite{KatMor:2020:OPODIS}.
    
    Recently, Chan and Woelfel have  proved a lower bound of $\bigOmega{\nicefrac{\log \n}{\log \log \n}}$ on the RMR complexity of any RME algorithm using currently available hardware instructions and practical word size of $\Theta(\log \n)$~\cite{ChaWoe:2021:PODC}.

        \section{Conclusions and Future Work}
        \label{sec:conclusion}

            In this work, we have described a general framework to transform any non-adaptive RME algorithm into a \sadaptive{} one without increasing its worst-case RMR complexity. In addition to the hardware instructions used by the underlying non-adaptive RME algorithm, our framework uses \CAS{} and \FAS{} RMW instructions, both of which are commonly available on most modern processors.  
            When applied to the non-adaptive RME algorithm proposed by \JJJ{} in~\cite{JayJay+:2019:PODC} or \KM{} in~\cite{KatMor:2020:OPODIS}, it yields a \sefficient{} RME algorithm that is simultaneously adaptive to 
            \begin{enumerate*}[label=(\alph*)]
            \item the number of processes competing for the lock, \emph{as well as}
            \item the number of failures that have occurred in the recent past,
            \end{enumerate*}
            while having the same (asymptotic) worst-case RMR complexity as that of the base RME algorithm.
            We have also shown that the RME algorithm obtained by applying our framework is fair and satisfies a variant of the FCFS property, even if the base RME algorithm is unfair.

            To make the framework practical, we have described an extension to our framework to reclaim the memory of shared objects when no longer needed. Our memory reclamation algorithm bounds the worst-case space complexity of the framework, while maintaining all its desirable properties at the same time.
            Our approach is general enough that it can be applied to other RME algorithms as well to bound their space complexity, such as \JJJ's algorithm in \cite{JayJay+:2019:PODC}.
        
            In our framework, a failed process, upon restarting, attempts to reacquire all the locks at every level it had advanced to, beginning from level one. As a result, the worst-case RMR complexity of a super-passage is  given by $\bigO{\F_0 \cdot \min \{\pc, \sqrt{\F+1}, \nicefrac{\log \n}{\log \log \n} \}}$, where $\F_0$ denotes the number of times the process fails while executing its (own) super-passage. However, we can modify our framework to allow a process to keep track of its last level. With this modification, the worst case RMR complexity of a super passage
            reduces to $\bigO{\F_0 + \min \{\pc, \sqrt{\F+1}, \nicefrac{\log \n}{\log \log \n} \}}$.
        
            In the future, we plan to explore ways to design a more \responsive{} and space-efficient \filter{} lock. We also plan to develop efficient recoverable algorithms for important variants of the ME problem including read-write mutual exclusion (RWME) and group mutual exclusion (GME).

\begin{acks}

This work was supported, in part, by the 
\grantsponsor{nsf}{National Science Foundation (NSF)}{} under 
grant number \grantnum{nsf}{CNS-1619197}.

\end{acks}

% Bibliography
\bibliographystyle{ACM-Reference-Format}
\bibliography{Citations,References}


\begin{thebibliography}{27}

%%% the Web address.

\ifx \showCODEN    \undefined \def \showCODEN     #1{\unskip}     \fi
\ifx \showDOI      \undefined \def \showDOI       #1{#1}\fi
\ifx \showISBNx    \undefined \def \showISBNx     #1{\unskip}     \fi
\ifx \showISBNxiii \undefined \def \showISBNxiii  #1{\unskip}     \fi
\ifx \showISSN     \undefined \def \showISSN      #1{\unskip}     \fi
\ifx \showLCCN     \undefined \def \showLCCN      #1{\unskip}     \fi
\ifx \shownote     \undefined \def \shownote      #1{#1}          \fi
\ifx \showarticletitle \undefined \def \showarticletitle #1{#1}   \fi
\ifx \showURL      \undefined \def \showURL       {\relax}        \fi
% invisible to TeX
\providecommand\bibfield[2]{#2}
\providecommand\bibinfo[2]{#2}
\providecommand\natexlab[1]{#1}
\providecommand\showeprint[2][]{arXiv:#2}

\bibitem[\protect\citeauthoryear{AMD}{AMD}{2019}]%
        {AMD64Manual}
AMD \bibinfo{year}{2019}\natexlab{}.
\newblock \bibinfo{booktitle}{\emph{{AMD64 Architecture Programmer's Manual
  Volume 3: General Purpose and System Instructions}}}.
\newblock AMD.
\newblock
\urldef\tempurl%
\url{https://www.amd.com/system/files/TechDocs/24594.pdf}
\showURL{%
\tempurl}


\bibitem[\protect\citeauthoryear{Anderson and Kim}{Anderson and Kim}{2002}]%
        {AndKim:2002:DC}
\bibfield{author}{\bibinfo{person}{J.~H. Anderson} {and} \bibinfo{person}{Y.-J.
  Kim}.} \bibinfo{year}{2002}\natexlab{}.
\newblock \showarticletitle{{An Improved Lower Bound for the Time Complexity of
  Mutual Exclusion}}.
\newblock \bibinfo{journal}{\emph{Distributed Computing (DC)}}
  \bibinfo{volume}{15}, \bibinfo{number}{4} (\bibinfo{date}{Dec.}
  \bibinfo{year}{2002}), \bibinfo{pages}{221--253}.
\newblock
\showISSN{0178-2770}
\urldef\tempurl%
\url{https://doi.org/10.1007/s00446-002-0084-2}
\showDOI{\tempurl}


\bibitem[\protect\citeauthoryear{Arcangeli, Cao, McKenney, and Sarma}{Arcangeli
  et~al\mbox{.}}{2003}]%
        {ArcCao+:2003:ATC}
\bibfield{author}{\bibinfo{person}{A. Arcangeli}, \bibinfo{person}{M. Cao},
  \bibinfo{person}{P.~E. McKenney}, {and} \bibinfo{person}{D. Sarma}.}
  \bibinfo{year}{2003}\natexlab{}.
\newblock \showarticletitle{{Using Read-Copy-Update Techniques for System V IPC
  in the Linux 2.5 Kernel}}. In \bibinfo{booktitle}{\emph{USENIX Annual
  Technical Conference, FREENIX Track}}. \bibinfo{pages}{297--309}.
\newblock


\bibitem[\protect\citeauthoryear{Attiya, Hendler, and Woelfel}{Attiya
  et~al\mbox{.}}{2008}]%
        {AttHen+:2008:STOC}
\bibfield{author}{\bibinfo{person}{H. Attiya}, \bibinfo{person}{D. Hendler},
  {and} \bibinfo{person}{P. Woelfel}.} \bibinfo{year}{2008}\natexlab{}.
\newblock \showarticletitle{{Tight RMR Lower Bounds for Mutual Exclusion and
  Other Problems}}. In \bibinfo{booktitle}{\emph{Proceedings of the 40th Annual
  ACM Symposium on Theory of Computing (STOC)}}. \bibinfo{publisher}{ACM},
  \bibinfo{address}{New York, NY, USA}, \bibinfo{pages}{217--226}.
\newblock
\showISBNx{978-1-60558-047-0}
\urldef\tempurl%
\url{https://doi.org/10.1145/1374376.1374410}
\showDOI{\tempurl}


\bibitem[\protect\citeauthoryear{Bohannon, Lieuwen, and Silberschatz}{Bohannon
  et~al\mbox{.}}{1996}]%
        {BohLie+:1996:SPDP}
\bibfield{author}{\bibinfo{person}{P. Bohannon}, \bibinfo{person}{D. Lieuwen},
  {and} \bibinfo{person}{A. Silberschatz}.} \bibinfo{year}{1996}\natexlab{}.
\newblock \showarticletitle{{Recovering Scalable Spin Locks}}. In
  \bibinfo{booktitle}{\emph{Proceedings of the 7th IEEE Symposium on Parallel
  and Distributed Processing (SPDP)}}. \bibinfo{publisher}{IEEE Computer
  Society}, \bibinfo{address}{Washington, DC, USA}, \bibinfo{pages}{314--322}.
\newblock
\showISBNx{0-8186-7683-3}
\urldef\tempurl%
\url{http://dl.acm.org/citation.cfm?id=829517.830751}
\showURL{%
\tempurl}


\bibitem[\protect\citeauthoryear{Bohannon, Lieuwen, Silberschatz, Sudarshan,
  and Gava}{Bohannon et~al\mbox{.}}{1995}]%
        {BohLie+:1995:SPDP}
\bibfield{author}{\bibinfo{person}{P. Bohannon}, \bibinfo{person}{D. Lieuwen},
  \bibinfo{person}{A. Silberschatz}, \bibinfo{person}{S. Sudarshan}, {and}
  \bibinfo{person}{J. Gava}.} \bibinfo{year}{1995}\natexlab{}.
\newblock \showarticletitle{{Recoverable User-level Mutual Exclusion}}. In
  \bibinfo{booktitle}{\emph{Proceedings of the 7th IEEE Symposium on Parallel
  and Distributed Processing (SPDP)}}. \bibinfo{publisher}{IEEE Computer
  Society}, \bibinfo{address}{Washington, DC, USA}, \bibinfo{pages}{293--301}.
\newblock
\showISBNx{0-8186-7195-5}
\urldef\tempurl%
\url{http://dl.acm.org/citation.cfm?id=829516.830651}
\showURL{%
\tempurl}


\bibitem[\protect\citeauthoryear{Chan and Woelfel}{Chan and Woelfel}{2021}]%
        {ChaWoe:2021:PODC}
\bibfield{author}{\bibinfo{person}{D. Chan} {and} \bibinfo{person}{P.
  Woelfel}.} \bibinfo{year}{2021}\natexlab{}.
\newblock \showarticletitle{{A Tight Lower Bound for the RMR Complexity of
  Recoverable Mutual Exclusion}}. In \bibinfo{booktitle}{\emph{Proceedings of
  the 40th ACM Symposium on Principles of Distributed Computing (PODC)}}.
  \bibinfo{publisher}{Association for Computing Machinery (ACM)}.
\newblock


\bibitem[\protect\citeauthoryear{Chan and Woelfel}{Chan and Woelfel}{2020}]%
        {ChaWoe:2020:PODC}
\bibfield{author}{\bibinfo{person}{D.~Y.~C. Chan} {and} \bibinfo{person}{P.
  Woelfel}.} \bibinfo{year}{2020}\natexlab{}.
\newblock \showarticletitle{{Recoverable Mutual Exclusion with Constant
  Amortized RMR Complexity from Standard Primitives}}. In
  \bibinfo{booktitle}{\emph{Proceedings of the 39th ACM Symposium on Principles
  of Distributed Computing (PODC)}}. \bibinfo{publisher}{Association for
  Computing Machinery (ACM)}, \bibinfo{address}{New York, NY, USA}, 10.
\newblock


\bibitem[\protect\citeauthoryear{Dijkstra}{Dijkstra}{1965}]%
        {Dij:1965:CACM}
\bibfield{author}{\bibinfo{person}{E.~W. Dijkstra}.}
  \bibinfo{year}{1965}\natexlab{}.
\newblock \showarticletitle{{Solution of a Problem in Concurrent Programming
  Control}}.
\newblock \bibinfo{journal}{\emph{Communications of the ACM (CACM)}}
  \bibinfo{volume}{8}, \bibinfo{number}{9} (\bibinfo{year}{1965}),
  \bibinfo{pages}{569}.
\newblock


\bibitem[\protect\citeauthoryear{Dvir and Taubenfeld}{Dvir and
  Taubenfeld}{2017}]%
        {DviTau:2018:DISC}
\bibfield{author}{\bibinfo{person}{R. Dvir} {and} \bibinfo{person}{G.
  Taubenfeld}.} \bibinfo{year}{2017}\natexlab{}.
\newblock \showarticletitle{{Mutual Exclusion Algorithms with Constant RMR
  Complexity and Wait-Free Exit Code}}. In
  \bibinfo{booktitle}{\emph{Proceedings of the 21st International Conference on
  Principles of Distributed Systems (OPODIS)}},
  \bibfield{editor}{\bibinfo{person}{James Aspnes}, \bibinfo{person}{Alysson
  Bessani}, \bibinfo{person}{Pascal Felber}, {and} \bibinfo{person}{Jo{\~a}o
  Leit{\~a}o}} (Eds.), Vol.~\bibinfo{volume}{95}. \bibinfo{publisher}{Schloss
  Dagstuhl--Leibniz-Zentrum fuer Informatik}, \bibinfo{address}{Dagstuhl,
  Germany}, \bibinfo{pages}{17:1--17:16}.
\newblock
\showISBNx{978-3-95977-061-3}
\showISSN{1868-8969}
\urldef\tempurl%
\url{https://doi.org/10.4230/LIPIcs.OPODIS.2017.17}
\showDOI{\tempurl}


\bibitem[\protect\citeauthoryear{Fraser}{Fraser}{2004}]%
        {Fra:2004:PhD}
\bibfield{author}{\bibinfo{person}{K. Fraser}.}
  \bibinfo{year}{2004}\natexlab{}.
\newblock \emph{\bibinfo{title}{{Practical Lock-Freedom}}}.
\newblock \bibinfo{thesistype}{Ph.D. Dissertation}. \bibinfo{school}{University
  of Cambridge}.
\newblock


\bibitem[\protect\citeauthoryear{Golab and Hendler}{Golab and Hendler}{2017}]%
        {GolHen:2017:PODC}
\bibfield{author}{\bibinfo{person}{W. Golab} {and} \bibinfo{person}{D.
  Hendler}.} \bibinfo{year}{2017}\natexlab{}.
\newblock \showarticletitle{{Recoverable Mutual Exclusion in Sub-Logarithmic
  Time}}. In \bibinfo{booktitle}{\emph{Proceedings of the ACM Symposium on
  Principles of Distributed Computing (PODC)}}. \bibinfo{publisher}{ACM},
  \bibinfo{address}{New York, NY, USA}, \bibinfo{pages}{211--220}.
\newblock
\showISBNx{978-1-4503-4992-5}
\urldef\tempurl%
\url{https://doi.org/10.1145/3087801.3087819}
\showDOI{\tempurl}


\bibitem[\protect\citeauthoryear{Golab and Hendler}{Golab and Hendler}{2018}]%
        {GolHen:2018:PODC}
\bibfield{author}{\bibinfo{person}{W. Golab} {and} \bibinfo{person}{D.
  Hendler}.} \bibinfo{year}{2018}\natexlab{}.
\newblock \showarticletitle{{Recoverable Mutual Exclusion Under System-Wide
  Failures}}. In \bibinfo{booktitle}{\emph{Proceedings of the ACM Symposium on
  Principles of Distributed Computing (PODC)}}. \bibinfo{publisher}{ACM},
  \bibinfo{address}{New York, NY, USA}, \bibinfo{pages}{17--26}.
\newblock
\showISBNx{978-1-4503-5795-1}
\urldef\tempurl%
\url{https://doi.org/10.1145/3212734.3212755}
\showDOI{\tempurl}


\bibitem[\protect\citeauthoryear{Golab, Hendler, and Woelfel}{Golab
  et~al\mbox{.}}{2006}]%
        {GolHen+:2006:PODC}
\bibfield{author}{\bibinfo{person}{W. Golab}, \bibinfo{person}{D. Hendler},
  {and} \bibinfo{person}{P. Woelfel}.} \bibinfo{year}{2006}\natexlab{}.
\newblock \showarticletitle{An $O(1)$ RMRs Leader Election Algorithm}. In
  \bibinfo{booktitle}{\emph{Proceedings of the 25th ACM Symposium on Principles
  of Distributed Computing (PODC)}}. \bibinfo{publisher}{Association for
  Computing Machinery (ACM)}, \bibinfo{address}{New York, NY, USA},
  \bibinfo{pages}{238–--247}.
\newblock
\showISBNx{1595933840}
\urldef\tempurl%
\url{https://doi.org/10.1145/1146381.1146417}
\showDOI{\tempurl}


\bibitem[\protect\citeauthoryear{Golab and Ramaraju}{Golab and
  Ramaraju}{2016}]%
        {GolRam:2016:PODC}
\bibfield{author}{\bibinfo{person}{W. Golab} {and} \bibinfo{person}{A.
  Ramaraju}.} \bibinfo{year}{2016}\natexlab{}.
\newblock \showarticletitle{{Recoverable Mutual Exclusion: [Extended
  Abstract]}}. In \bibinfo{booktitle}{\emph{Proceedings of the ACM Symposium on
  Principles of Distributed Computing (PODC)}}. \bibinfo{publisher}{ACM},
  \bibinfo{address}{New York, NY, USA}, \bibinfo{pages}{65--74}.
\newblock
\showISBNx{978-1-4503-3964-3}
\urldef\tempurl%
\url{https://doi.org/10.1145/2933057.2933087}
\showDOI{\tempurl}


\bibitem[\protect\citeauthoryear{Golab and Ramaraju}{Golab and
  Ramaraju}{2019}]%
        {GolRam:2019:DC}
\bibfield{author}{\bibinfo{person}{W. Golab} {and} \bibinfo{person}{A.
  Ramaraju}.} \bibinfo{year}{2019}\natexlab{}.
\newblock \showarticletitle{{Recoverable Mutual Exclusion}}.
\newblock \bibinfo{journal}{\emph{Distributed Computing (DC)}}
  \bibinfo{volume}{32}, \bibinfo{number}{6} (\bibinfo{date}{Nov.}
  \bibinfo{year}{2019}), \bibinfo{pages}{535--564}.
\newblock


\bibitem[\protect\citeauthoryear{Hart, McKenney, Brown, and Walpole}{Hart
  et~al\mbox{.}}{2007}]%
        {HarMcK+:2007:JPDC}
\bibfield{author}{\bibinfo{person}{T.~E. Hart}, \bibinfo{person}{P.~E.
  McKenney}, \bibinfo{person}{A.~D. Brown}, {and} \bibinfo{person}{J.
  Walpole}.} \bibinfo{year}{2007}\natexlab{}.
\newblock \showarticletitle{{Performance of Memory Reclamation for Lockless
  Synchronization}}.
\newblock \bibinfo{journal}{\emph{Journal of Parallel and Distributed Computing
  (JPDC)}} \bibinfo{volume}{67}, \bibinfo{number}{12} (\bibinfo{year}{2007}),
  \bibinfo{pages}{1270--1285}.
\newblock


\bibitem[\protect\citeauthoryear{Intel}{Intel}{2016}]%
        {Intel64Manual}
Intel \bibinfo{year}{2016}\natexlab{}.
\newblock \bibinfo{booktitle}{\emph{{Intel 64 and IA-32 Architectures Software
  Developer's Manual, Volume 2A: Instruction Set Reference, A-M}}}.
\newblock Intel.
\newblock
\urldef\tempurl%
\url{https://software.intel.com/sites/default/files/managed/a4/60/325383-sdm-vol-2abcd.pdf}
\showURL{%
\tempurl}


\bibitem[\protect\citeauthoryear{Jayanti, Jayanti, and Joshi}{Jayanti
  et~al\mbox{.}}{2019}]%
        {JayJay+:2019:PODC}
\bibfield{author}{\bibinfo{person}{P. Jayanti}, \bibinfo{person}{S. Jayanti},
  {and} \bibinfo{person}{A. Joshi}.} \bibinfo{year}{2019}\natexlab{}.
\newblock \showarticletitle{{A Recoverable Mutex Algorithm with Sub-logarithmic
  RMR on Both CC and DSM}}. In \bibinfo{booktitle}{\emph{Proceedings of the ACM
  Symposium on Principles of Distributed Computing (PODC)}}.
  \bibinfo{publisher}{ACM}, \bibinfo{address}{New York, NY, USA},
  \bibinfo{pages}{177--186}.
\newblock
\showISBNx{978-1-4503-6217-7}
\urldef\tempurl%
\url{https://doi.org/10.1145/3293611.3331634}
\showDOI{\tempurl}


\bibitem[\protect\citeauthoryear{Jayanti and Joshi}{Jayanti and Joshi}{2017}]%
        {JayJos:2017:DISC}
\bibfield{author}{\bibinfo{person}{P. Jayanti} {and} \bibinfo{person}{A.
  Joshi}.} \bibinfo{year}{2017}\natexlab{}.
\newblock \showarticletitle{{Recoverable FCFS Mutual Exclusion with Wait-Free
  Recovery}}. In \bibinfo{booktitle}{\emph{Proceedings of the 31st Symposium on
  Distributed Computing (DISC)}},
  \bibfield{editor}{\bibinfo{person}{Andr{\'e}a~W. Richa}} (Ed.),
  Vol.~\bibinfo{volume}{91}. \bibinfo{publisher}{Schloss
  Dagstuhl--Leibniz-Zentrum fuer Informatik}, \bibinfo{address}{Dagstuhl,
  Germany}, \bibinfo{pages}{30:1--30:15}.
\newblock
\showISBNx{978-3-95977-053-8}
\showISSN{1868-8969}
\urldef\tempurl%
\url{https://doi.org/10.4230/LIPIcs.DISC.2017.30}
\showDOI{\tempurl}


\bibitem[\protect\citeauthoryear{Jayanti and Joshi}{Jayanti and Joshi}{2019}]%
        {JayJos:2019:NETSYS}
\bibfield{author}{\bibinfo{person}{P. Jayanti} {and} \bibinfo{person}{A.
  Joshi}.} \bibinfo{year}{2019}\natexlab{}.
\newblock \showarticletitle{Recoverable Mutual Exclusion with Abortability}. In
  \bibinfo{booktitle}{\emph{Proceedings of the International Conference on
  Networked Systems (NETSYS)}}. \bibinfo{publisher}{Springer International
  Publishing}, \bibinfo{pages}{217--232}.
\newblock


\bibitem[\protect\citeauthoryear{Katzan and Morrison}{Katzan and
  Morrison}{2021}]%
        {KatMor:2020:OPODIS}
\bibfield{author}{\bibinfo{person}{D. Katzan} {and} \bibinfo{person}{A.
  Morrison}.} \bibinfo{year}{2021}\natexlab{}.
\newblock \showarticletitle{{Recoverable, Abortable, and Adaptive Mutual
  Exclusion with Sublogarithmic RMR Complexity}}. In
  \bibinfo{booktitle}{\emph{Proceedings of the 24th International Conference on
  Principles of Distributed Systems (OPODIS)}} \emph{(\bibinfo{series}{Leibniz
  International Proceedings in Informatics (LIPIcs)})},
  \bibfield{editor}{\bibinfo{person}{Q.~Bramas}, \bibinfo{person}{R.~Oshman},
  {and} \bibinfo{person}{P.~Romano}} (Eds.), Vol.~\bibinfo{volume}{184}.
  \bibinfo{publisher}{Schloss Dagstuhl--Leibniz-Zentrum f{\"u}r Informatik},
  \bibinfo{address}{Dagstuhl, Germany}, \bibinfo{pages}{15:1--15:16}.
\newblock
\showISBNx{978-3-95977-176-4}
\showISSN{1868-8969}
\urldef\tempurl%
\url{https://doi.org/10.4230/LIPIcs.OPODIS.2020.15}
\showDOI{\tempurl}


\bibitem[\protect\citeauthoryear{McKenney and Slingwine}{McKenney and
  Slingwine}{1998}]%
        {McKSli:1998:ICPDCS}
\bibfield{author}{\bibinfo{person}{P.~E. McKenney} {and} \bibinfo{person}{J.~D.
  Slingwine}.} \bibinfo{year}{1998}\natexlab{}.
\newblock \showarticletitle{{Read-Copy Update: Using Execution History to Solve
  Concurrency Problems}}. In \bibinfo{booktitle}{\emph{Proceedings of the
  IASTED International Conference on Parallel and Distributed Computing and
  Systems}}. \bibinfo{pages}{509--518}.
\newblock


\bibitem[\protect\citeauthoryear{Mellor-Crummey and Scott}{Mellor-Crummey and
  Scott}{1991}]%
        {MelSco:1991:TrCS}
\bibfield{author}{\bibinfo{person}{J.~M. Mellor-Crummey} {and}
  \bibinfo{person}{M.~L. Scott}.} \bibinfo{year}{1991}\natexlab{}.
\newblock \showarticletitle{{Algorithms for Scalable Synchronization on
  Shared-Memory Multiprocessors}}.
\newblock \bibinfo{journal}{\emph{ACM Transactions on Computer Systems}}
  \bibinfo{volume}{9}, \bibinfo{number}{1} (\bibinfo{date}{Feb.}
  \bibinfo{year}{1991}), \bibinfo{pages}{21--65}.
\newblock
\showISSN{0734-2071}
\urldef\tempurl%
\url{https://doi.org/10.1145/103727.103729}
\showDOI{\tempurl}


\bibitem[\protect\citeauthoryear{Narayanan and Hodson}{Narayanan and
  Hodson}{2012}]%
        {NarHod:2010:ASPLOS}
\bibfield{author}{\bibinfo{person}{D. Narayanan} {and} \bibinfo{person}{O.
  Hodson}.} \bibinfo{year}{2012}\natexlab{}.
\newblock \showarticletitle{{Whole-System Persistence}}. In
  \bibinfo{booktitle}{\emph{Proceedings of the International Conference on
  Architectural Support for Programming Languages and Operating Systems
  (ASPLOS)}}. \bibinfo{publisher}{ACM}, \bibinfo{address}{New York, NY, USA},
  \bibinfo{pages}{401--410}.
\newblock


\bibitem[\protect\citeauthoryear{Ramaraju}{Ramaraju}{2015}]%
        {Ram:2015:Thesis}
\bibfield{author}{\bibinfo{person}{A. Ramaraju}.}
  \bibinfo{year}{2015}\natexlab{}.
\newblock \emph{\bibinfo{title}{{RGLock: Recoverable Mutual Exclusion for
  Non-Volatile Main Memory Systems}}}.
\newblock \bibinfo{thesistype}{Master's\ thesis}. \bibinfo{school}{Electrical
  and Computer Engineering Department, University of Waterloo}.
\newblock
\urldef\tempurl%
\url{http://hdl.handle.net/10012/9473}
\showURL{%
\tempurl}


\bibitem[\protect\citeauthoryear{Yang and Anderson}{Yang and Anderson}{1995}]%
        {YanAnd:1995:DC}
\bibfield{author}{\bibinfo{person}{J.-H. Yang} {and} \bibinfo{person}{J.~H.
  Anderson}.} \bibinfo{year}{1995}\natexlab{}.
\newblock \showarticletitle{{A Fast, Scalable Mutual Mxclusion Algorithm}}.
\newblock \bibinfo{journal}{\emph{Distributed Computing (DC)}}
  \bibinfo{volume}{9}, \bibinfo{number}{1} (\bibinfo{date}{March}
  \bibinfo{year}{1995}), \bibinfo{pages}{51--60}.
\newblock
\showISSN{1432-0452}
\urldef\tempurl%
\url{https://doi.org/10.1007/BF01784242}
\showDOI{\tempurl}


\end{thebibliography}

\end{document}